\documentclass[traditabstract]{aa} 
\usepackage[pdftex]{graphicx}
\usepackage{breqn}
\usepackage[flushleft]{threeparttable}
\usepackage{subfig}

\def\la{\lower.5ex\hbox{$\; \buildrel < \over \sim \;$}}
\def\ga{\lower.5ex\hbox{$\; \buildrel > \over \sim \;$}}

\newcommand{\HI}{\mbox{H{\sc i}}}

\begin{document}
   \title{A Virgo Environmental Survey Tracing Ionised Gas Emission (VESTIGE). VIII. Modeling ram pressure stripping of diffuse gas in the Virgo cluster spiral galaxy NGC~4330\thanks{Based on observations obtained with MegaPrime/MegaCam, a joint project of CFHT and CEA/DAPNIA, at the Canada-French-Hawaii Telescope (CFHT) which is operated by the National Research Council(NRC) of Canada, the Institut National des Sciences de l’Univers of the Centre National de la Recherche Scientifique (CNRS) of France and the University of Hawaii.}}

   \author{B.~Vollmer\inst{1}, M.~Fossati\inst{2, 3}, A.~Boselli\inst{4}, M.~Soida\inst{5}, S.~Gwyn\inst{6}, J.C.~Cuillandre\inst{7}, Ph.~Amram\inst{4}, S.~Boissier\inst{4}, M.~Boquien\inst{8}, G.~Hensler\inst{9}}

   \institute{Universit\'e de Strasbourg, CNRS, Observatoire astronomique de Strasbourg, UMR 7550, F-67000 Strasbourg, France \and 
          Dipartimento di Fisica G. Occhialini, Universit{\`a} degli Studi di Milano Bicocca, Piazza della Scienza 3, I-20126 Milano, Italy \and
          Institute for Computational Cosmology and Center for Extragalactic Astronomy, Durham University, South Road, Durham DH1 3LE, UK \and
          Aix Marseille Universit\'e, CNRS, CNES, LAM, Marseille, France \and
          Astronomical Observatory, Jagiellonian University, Krak\'ow, Poland \and
          NRC Herzberg Astronomy and Astrophysics, 5071 West Saanich Road, Victoria, BC V9E 2E7, Canada \and
          AIM, CEA, CNRS, Universit\'e Paris-Saclay, Universit\'e Paris Diderot, Sorbonne Paris Cit\'e, Observatoire de Paris, PSL University, 91191 Gif-sur-Yvette Cedex, France \and
          Centro de Astronomía (CITEVA), Universidad de Antofagasta, Avenida Angamos 601, Antofagasta 1270300, Chile \and
          Department for Astrophysics, University of Vienna, T\"urkenschanzstrasse 17, A-1180 Vienna, Austria
          }

   \date{Received ; accepted }


  \abstract
{
NGC~4330 is one of the Virgo galaxies whose UV emission distribution shows a tail structure. An associated tail structure is also
observed in the H{\sc i} and H$\alpha$ emission distributions. Previous dynamical modeling showed that the galaxy is approaching
the cluster center and is therefore undergoing increasing ram pressure stripping. 
Recent stellar population fitting of deep optical spectra together with multiband photometry lead to the determination of the time
when star formation was quenched in the galactic disk. We introduce a new version of the dynamical model that includes not only the dense neutral gas,
but also the diffuse ionized gas and aim to reproduce the H{\sc i}, H$\alpha$, UV distributions together with
the star formation histories of the outer gas-free parts of the galactic disk.
The results of $50$ simulations with five different Lorentzian temporal ram-pressure profiles and five different delays between the simulation onset and
peak ram pressure are presented. The delays were introduced to study the influence of galactic structure on the outcome of the simulations.
The inclusion of diffuse gas stripping changes significantly the H{\sc i}, UV, and H$\alpha$ emission distributions.
The simulations with diffuse gas stripping naturally lead to vertical low surface density filaments in the downwind region of the galactic disk. 
These filaments occur less frequently in the simulations without diffuse gas stripping. The simulations with diffuse gas stripping lead to better 
joint fits to the SEDs and optical spectra. The H{\sc i}, NUV, and H$\alpha$ morphologies of the model snapshots which best reproduce the SEDs and optical 
spectra are sufficiently different to permit a selection of best-fit models. We conclude that the inclusion of diffuse gas stripping
significantly improves the resemblance between the model and observations. Our preferred model yields a time to peak ram pressure of $140$~Myr in the future.
The spatial coincidence of the radio continuum and diffuse H$\alpha$ tails suggests that both gas phases are stripped together.
We suggest that the star formation in the outer tail is sporadic and low level and this explains the absence of a significant amount of cosmic ray electrons there.
Furthermore, we suggest that the mixed ISM is ionized by collisions with the thermal electrons of the ambient ICM which confines the filaments.
}


   \authorrunning{Vollmer et al.}
   \titlerunning{Modeling ram pressure stripping of diffuse gas in NGC~4330}

   \maketitle
%

\section{Introduction\label{sec:introduction}}

Once a galaxy has entered a galaxy cluster, its evolution will change, sometimes drastically.
Multiple gravitational interactions together with the influence of the cluster potential
can remove the stellar and gaseous content of its outer disk (galaxy harassment; Moore et al. 1996).
The rapid motion of the galaxy within the cluster atmosphere, the hot (temperature $T \sim 10^7$~K) and tenuous 
(density $n \sim 10^{-4}$~cm$^{-3}$) intracluster medium (ICM), causes the removal of the outer gas
disk via ram pressure. In contrast to galaxy harassment, ram pressure stripping does not
affect the stellar content of the galaxy. Spiral galaxies which underwent or undergo ram pressure stripping 
show a truncated gas disk together with a symmetric stellar disk (e.g., VIVA Chung et al. 2009).
If the interaction is ongoing, a gas tail mainly detected in H{\sc i} is present (e.g., Chung et al. 2007).
The gas truncation radius is set by the galaxy's closest passage to the cluster center via the
criterion introduced by Gunn \& Gott (1972):
\begin{equation}
\rho_{\rm ICM} v_{\rm gal}^2 = \pi \Sigma v_{\rm rot}^2/R \ ,
\label{eq:rps}
\end{equation}
where $\rho_{\rm ICM}$ is the ICM density, $v_{\rm gal}$ the galaxy velocity, $\Sigma$ the
surface density of the interstellar medium (ISM), $v_{\rm rot}$ the rotation velocity of the galaxy,
and $R$ the stripping radius. If peak ram pressure occurred more than $400$-$500$~Myr ago,
the gas tails have disappeared and the truncated gas disk has become symmetric again (Vollmer 2009).

Under extreme conditions, ram-pressure stripped galaxies often show extraplanar, one-sided optical and UV emission 
and important tails of ionized gas (e.g., Yagi et al. 2010, 2017; Poggianti et al. 2017). Because of their optical appearance these
objects are called jellyfish-galaxies. 
Most of the H$\alpha$ emission in these tails is due to photoionization by massive stars born in situ in the 
tails (Poggianti et al. 2019). Often extraplanar molecular gas is found in jellyfish galaxies (Jachym et al. 2014, 2019;
Moretti et al. 2018), from which the ionizing stars are formed.

Once the gas has been removed from the outer galactic disk under the action of ram pressure, the corresponding stellar
content remains gas and dust-free and gradually stops star formation. Optical spectroscopy together with multiwavelength
photometry can be used to determine the stellar populations and thus the star formation history of the
ram-pressure stripped parts of the galactic disk (Boselli et al. 2006, Crowl et al. 2008). 
These star formation histories can then be compared to those of dynamical models.
For the Virgo cluster galaxy NGC~4388 an observed quenching timescale of $\sim 200$~Myr was determined by
Pappalardo et al. (2010) based on a VLT spectrum and optical/UV photometry. An updated dynamical model
including ram pressure stripping was able to reproduce the observed quenching timescale together with
additional observational characteristics (Vollmer et al. 2018): H{\sc i} distribution and velocity field, FUV, H$\alpha$,
and polarized radio continuum emission distribution. A realistic model has to fulfill all these
observational constraints. From the model the ram pressure temporal profile, time to peak ram 
pressure, and the angle between the galaxy velocity within the cluster and its disk plane can be determined.  
A model-based time series of ram pressure-stripped galaxies was established by Vollmer (2009).
The Virgo spiral galaxy NGC~4330 is part of this time series. 

The interstellar medium (ISM) of a spiral galaxy consists of different phases: (i) cool dense molecular gas  ($T \sim 10$-$30$~K, $n \ga 100$~cm$^{-1}$),
(ii) cold neutral hydrogen ($T \sim 100$~K, $n \sim 100$~cm$^{-1}$), (iii) warm neutral hydrogen ($T \sim 8000$~K, $n \sim 1$~cm$^{-1}$), and
(iv) warm diffuse ionized hydrogen ($T \sim 8000$~K, $n \sim 0.4$~cm$^{-1}$). Since the acceleration of a gas cloud by ram pressure
is inversely proportional to the gas surface density (Eq.~\ref{eq:rps}), it is expected that the different phases are stripped with different efficiencies.
Indeed, the stripped diffuse ionized gas is accelerated to higher velocities and is therefore observed at larger distances from the host galaxy
than the dense gas (Vollmer et al. 2009, Moretti et al. 2018). On the other hand, the stripped dense CO-emitting gas seems to be 
well mixed with the warm H{\sc i} (Vollmer et al. 2008, 2012; J{\'a}chym et al. 2014; Nehlig et al. 2016), 
which is consistent with a scenario where the neutral turbulent 
ISM is stripped as an entity. We therefore expect that the distribution of the stripped diffuse ionized gas is more extended than that
of the neutral gas. The detection of this gas in deep H$\alpha$ observations represents an important additional constraint on
simulations of ram-pressure stripped galaxies. An additional constraint on the simulations comes from the star formation history of stripped 
gas-free regions of the stellar disk.
Detailed simulations of a ram-pressure stripped galaxy can thus not only give access to the interaction parameters (see Vollmer 2009),
but also give insight into the reaction of the different gas phases to ram pressure.

In this article we (i) test a new version of the dynamical model that includes not only the dense neutral gas,
but also the diffuse ionized gas and apply it to NGC~4330 and (ii) aim to reproduce the observed H{\sc i}, H$\alpha$, NUV distributions of NGC~4330
together with the star formation histories of the outer gas-free parts of the galactic disk. This article represents a continuation and update of
the work of Vollmer et al. (2012).

\section{NGC~4330}

\begin{figure*}[!ht]
  \centering
   \subfloat[][]{\includegraphics[width=.4\textwidth]{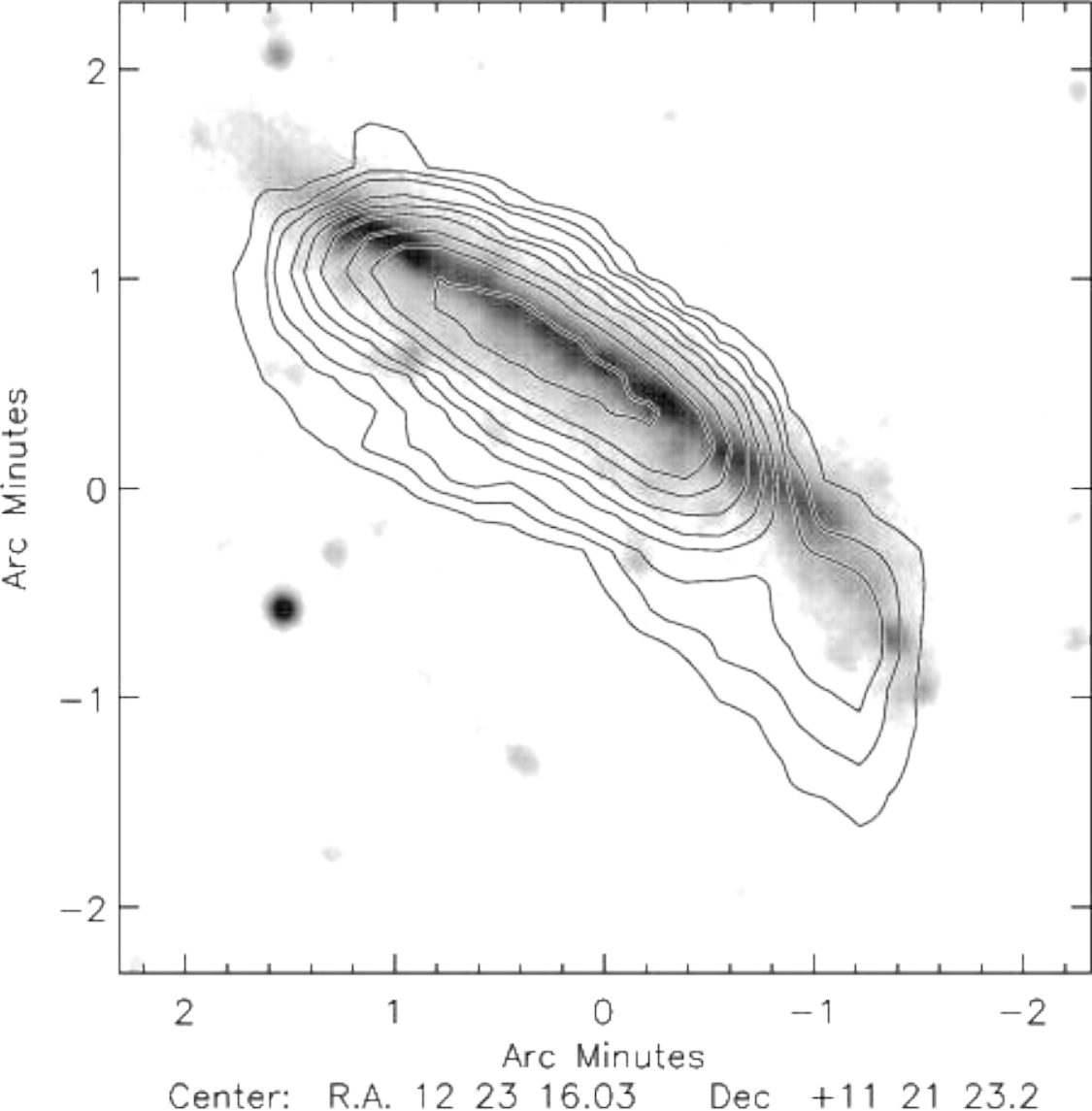}}\quad
   \put(-70,190){\large HI on NUV}
   \subfloat[][]{\includegraphics[width=.4\textwidth]{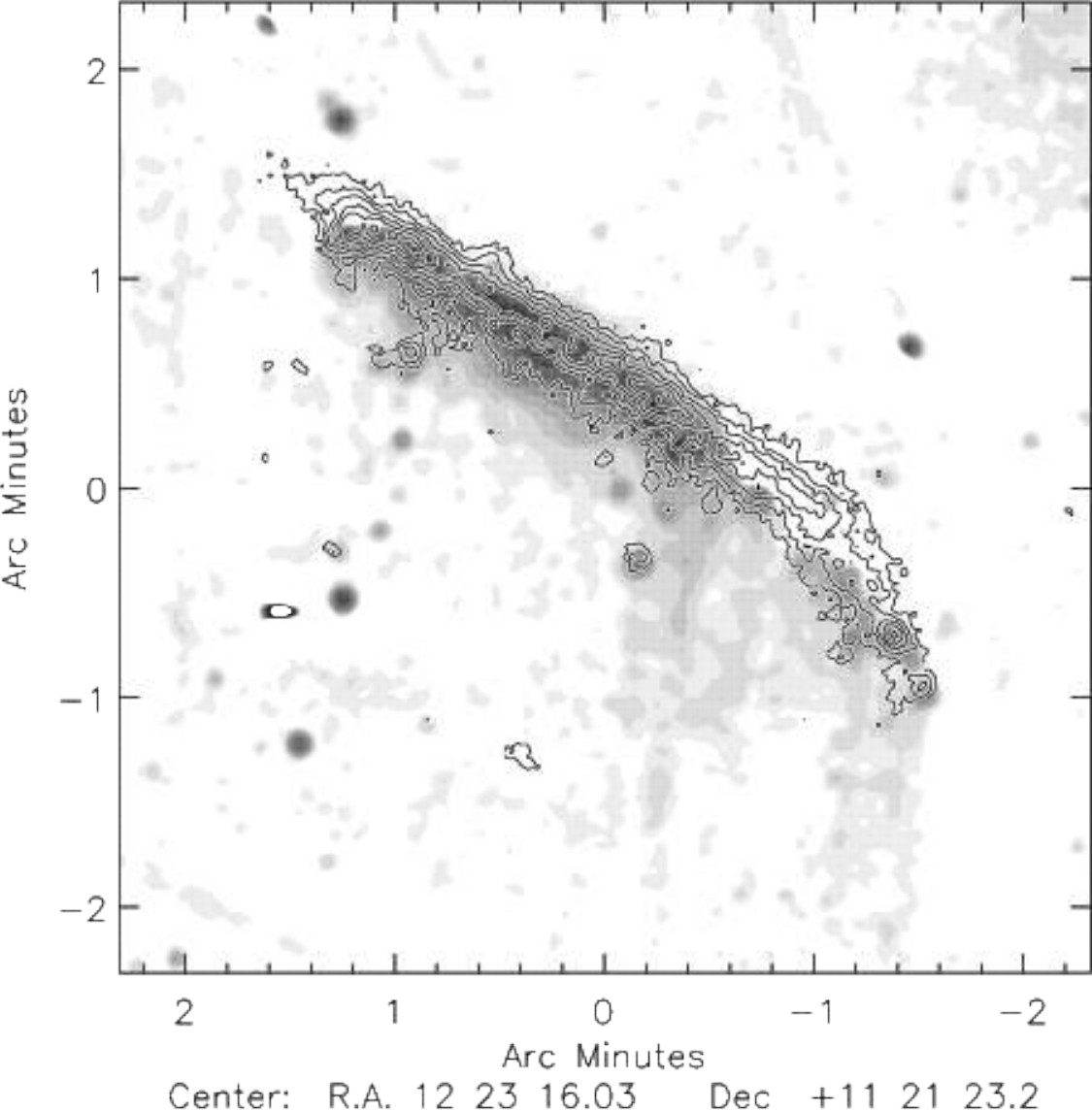}}\quad
   \put(-70,190){\large FUV on H$\alpha$}\\
   \subfloat[][]{\includegraphics[width=.4\textwidth]{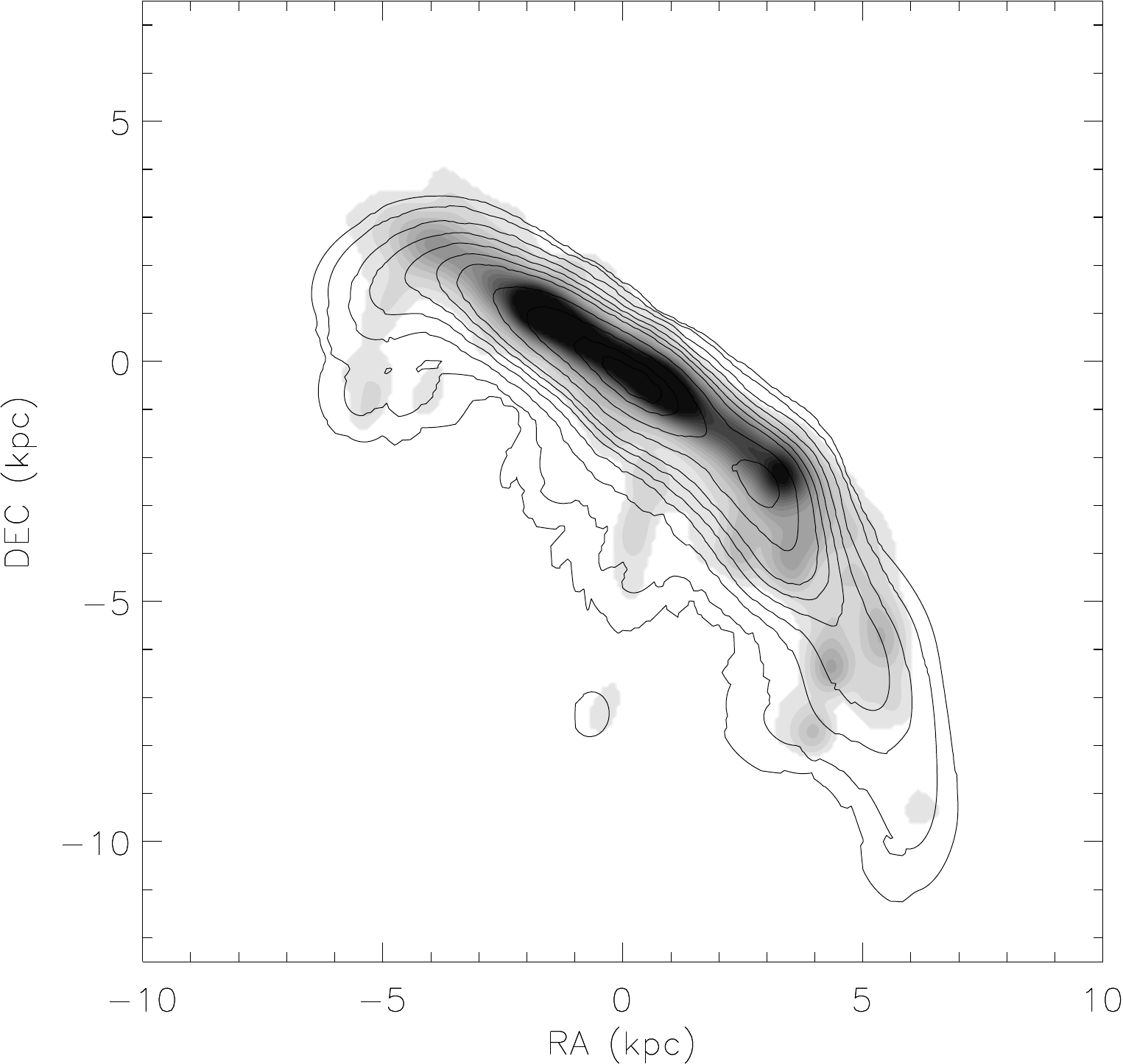}}
   \put(-100,180){\large model HI on NUV}
   \subfloat[][]{\includegraphics[width=.4\textwidth]{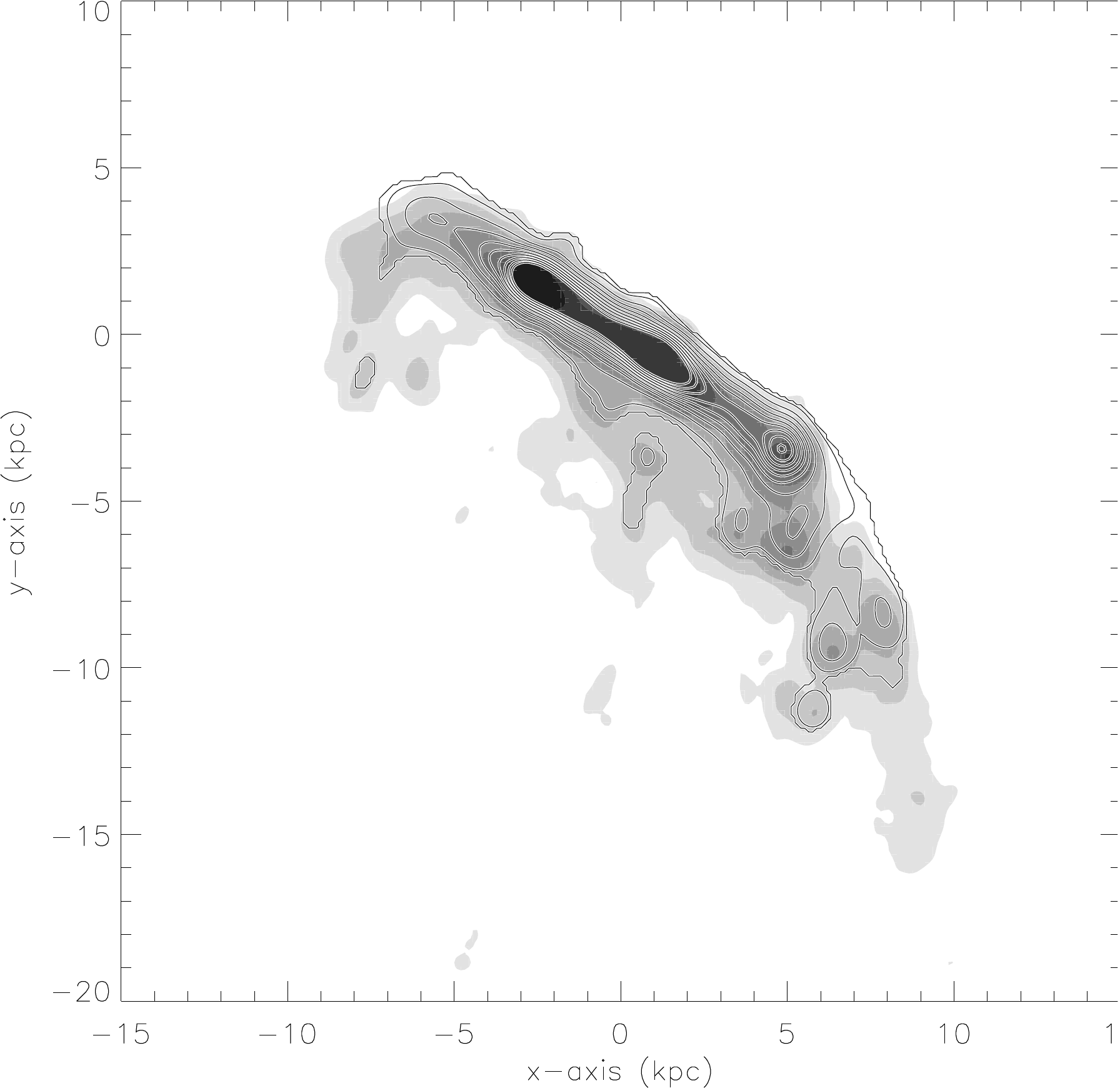}}
   \put(-100,180){\large model FUV on H$\alpha$}
  \caption{NGC~4330. Upper left panel: VIVA H{\sc i} emission distribution (contours) on the GALEX NUV image.
    Upper right panel: GALEX FUV emission distribution (contours) on the VESTIGE continuum subtracted H$\alpha$+[N{\sc ii}] image.
    Lower left panel: model H{\sc i} emission distribution (contours) on the model extinction-free NUV image. 
    Lower right panel: model FUV emission distribution (contours) on the model extinction-free H$\alpha$ image.
  \label{fig:Halpha_HI}}
\end{figure*}

NGC~4330 is one of the Virgo spiral galaxies with a long H{\sc i} tail observed by Chung et al. (2007).
It has a maximum rotation velocity of $v_{\rm rot} \sim 180$~km\,s$^{-1}$ and a total H{\sc i} mass of
$M_{\rm HI}=4.5 \times 10^{8}$~M$_{\odot}$ (Chung et al. 2009). NGC~4330 is located at a projected distance of
$\sim 2^{\circ}$ (600~kpc) from the cluster center, i.e. it is relatively close to M~87, and has
a radial velocity of 400~km\,s$^{-1}$ with respect to the Virgo cluster mean.   
The H{\sc i} deficiency of NGC~4330, which is defined as the logarithm of the ratio of
the \HI\ content of a field galaxy of same morphological type and diameter
to the observed \HI\ mass, is $0.8$ (Chung et al. 2007), i.e. the galaxy has lost about $80$\,\% of its atomic hydrogen.
The H{\sc i} distribution in the galactic disk (upper left panel of Fig.~\ref{fig:Halpha_HI}) is truncated at about half the optical radius. 
In addition, NGC~4330 is one of the rare Virgo galaxies showing an extended UV tail (Abramson et al. 2011). 
The H{\sc i} and UV tails show a significant offset, with the H{\sc i} tail being downwind of the UV tail.
At the leading edge of the interaction, the H$\alpha$ emission and dust extinction distribution is bent sharply
out of the galactic disk. These features are signs of active ram pressure stripping. 
Roediger \& Hensler (2005) elaborated the subsequent stripping phases for face-on ram pressure stripping.
NGC~4330 undergoes the instantaneous stripping phase where the outer part of the gas disk is displaced but only partially unbound.
This leads to straightly bent but coherent outer gas disk.

Vollmer et al. (2012) presented a detailed dynamical model for NGC~4330.
Their best-fit model qualitatively reproduced the observed projected position, the radial velocity of the galaxy, the molecular and 
atomic gas distribution and velocity field, and the UV distribution in the region where a gas tail is present. 
In contrast to other Virgo spiral galaxies affected by ram pressure stripping, NGC~4330 does not show an asymmetric ridge of polarized radio continuum 
emission.  This is due to the particular projection of NGC~4330 where the compressed large-scale magnetic fields are 
oriented along the light-of-sight. 
Vollmer et al. (2012) concluded that  the  ISM is  stripped  as  a  whole  entity  and  that  only  a  tiny  fraction  of dense clouds can be left behind.
Extraplanar star formation proceeds in dense gas arms pushed by ram pressure. The collapsing clouds decouple rapidly from the ram pressure wind and the young, 
UV-emitting stars stay behind the gas.

On the basis of dynamical models, the galaxy moves to the north and still approaches the cluster center with the closest approach occurring 
in $\sim 100$~Myr. Despite the success of their model, Vollmer et al. (2012) noted that the observed red UV color on the windward side was 
not reproduced by the model. Moreover, their simulations could only produce the model distribution of H{\sc ii} regions, i.e. compact
regions of high ionized gas density and high H$\alpha$ surface brightness (Fig.~9 of Vollmer et al. 2012). 
Since the diffuse ionized gas is much less dense than the gas in the H{\sc ii} regions, it has a much lower surface brightness and can
only be detected through deep narrow-band observations. Such observations for NGC~4330 are only available now.
Therefore, the model of Vollmer et al. (2012) was upgraded to enable the simulation of diffuse ionized gas stripping.

Within the Virgo Environmental Survey Tracing Ionised Gas Emission (VESTIGE; Boselli et al. 2018), Fossati et al. (2018) detected a low surface brightness 
$10$~kpc H$\alpha$ tail exhibiting a peculiar filamentary structure. In addition, these authors
collected literature photometry in 15~bands from the far-UV to the far-IR and VLT FORS2 deep optical long-slit spectroscopy. 
Using a Monte Carlo code that jointly fits spectroscopy and photometry, they reconstructed the star formation histories in apertures along the major 
axis of NGC~4330. In this way they found a clear outside-in gradient with radius of the time when the ram pressure induced star formation quenching 
started: the outermost radii were stripped about $500$~Myr ago, while the stripping reached the inner $5$~kpc from the center in the last $100$~Myr. 
In addition, Fossati et al. (2018) discovered a low surface brightness $10$~kpc tail of ionized gas exhibiting a peculiar filamentary structure
(upper right panel of Fig.~\ref{fig:Halpha_HI}). 
The ionized gas tail is associated to the H{\sc i} tail, but extends further to the south.

\section{The dynamical model\label{sec:model}}

We used the N-body code described in Vollmer et al. (2001), which consists of 
two components: a non-collisional component
that simulates the stellar bulge/disk and the dark halo, and a
collisional component that simulates the ISM as an ensemble of gas clouds.
A scheme for star formation was implemented, where stars were formed
during cloud collisions and then evolved as non-collisional particles.

\subsection{Halo, stars, and gas} 

The non--collisional component consists of 81\,920 particles, which simulate
the galactic halo, bulge, and disk.
The characteristics of the different galactic components are shown in
Table~\ref{tab:param}.
\begin{table}
      \caption{Total mass, number of collisionless particles $N$, particle mass $M$, and smoothing
        length $l$ for the different galactic components.}
         \label{tab:param}
      \[
         \begin{array}{lllll}
           \hline
           \noalign{\smallskip}
           {\rm component} & M_{\rm tot}\ ({\rm M}$$_{\odot}$$)& N & M\ ({\rm M}$$_{\odot}$$) & l\ ({\rm pc}) \\
           \hline
           {\rm halo} & 1.7 \times 10$$^{11}$$ & 32768 & $$5.0 \times 10^{6}$$ & 1200 \\
           {\rm bulge} & 5.7 \times 10$$^{9}$$ & 16384 & $$3.5 \times 10^{5}$$ & 180 \\
           {\rm disk} & 2.8 \times 10$$^{10}$$ & 32768 & $$8.7 \times 10^{5}$$ & 240 \\
           \noalign{\smallskip}
        \hline
        \end{array}
      \]
\end{table}
The resulting rotation velocity is $\sim$180~km\,s$^{-1}$ and the rotation curve
becomes flat at a radius of about 5~kpc. 

We adopted a model where the ISM is simulated as a collisional component,
i.e. as discrete particles that possess a mass and a radius and 
can have inelastic collisions (sticky particles).
The advantage of our approach is that ram pressure can be easily included as an additional
acceleration on particles that are not protected by other particles (see Vollmer et al. 2001).

The 20\,000 particles of the collisional component represent gas cloud complexes that
evolve in the gravitational potential of the galaxy.
The total assumed gas mass is $M_{\rm gas}^{\rm tot}=4.3 \times 10^{9}$~M$_{\odot}$,
which corresponds to the total neutral gas mass before stripping, i.e.
to an \HI\ deficiency of 0.
A radius is attributed to each particle which depends on its mass assuming a constant surface density.
The normalization of the mass-size relation was taken from Vollmer et al. (2012a).
During each cloud-cloud collision the overlapping parts of the clouds are calculated. 
Let $b$ be impact parameter and $r_1$ and $r_2$ the radii of the larger and smaller clouds. 
If $r_1+r_2 > b > r_1-r_2$ the collision can result into fragmentation (high-speed encounter) or
mass exchange. If $b < r_1-r_2$ mass exchange or coalescence (low speed encounter) can occur.
The outcome of the collision is simplified following Wiegel (1994). 
If the maximum number of gas particles/cloud ($20000$) is reached, only coalescent or mass 
exchanging collisions are allowed. In this way a cloud mass distribution is naturally
produced. The cloud-cloud collisions result in an effective gas viscosity in the disk.

As the galaxy moves through the ICM, its clouds are accelerated by
ram pressure $a=\rho_{\rm ICM}v_{\rm gal}^2/\Sigma$. In addition, the gas clouds are accelerated by the 
gradients of the gravitational potential $a=- \nabla \phi$.
The ISM surface density is assumed to be $\Sigma(r)=(1+\exp(-r/2~{\rm kpc})) \times 10$~M$_{\odot}$pc$^{-2}$.
The constant surface density at the outer disk is motivated by the observed saturation of
the H{\sc i} surface density (e.g., Leroy et al. 2008). The radial dependence takes into account
the increased surface density due to the presence of molecular gas.
Within the galaxy's inertial system, its clouds
are exposed to a wind coming from a direction opposite to that of the galaxy's 
motion through the ICM. 
The temporal ram pressure profile has the form of a Lorentzian,
which is realistic for galaxies on highly eccentric orbits within the
Virgo cluster (Vollmer et al. 2001).
The effect of ram pressure on the clouds is simulated by an additional
force on the clouds in the wind direction. Only clouds that
are not protected by other clouds against the wind are affected.
This results in a finite penetration length of the intracluster medium into the ISM (see Appendix~\ref{sec:plength}).
Since the gas cannot develop instabilities, the influence of turbulence 
on the stripped gas is not included in the model. The mixing of the
intracluster medium into the ISM is very crudely approximated by the finite
penetration length of the intracluster medium into the ISM, i.e. up to this penetration
length the clouds undergo an additional acceleration caused by ram pressure.

\subsection{Star formation \label{sec:sfr}}

We assume that the star formation rate is proportional to the cloud collision rate.
A mass-size relation based on a constant gas surface density is assumed for the collisional particles.
The collisions are explicitly calculated for each timestep ($\sim 10^4$~yr) via the simplified
geometrical recipes of Wiegel (1994). We verified that the star formation rate of an unperturbed galaxy
is constant within $1$~Gyr.

Vollmer \& Beckert (2003) elaborated an analytical model for a galactic gas disk which considers the warm, cold, and molecular phases 
of the ISM as a single, turbulent gas. This gas is assumed to be in vertical hydrostatic equilibrium, with the midplane pressure balancing 
the weight of the gas and stellar disk. The gas is taken to be clumpy, so that the local density is enhanced relative to the average 
density of the disk. Using this local density, the free-fall time of an individual gas clump, the governing timescale for star formation,
can be determined. The star formation rate is used to calculate the rate of energy injection by supernovae. This rate is related to the 
turbulent velocity dispersion and the driving scale of turbulence. These quantities in turn provide estimates of the clumpiness of gas 
in the disk (i.e., the contrast between local and average density) and the rate at which viscosity moves matter inward.
Vollmer \& Leroy (2011) applied the model successfully to a sample of local spiral galaxies.
Within the framework of the model the star formation rate per unit volume is given by
\begin{equation}
\dot{\rho}_*=\Phi_{\rm V} \rho\,t_{\rm ff,cl}^{-1} \ ,
\end{equation}
where $\Phi_{\rm V}$ and $t_{\rm ff,\ cl}^{-1}$ are the volume filling factor and free-fall time of selfgravitating gas clouds and
$\rho$ is the overall gas density. The volume filling factor is $\Phi_{\rm V}=n_{\rm cl}\,4 \pi / 3 \,r_{\rm cl}^3$, where
$n_{\rm cl}$ is the internal number density of clouds and $r_{\rm cl}$ is the cloud radius. For a selfgravitating cloud the free-fall time equals
the turbulent crossing time $t_{\rm ff,cl}=2\,r_{\rm cl}/v_{\rm turb,cl}$. With the collision time given by
$t_{\rm coll}=(n_{\rm cl}\,\pi\,r_{\rm cl}^2 v_{\rm turb})^{-1}$, this leads to $\dot{\rho}_*= 2/3\,\rho\,t_{\rm coll}^{-1}$.
Thus, the star formation rates of the two recipes are formally equivalent.

In numerical simulations, the star formation recipe usually involves the gas density $\rho$ and free-fall time 
$t_{\rm ff}=\sqrt{3\,\pi/(32\,G \rho)}$: $\dot{\rho}_* \propto \rho\, t_{\rm ff}^{-1} \propto \rho^{1.5}$.
We verified that our star formation recipe based on cloud-cloud collisions leads to the same exponent
of the gas density in a simulation of an isolated spiral galaxy.
As a consequence, our code reproduces the observed SFR-total gas surface 
density, SFR-molecular gas surface density, and SFR-stellar surface density relations (Vollmer et al. 2012).

During the simulations, stars are formed in cloud-cloud collisions. At each collision,
a collisionless particle is created, which is added to the ensemble of collisional and
collisionless particles. The newly created collisionless particles have zero mass
(they are test particles) and the positions and velocities of the colliding clouds after the collision. 
These particles are then evolved passively with the whole system. 
Since in our sticky-particle scheme there is mass exchange, coalescence, or fragmentation at the
end of a collision, the same clouds do not collide infinitely.
The local collision rate traces the cloud density and the velocity dispersion of the collisional component.

The information about the time of creation is attached to each newly created star particle.
In this way, the H$\alpha$ emission distribution caused by H{\sc ii} regions can be modeled by the distribution of
star particles with ages younger than $20$~Myr. The UV emission of a star particle in the two GALEX bands
is modeled by the UV flux from single stellar population models from STARBURST99 (Leitherer et al. 1999).
The age of the stellar population equals the time since the creation of the star particle.
The total UV distribution is then the extinction-free distribution of the UV emission of the newly created
star particles. 

\subsection{Diffuse gas stripping \label{sec:iong}}

Our numerical code is not able to treat diffuse gas in a consistent way. For a realistic treatment 3D hydrodynamical simulations
should be adopted. Nevertheless, we can mimic the action of ram pressure on diffuse gas by applying
very simple recipes based on the fact that the acceleration by ram pressure is inversely proportional to the gas
surface density which in turn depends on the gas density for a gas cloud of constant mass.
For simplicity we do not modify the radii of the diffuse gas clouds for the calculation of the cloud-cloud collisions.
The diffuse clouds are thus mostly ballistic particles under the influence of a ram-pressure induced acceleration.
The recipes only take into account gas cooling of the hot phase.

We assume that the warm ($\sim 10^4$-$10^5$~K) gas clouds become diffuse, i.e. their sizes and volume filling factor increase and 
their densities and column densities decrease, 
if it is stripped out of the galactic plane and if its density falls below the critical density
of $n_{\rm crit}^{\rm warm}=5 \times 10^{-3}$~cm$^{-3}$. We further assume that the first condition is fulfilled if the
stellar density at the location of the gas particle is lower than $\rho_*^{\rm crit}=2.5 \times 10^{-4}$~M$_{\odot}$pc$^{-3}$. 
For our model galaxy this density is reached at a disk height of $\sim 3.5$-$4$~kpc.

Once the ionized gas has a high volume filling factor, i.e. it becomes diffuse, its volume and surface densities decrease.
The decrease of the surface density is taken into account by considering gas cloud of constant mass:
for $\rho_* <  \rho_*^{\rm crit}$ the cloud size is proportional to $n^{-1/3}$, the cloud surface density 
is proportional to $n^{2/3}$, and the acceleration caused by ram pressure $a=\rho_{\rm ICM}v_{\rm gal}^2/\Sigma$ is increased by
a factor $(0.044~{\rm cm}^{-3}/n_{\rm ISM})^{2/3}$.
The latter normalization and the critical densities were chosen such that they led to acceptable results for several observed 
galaxies undergoing ram pressure stripping. The critical stellar density is motivated
by the fact that a significant change of the ISM properties only occurs once the ISM has entirely left the galactic disk. 
It turned out that this condition is necessary to avoid excessive gas stripping.
The gas and stellar densities are calculated via the $50$ nearest neighbouring particles.

The hot ($>10^6$~K) diffuse gas is taken into account in the following way:
we assume that once the stripped gas has left the galactic disk, it is mixing with the ambient ICM.
The initial temperature of the ISM is $T_{\rm ISM}=10^4$~K. If (i) the gas density falls below the critical
value of $n_{\rm crit}^{\rm hot}=5 \times 10^{-4}$~cm$^{-3}$, (ii) the stellar density is below $\rho_*^{\rm crit}$,
and (iii) the ISM temperature is below $9 \times 10^6$~K,
ICM-ISM mixing begins to act. On the other hand, if the stellar density exceeds its critical value, i.e. the gas is located within or close
to the galactic disk,
the gas is assumed to cool rapidly and its temperature is set to $T_{\rm ISM}=10^5$~K\footnote{A temperature of $T_{\rm ISM}=10^4$~K does not
change the results of the simulations.}. 
We assume that mixing occurs instantaneously and raises the temperature of the mixed gas clouds to
\begin{equation}
T=\frac{n_{\rm ICM}\,3 \times 10^7\ {\rm K}+\sqrt{n_{\rm ICM}\,n_{\rm ISM}}\,T_{\rm ISM}}{n_{\rm ICM}+\sqrt{n_{\rm ICM}\,n_{\rm ISM}}}=9 \times 10^6~{\rm K}\ ,
\end{equation}
where $n_{\rm ICM}$ is the density of the mixed ICM which is calculated for each particle.
If an exceedingly high mixed ICM density ($n_{\rm ICM} > 2 \times 10^{-3}$~cm$^{-3}$) is needed to heat the gas cloud to $9 \times 10^6$~K, 
ICM-ISM heating is not allowed.
The temperature of the mixed hot gas clouds $T_{\rm ISM}$, if not heated, decreases with a cooling timescale of $t_{\rm cool}=1.5 \times 10^7 n_{\rm ISM}^{-1}$~yr,
where the ISM density $n_{\rm ISM}$ is given in cm$^{-3}$. Once mixed, the gas cloud is not allowed to cool below a
temperature of $10^6$~K\footnote{A temperature limit of $T_{\rm ISM}=10^5$~K does not
change the results of the simulations because the cooling times of the X-ray emitting gas are of the order of several Gyr. Cooling of the hot gas does not
occur in these simulations.}. This algorithm ensures that stripped gas below the critical density $n_{\rm crit}$ is kept at
a temperature of $9 \times 10^6$~K, whereas denser gas can cool to a temperature of $10^6$~K.

For the stripped hot gas clouds ($>10^6$~K) we modified the equation of motion by adding the acceleration caused by the gas pressure gradient
$a=- \nabla p_{\rm ISM}/\rho_{\rm ISM}$ using an SPH formalism. 
For simplicity an isothermal stripped ISM with a sound speed of $c_{\rm s}=235$~km\,s$^{-1}$ is assumed for this purpose.
We thus do not solve an explicit energy equation and the calculation of hydrodynamic effects is approximate.
Its main purpose is the avoiding of clumping of the hot diffuse ISM.

For the calculation of the acceleration caused by ram pressure $a=\rho_{\rm ICM}v_{\rm gal}^2/\Sigma$, the acceleration is further increased
by a heuristic factor of $10$ if the ISM temperature exceeds $10^5$~K and the 
stellar density is below $\rho_*=2.5 \times 10^{-4}$~M$_{\odot}$pc$^{-3}$.   
We note that without the inclusion of hydrodynamical effects thin gas filaments cannot be produced by our simple model.  
The diffuse gas is assumed to emit in the H$\alpha$ line if its temperature is below $10^6$~K.

\subsection{Parameters of the ram pressure stripping event \label{sec:param}}

\begin{figure}[!ht]
  \centering
  \resizebox{\hsize}{!}{\includegraphics{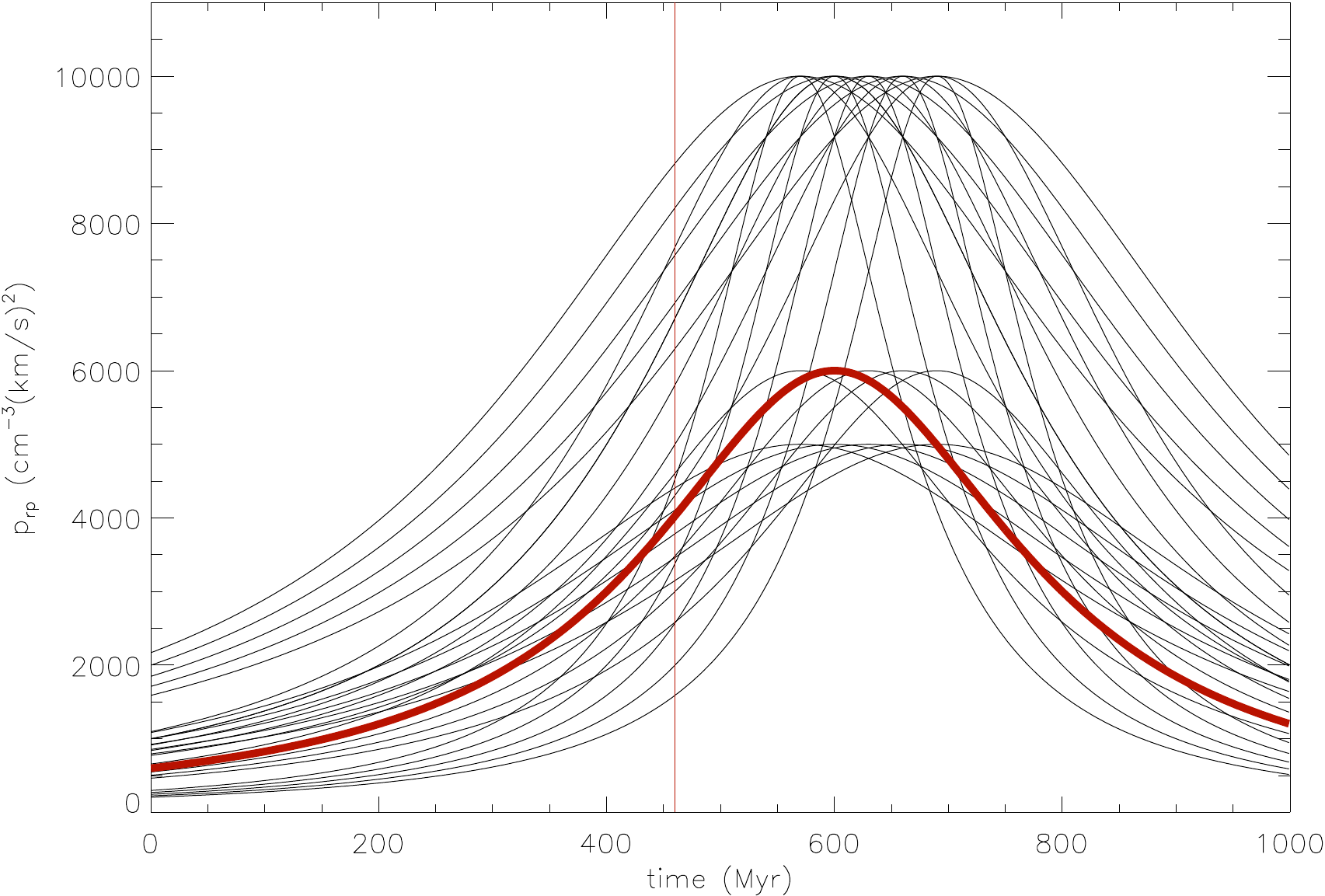}}
  \caption{Model ram pressure profiles. Thick red line: ``best-fit'' model. Vertical red line: timestep of the ``best-fit'' model.
  \label{fig:profiles}}
\end{figure}

Following Vollmer et al. (2001) we used Lorentzian profile for the time evolution of ram pressure stripping:
\begin{equation}
p_{\rm rp}=p_{\rm max} \frac{t_{\rm HW}^2}{(t-t_{\rm peak})^2+t_{\rm HW}^2}\ ,
\end{equation}
where $p_{\rm max}$ is the maximum ram pressure occurring at the galaxy's closest passage to the cluster center,
$t_{\rm HW}$ is the width of the profile, and $t_{\rm peak}$ is the time of peak ram pressure.
The simulations were calculated from $t=0$ to $t=800$~Myr.
We set $p_{\rm max}=5000,\ 6000,\ 10000~{\rm cm}^{-3}({\rm km\,s}^{-1})^2$ and $t_{\rm HW}=100,\ 200,\ 300$~Myr.
Following Nehlig et al. (2016) and Vollmer et al. (2018), we investigated the influence of galactic structure,
i.e. the position of spiral arms, on the results of ram pressure stripping by varying the time of peak ram pressure
between $t_{\rm peak}=570$~Myr and $690$~Myr. The ram pressure profiles are specified in Table~\ref{tab:molent} and shown in Fig.~\ref{fig:profiles}.
The parameters of the Vollmer et al. (2012) model were  $p_{\rm max}=5000~{\rm cm}^{-3}({\rm km\,s}^{-1})^2$ and $t_{\rm HW}=100$~Myr.

\begin{table}[!ht]
      \caption{Model ram pressure stripping profiles.}
         \label{tab:molent}
      \[
         \begin{array}{lrcc}
           \hline
           {\rm model} & {\rm amplitude}\ p_{\rm max} & {\rm width}\ t_{\rm HW} & {\rm peak\ time}\ t_{\rm peak} \\
            & ({\rm cm}^{-3}({\rm km\,s}^{-1})^2) & ({\rm Myr}) & ({\rm Myr}) \\
           \hline
           {\rm 1,\ 1new} & 10000 & 200 & 600 \\
           {\rm 1a,\ 1anew} & 10000 & 200 & 570 \\
           {\rm 1b,\ 1bnew} & 10000 & 200 & 630 \\
           {\rm 1c,\ 1cnew} & 10000 & 200 & 660 \\
           {\rm 1d,\ 1dnew} & 10000 & 200 & 690 \\
           {\rm 2,\ 2new} & 5000 & 300 & 600 \\
           {\rm 2a,\ 2anew} & 5000 & 300 & 570 \\
           {\rm 2b,\ 2bnew} & 5000 & 300 & 630 \\
           {\rm 2c,\ 2cnew} & 5000 & 300 & 660 \\
           {\rm 2d,\ 2dnew} & 5000 & 300 & 690 \\
           {\rm 3,\ 3new} & 10000 & 300 & 600 \\
           {\rm 3a,\ 3anew} & 10000 & 300 & 570 \\
           {\rm 3b,\ 3bnew} & 10000 & 300 & 630 \\
           {\rm 3c,\ 3cnew} & 10000 & 300 & 660 \\
           {\rm 3d,\ 3dnew} & 10000 & 300 & 690 \\
           {\rm 4,\ 4new} & 6000 & 200 & 600 \\
           {\rm 4a,\ 4anew} & 6000 & 200 & 570 \\
           {\rm 4b,\ 4bnew} & 6000 & 200 & 630 \\
           {\rm 4c,\ 4cnew} & 6000 & 200 & 660 \\
           {\rm 4d,\ 4dnew} & 6000 & 200 & 690 \\
           {\rm 5,\ 5new} & 10000 & 100 & 600 \\
           {\rm 5a,\ 5anew} & 10000 & 100 & 570 \\
           {\rm 5b,\ 5bnew} & 10000 & 100 & 630 \\
           {\rm 5c,\ 5cnew} & 10000 & 100 & 660 \\
           {\rm 5d,\ 5dnew} & 10000 & 100 & 690 \\
        \noalign{\smallskip}
        \hline
        \noalign{\smallskip}
        \hline
        \end{array}
      \]
\begin{list}{}{}
\item[Nomenclature:]
\item[Models {\it Na}: without diffuse ISM component.]
\item[Models {\it Na}new: with diffuse ISM component.]
\end{list}
\end{table}
We adopted the Vollmer et al. (2012) values for the angle between the ram pressure wind and the disk plane ($75^{\circ}$).
The inclination of the model galaxy is $i=90^{\circ}$ to the line-of-sight.

\section{Observations and model fitting \label{sec:observations}}

The deep optical spectrum was obtained with the FOcal Reducer and low dispersion Spectrograph (FORS2; Appenzeller et al. 1998) 
mounted on UT1 of the ESO Very Large Telescope.
With a length of $6.8'$, the FORS2 slit covered the full extent of the galaxy. A slit aperture of $1.3''$ was used to optimize the covered  
area without compromising the spectral resolution of $R \sim 1000$. Six exposures of $900$~s each using the 600B+22 grism which covers the 
wavelength range $3300$--$6210$~$\AA$, and three exposures of $900$~s each using the 600RI+19 grism covering the wavelength range 
$5120$--$8450$~$\AA$. 
The deep continuum subtracted H$\alpha$+[N{\sc ii}] observations were obtained within the VESTIGE survey (Boselli et al. 2018) 
using  MegaCam  at  the  CFHT  with the NB filter (MP9603) and the broad-band $r$ filter (MP9602).
The ancillary GALEX, CFHT, Calar Alto 2.2m telescope, {\it Spitzer}, and {\it Herschel} photometric data are described in Fossati et al. (2018).
In addition, for the comparison with our models we use VIVA H{\sc i} and $20$~cm continuum data (Chung et al. 2009) and $6$~cm radio continuum data
from Vollmer et al. (2013).

\begin{figure*}[!ht]
  \centering
  \resizebox{\hsize}{!}{\includegraphics{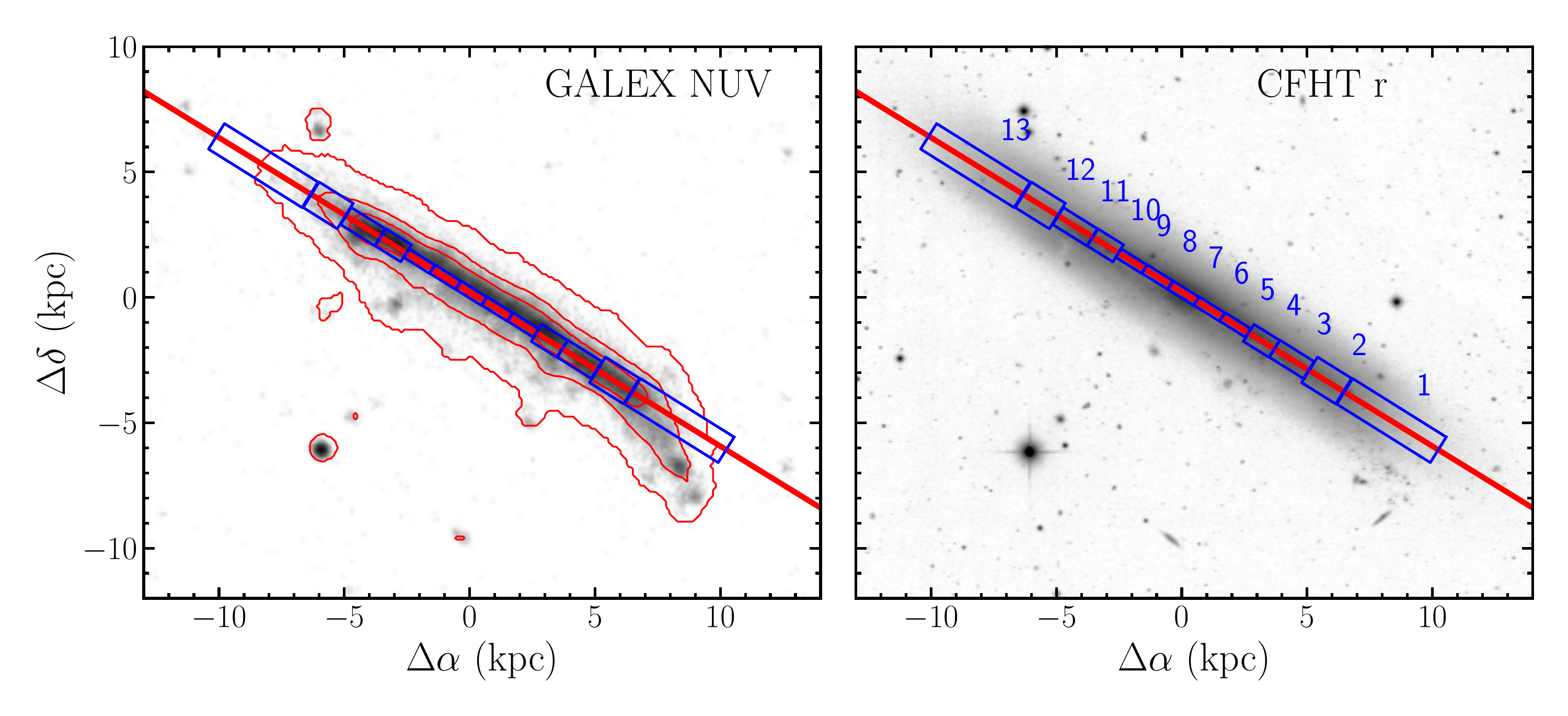}}
  \put(-470,150){\Large downturn}
  \put(-300,50){\Large tail}
  \caption{NGC~4330. Left panel: VLT slit position on the GALEX NUV image. Right panel: VLT slit position on the CFHT $r$ band image
    adapted from Fossati et al. (2018). The downturn region includes the slit regions $11$ to $13$, the tail region includes the slit regions $1$ to $3$.
    The downturn+tail region includes the slit regions $1$ to $3$ and $11$ to $13$. 
  \label{fig:N4330_fig3_NUV}}
\end{figure*}

Fossati et al. (2018) reconstructed the quenching histories of NGC~4330 in various regions along the major axis by means of stellar population 
fitting. They performed a joint fitting of spectra and photometry using Monte Carlo techniques to optimally explore the parameter space.
The emission of the galactic disk was divided into $13$ apertures along the major axis of NGC~4330, as shown in Fig.~\ref{fig:N4330_fig3_NUV}.
Aperture photometry was obtained by summing the calibrated flux values in each pixel contributing to a given aperture minus the background 
value in the same aperture. To estimate the background  each  aperture  was randomly placed in  empty  regions  of the  images  and the  sum  
of  the  counts was taken.  This  procedure was repeated $1000$ times and the median value was adopted as the best estimator for the background value. 
For the UV, optical, and near infrared bands, the measured fluxes were corrected for the Galactic attenuation.
The spectra from the FORS2 reduced 2D spectral images were extracted by summing the flux in the spatial pixels contributing to each region along 
the slit direction.

For the reconstruction of the quenching history of each radial bin a model of its unperturbed radial star formation history  (SFH) is needed.
For this purpose Fossati et al. (2018) adopted  the  multizone  models  for  the  chemical and spectrophotometric 
unperturbed disc evolution originally presented  in  Boissier  \&  Prantzos  (2000)  and  later  updated  by Mu\~{n}oz-Mateos et al. (2011).
The model was projected to fit the high inclination of the disc of NGC~4330 and the SFH were derived in each of the apertures.
Fossati et al. (2018) further  modified  the unperturbed SFHs by introducing an exponential decline of the SFR to parametrize the quenching event.

Parametric models to the spectro-photometric  data  were produced using  MULTINEST (Feroz  \&  Hobson  2008; Feroz et al. 2009, 2013).
A  grid  of  stellar  spectra was constructed using 
the quenched SFHs coupled to Bruzual \& Charlot  (2003)  high-resolution  models.  Nebular  emission  lines  were added to  the  stellar  template.
To  do  so  Fossati et al. (2018)  first  computed  the  number  of  Lyman  continuum ($\lambda < 912$~$\AA$) photons $Q_{{\rm H}_0}$ for a given 
stellar spectrum. They then assumed that all the ionizing radiation is absorbed by the gas and that  the  ionizing  photons do  not  contribute  
to  the  heating  of  dust.  However,  they  modeled the uncertainty in the conversion of $Q_{{\rm H}_0}$ into emission line flux  by  adding  a  nuisance  
parameter  (Ly$_{\rm scale}$)  in  their  MC-SPF code. This parameter also takes into account differential aperture effects  between  the  $1.3''$
spectral  slit  and  the  larger  photometric  apertures.  A  double  Calzetti  
et  al.  (2000)  attenuation law  was assumed to  include  the  effects  of  dust  attenuation  on  the  stellar and  nebular  emission  line  
spectrum.  Fossati et al. (2018) coupled the stellar spectrum with dust emission  models  from  the  mid-  to  the  far-infrared  by  means  of  
an energy  balance  between  the  stellar  flux  absorbed  by  the  dust and re-emitted in the infrared. 
They  jointly  fitted the photometric data points and the FORS2 spectra  
at  their  native  resolution.  The multidimensional likelihood space was sampled by using PYMULTINEST (Buchner et al. 2014), 
a python wrapper for the MULTINEST code. For direct comparison with the results obtained from our model SFHs, we reproduce the
results of Fossati et al. (2018) in Fig.~\ref{fig:Region1_vollmer_summary_4_47} to \ref{fig:Region13_vollmer_summary_4_47}.

For our purpose we use the modeling of Fossati et al. (2018) to obtain the synthetic photometry
and the optical spectra for our model SFHs. The model photometry and spectrum is then compared to observations using MULTINEST
and the total evidence is calculated.
The goodness is evaluated using the Bayesian evidence. This value is formally defined as the integral of the likelihood of a given model 
over the entire parameter space. When two models are compared, their goodness in describing the data can be summarized in the ratio of their evidences 
(or the difference in log space). The Bayesian evidence automatically penalises models with more free parameters thus optimally highlighting the best 
model even in case the models under considerations have a different number of free parameters.
In our case, these free parameters are the temporal ram pressure profile, the chosen timestep, and the inclusion of diffuse gas stripping.
The variation of parameters which are used to compare the model to the data (stellar libraries, attenuation model, Ly$_{\rm scale}$ parameter)
can change the goodnesses and thus the order of the preferred models.
A better model yields a higher goodness.

\section{Results\label{sec:results}}

We calculated the model photometry and spectrum for all timesteps ($\Delta t=10$~Myr) of the $50$ simulations presented in 
Table~\ref{tab:molent} with (format {\it Na}new) and without (format {\it Na}) a diffuse gas component. In addition, we decided to handle the 
Ly$_{\rm scale}$ nuisance parameter in two different ways: (i) we set it to the fixed values found by Fossati et al. (2018) and
(ii) we treated it as a free parameter and kept the value that lead to the best fit to observations.
The distributions of the total goodnesses for all model timesteps are shown in Fig.~\ref{fig:goodnessdist}.
The distributions show a tail to high values containing about $10$-$20$ timesteps.
The resulting total goodnesses are presented in Table~\ref{tab:goodness}. The last number of the model names is the simulation timestep $N$.
The time to peak ram pressure is $((N-1) \times 10) - t_{\rm peak}$~Myr, where $t_{\rm peak}$ is given in Table~\ref{tab:molent}.

We then searched for the $12$ timesteps with the highest goodnesses for the downturn, tail, and total (downturn+tail) regions.
These different regions are generally not best fit by a single timestep (Table~\ref{tab:goodness1}).
Fig.~\ref{fig:goodcomp} compared to Fig.~\ref{fig:Region3_vollmer_summary_4_47} gives an impression of the meaning of the total goodness $g$: 
the model and observed spectra and SEDs of region~3 for two timesteps (models 4new\_47 and 1cnew\_46) with a goodness difference of 
$\Delta g \sim 200$ show that the FUV and NUV fluxes are significantly better reproduced by
model 4new\_47. We therefore presume that goodness differences in excess of $\Delta g \ga 200$ are significant.
\begin{table*}[!ht]
      \caption{Model fitting results.}
         \label{tab:goodness}
      \[
         \begin{array}{lcccccccr}
           \hline
               {\rm downturn+tail} & {\rm total} & {\rm down} & {\rm turn} & & {\rm tail} & & & {\rm time\ to\ } t_{\rm peak} \\
               {\rm model} & {\rm goodness} & {\rm HI} & {\rm NUV} & {\rm H}\alpha & {\rm HI} & {\rm NUV} & {\rm H}\alpha & {\rm (Myr)} \\
           \hline
           {\rm Fossati\ et\ al.} & 32621 & & & & & & &  \\
           \hline
           {\rm With\ diffuse\ ISM} & {\rm component} & & & & & & &  \\
           \hline
           {\rm variable\ Ly}_{\rm scale} & {\rm parameter} & & & & & & & \\
           \hline
           {\rm 4new\_47} &        32209  & + & \sim & +  & + & \sim & +  & -140 \\
           {\rm 1cnew\_47} &        32003  & - & + & \sim  & + & + & \sim  & -200 \\
           {\rm 3dnew\_47} &        31719  & - & - & -  & - & - & -  & -230 \\
           {\rm 4anew\_47} &        31658  & - & \sim & \sim  & + & \sim & - & -110 \\
           {\rm 2anew\_64} &        31647  & - & - & -  & \sim & - & -  & +60 \\
           {\rm 2anew\_65} &        31605  & - & - & \sim  & - & - & -  & +70 \\
           {\rm 4new\_59} &        31589  & - & - & -  & - & \sim & -  & -20 \\
           {\rm 4anew\_60} &        31557  & - & - & -  & - & - & -  & +20 \\
           {\rm 4new\_61} &        31541  & \sim & - & +  & + & \sim & -  & 0 \\
           {\rm 1cnew\_46} &        31467  & - & - & -  & \sim & + & \sim  & -210 \\
           {\rm 1dnew\_47} &        31433  & - & - & - & + & + & \sim  & -230 \\
           {\rm 4anew\_59} &        31423  & - & - & \sim  & - & \sim & -  & +10 \\
           \hline
           {\rm constant\ Ly}_{\rm scale} & {\rm parameter} & & & & & & \\
           \hline
           {\rm 4anew\_60} &        31248  & \sim & - & +  & \sim & \sim & - & +20 \\
           {\rm 2anew\_65} &        30771  & - & - & \sim  & - & - & -  & +70 \\
           {\rm 4dnew\_58} &        30373  & + & - & \sim  & + & \sim & -  & -120 \\
           {\rm 4dnew\_59} &        30240  & + & - & -  & + & \sim & -  & -110 \\
           {\rm 4cnew\_57} &        29978  & - & - & +  & \sim & \sim & -  & -100 \\
           {\rm 4dnew\_57} &        29969  & + & - & \sim  & + & \sim & -  & -130 \\
           {\rm 4new\_47} &        29961  & +  & \sim & +  & + & \sim & +  & -140 \\
           {\rm 2anew\_56} &        29898  & - & - & -  & - & - & -  & -20 \\
           {\rm 2anew\_55} &        29880  & - & - & \sim  & - & - & \sim  & -30 \\
           {\rm 2anew\_64} &        29868  & - & - & \sim  & \sim & - & -  & +60 \\
           {\rm 2bnew\_73} &        29845  & - & - & \sim  & \sim & - & \sim  & +90 \\
           {\rm 4cnew\_56} &        29575  & \sim & - & +  & - & - & -  & -110 \\
           \hline
           {\rm Without\ diffuse\ ISM} & {\rm component} & & & & & &  & \\
           \hline
           {\rm variable\ Ly}_{\rm scale} & {\rm parameter} & & & & & &  & \\
           \hline
           {\rm 4b\_67} &        32103  & - & - & +  & \sim & \sim & -  & +30 \\
           {\rm 2b\_80} &        32004  & - & \sim & +  & - & - & \sim  & +160 \\
           {\rm 3c\_47} &        31904  & - & \sim & +  & \sim & \sim & -  & -200 \\
           {\rm 4b\_68} &        31897  & - & - & -  & \sim & \sim & -  & +40 \\
           {\rm 2d\_70} &        31816  & \sim & \sim & +  & \sim & \sim & +  & 0 \\
           {\rm 3\_47} &        31795  & \sim & + & \sim  & \sim & \sim & \sim  & -140 \\
           {\rm 4c\_70} &        31795 & - & - & \sim  & \sim & - & \sim  & +30 \\
           {\rm 3b\_47} &        31760  & - & \sim & +  & \sim & \sim & -  & -170 \\
           {\rm 4d\_70} &        31708  & + & + & \sim  & \sim & \sim & +  & 0 \\
           {\rm 2a\_68} &        31699  & - & \sim & -  & \sim & \sim & -  & +70 \\
           {\rm 2c\_80} &        31695  & \sim & - & -  & - & - & -  & +130 \\
           {\rm 3d\_48} &        31671  & - & - & +  & + & \sim & -  & -220 \\
           \hline
           {\rm constant\ Ly}_{\rm scale} & {\rm parameter} & & & & & &  & \\
           \hline
           {\rm 2b\_80} &        31854  & - & - & +  & - & \sim & -  & +160 \\
           {\rm 3d\_48} &        31263  & - & - & +  & + & \sim & -  & -220 \\
           {\rm 2d\_77} &        30802  & + & - & +  & + & + & -  & +70 \\
           {\rm 3c\_48} &        30689  & - & - & +  & + & \sim & -  & -190 \\
           {\rm 2d\_74} &        30685  & - & - & +  & \sim  & \sim & -  & +40 \\
           {\rm 3d\_57} &        30455  & \sim & - & +  & - & + & -  & -130 \\
           {\rm 2d\_70} &        30386  & \sim & \sim & +  & \sim & \sim & +  & 0 \\
           {\rm 2b\_65} &        30303  & - & - & -  & \sim & \sim & -  & +10 \\
           {\rm 2d\_66} &        30126  & \sim & - & \sim  & \sim & \sim & \sim  & -40 \\
           {\rm 2c\_80} &        30062  & \sim & - & -  & - & - & -  & +130 \\
           {\rm 4b\_67} &        30047  & - & - & +  & \sim & \sim & -  & +30 \\
           {\rm 3d\_56} &        29908  & - & - & -  & - & - & -  & -140 \\
        \noalign{\smallskip}
        \hline
        \noalign{\smallskip}
        \hline
        \end{array}
      \]
\begin{list}{}{}
\item[The time to peak ram pressure is $((N-1) \times 10) - t_{\rm peak}$~Myr, where $N$ is the number of the timestep]
\item[(last number of the model names).]
\item[The quality assessment of the reproduction of the observational characteristics is based on the joint fit of]
\item[the SEDs and optical spectra.]
\item[+: reproduced; $\sim$: approximately reproduced; -: not reproduced]
\item[Models {\it Na}: without diffuse ISM component. Models {\it N}new{\it a}: with diffuse ISM component.]
\item[The different regions (downturn, tail, and downturn+tail) are defined in Fig.~\ref{fig:N4330_fig3_NUV}.]
\end{list}
\end{table*}

\begin{table*}[!ht]
      \caption{Model fitting results.}
         \label{tab:goodness1}
      \[
         \begin{array}{lrclrc}
           \hline
           {\rm downturn} & \Delta t_{\rm peak} & & {\rm tail} & \Delta t_{\rm peak} & \\
            & \ {\rm (Myr)} & \ \ \ \ \ \ \  {\rm goodness}\ \ \ \   &  & {\rm (Myr)} & {\rm goodness}  \\
           \hline
           {\rm With\ diffuse\ ISM} & {\rm component} & & & &\\
           \hline
           {\rm variable\ Ly}_{\rm scale} & {\rm parameter} & & & & \\
           \hline
 {\rm 2bnew\_63} & -10  &     15120 & {\rm 5dnew\_58} & -120  &     17351 \\
 {\rm 3dnew\_51} & -190  &     15109 & {\rm 4new\_47} &  -140  &    17268 \\
 {\rm 2dnew\_61} & -90  &     15082 & {\rm 5cnew\_57} & -100  &     17261 \\
 {\rm 2dnew\_62} & -80  &     15071 & {\rm 5cnew\_58} & -90  &     17254 \\
 {\rm 4anew\_60} & +20  &     15066 & {\rm 4cnew\_57} & -100  &     17221 \\
 {\rm 1bnew\_48} & -160  &     15053 & {\rm 4cnew\_56} & -110  &     17214 \\
 {\rm 2new\_76} & +150   &    15049 & {\rm 5cnew\_66} &  -10  &    17196 \\
 {\rm 1cnew\_47} & -200  &     15048 & {\rm 5bnew\_63} & -10  &     17196 \\
 {\rm 2anew\_64} &  +60 &     15047 & {\rm 1cnew\_46} & -210  &     17175 \\
 {\rm 2bnew\_62} & -20  &     15045 & {\rm 5bnew\_69} & +50  &     17174 \\
 {\rm 1dnew\_55} &  -130 &     15043 & {\rm 1bnew\_47} & -170  &     17159 \\
 {\rm 2anew\_42} & -160  &     15039 &  {\rm 4dnew\_58} & -120 &      17132 \\
\hline
           {\rm constant\ Ly}_{\rm scale} & {\rm parameter} & & \\
           \hline
 {\rm 4anew\_60} & +20  &     15039 & {\rm 5cnew\_58} & -90   &    17135 \\
 {\rm 2dnew\_62} & -80  &     15036 & {\rm 4dnew\_58} & -120   &    17113 \\
 {\rm 1bnew\_48} & -160  &     14917 & {\rm 4cnew\_56} &  -110  &    17029 \\
 {\rm 1dnew\_55} & -150  &     14875 & {\rm 5cnew\_65} & -20   &    17023 \\
 {\rm 1dnew\_41} & -290  &     14857 & {\rm 4new\_61} &  0   &   17022 \\
 {\rm 1cnew\_54} & -130  &     14854 & {\rm 5cnew\_57} & -100   &    17001 \\
 {\rm 2cnew\_75} & +80  &     14814 & {\rm 4anew\_47} & -110   &    16968 \\
 {\rm 2anew\_47} & -110  &     14812 & {\rm 5dnew\_76} &  +60  &    16950 \\
 {\rm 2cnew\_78} & +110  &     14768 & {\rm 1dnew\_47} &  -230  &    16903 \\
 {\rm 2cnew\_76} & +90  &     14746 & {\rm 5new\_62} &  +10   &   16900 \\
 {\rm 2cnew\_66} & -10  &     14727 & {\rm 5bnew\_63} & -10   &    16823 \\
 {\rm 2dnew\_61} & -90  &     14720 & {\rm 4dnew\_57} &  -130  &    16820 \\
\hline
           {\rm Without\ diffuse\ ISM} & {\rm component} & & \\
           \hline
           {\rm variable\ Ly}_{\rm scale} & {\rm parameter} & & \\
           \hline
 {\rm 3d\_44} &  -260  &    15074 & {\rm 2\_58} &  -30  &    17461 \\
 {\rm 2b\_80} & +160   &    15030 & {\rm 4d\_70} &  0 &     17408 \\
 {\rm 3d\_41} &  -290  &    15012 & {\rm 2\_57} &  -40  &    17398 \\
 {\rm 1c\_42} & -250   &    15006 & {\rm 1a\_48} & -100  &     17390 \\
 {\rm 1\_43} &  -180   &   14983 & {\rm 2a\_58} & 0   &    17362 \\
 {\rm 1b\_42} &  -220  &    14973 & {\rm 3b\_48} & -160  &     17357 \\
 {\rm 3\_41} &  -200   &   14962 & {\rm 2d\_66} &  -40  &    17356 \\
 {\rm 2a\_66} &  +80  &    14961 & {\rm 4\_54} &  -70  &    17355 \\
 {\rm 4c\_68} &  +10  &    14960 & {\rm 2\_56} & -50   &    17355 \\
 {\rm 2a\_65} & +70   &    14958 & {\rm 2c\_74} & +70  &     17333 \\
 {\rm 3d\_42} & -280   &    14954 & {\rm 2b\_57} & -70  &     17324 \\
 {\rm 3d\_43} &  -270  &    14950 & {\rm 3\_47} & -140   &    17324 \\
\hline
           {\rm constant\ Ly}_{\rm scale} & {\rm parameter} & & \\
           \hline
 {\rm 2b\_80} &  +160  &    15008 & {\rm 2a\_58} &  0   &   17360 \\
 {\rm 3d\_44} &  -260  &    14860 & {\rm 2d\_66} &  -40   &   17339 \\
 {\rm 2a\_63} & +50   &    14733 & {\rm 2d\_70} &  0   &   17294 \\
 {\rm 2a\_66} &  +80  &    14717 & {\rm 3\_47} &  -140    &  17285 \\
 {\rm 2a\_65} & +70   &    14599 & {\rm 1d\_58} &  -120   &   17230 \\
 {\rm 1d\_41} & -290   &    14565 & {\rm 2c\_72} &  +50   &   17190 \\
 {\rm 3c\_46} & -210   &    14484 & {\rm 3d\_58} &  -120   &   17185 \\
 {\rm 2c\_67} & 0   &    14448 & {\rm 2d\_77} &  +70   &   17174 \\
 {\rm 2c\_80} & +130   &    14275 & {\rm 4d\_70} &  0   &   17157 \\
 {\rm 1c\_41} & -260   &    14245 & {\rm 4\_66} &  +50    &  17118 \\
 {\rm 2b\_79} &  +150  &    14244 & {\rm 2\_58} &  -30    &  17111 \\
 {\rm 2c\_68} &  +10  &    14211 & {\rm 3d\_48} &  -220   &   17094 \\
 \noalign{\smallskip}
        \hline
        \noalign{\smallskip}
        \hline
        \end{array}
      \]
\begin{list}{}{}
\item[The time to peak ram pressure is $\Delta t_{\rm peak}=((N-1) \times 10) - t_{\rm peak}$~Myr, where $N$ is the number of the timestep]
\item[(last number of the model names).]
\item[Models {\it Na}: without diffuse ISM component. Models {\it N}new{\it a}: with diffuse ISM component.]
\item[The different regions (downturn, tail) are defined in Fig.~\ref{fig:N4330_fig3_NUV}.]
\end{list}
\end{table*}

All of our total model goodnesses are smaller than that of parametric model of Fossati et al. (2018). This is expected because the parametric model had a higher degree of freedom.
The models with a diffuse stripped ISM lead to somewhat higher goodnesses than those without a diffuse stripped ISM.
Moreover, models with a variable Ly$_{\rm scale}$ parameter systematically yield higher goodnesses thanks to the physical 
relevance of this parameter in reproducing the observed emission lines in the spectra.
This is true for the downturn, tail, and total regions.

Model 4new\_47 with diffuse gas stripping has the highest total goodness of all tested models (Table~\ref{tab:goodness}). 
Given that a goodness difference of $\Delta g \ga 200$ is significant, models 1cnew\_47 (diffuse gas stripping), 
4b\_67, and 2b\_80 are also acceptable amongst the models with a variable Ly$_{\rm scale}$ parameter. 
For a constant Ly$_{\rm scale}$ parameter, only model 2b\_80 is acceptable.

If we look at the goodnesses of the downturn and tail regions separately (Table~\ref{tab:goodness1}), the results change:
in the case of a variable Ly$_{\rm scale}$ parameter all $12$ best-fit models of the downturn region are acceptable if diffuse
gas stripping is included or not. For the tail region all models without diffuse gas stripping are acceptable. If the stripping 
of diffuse gas is included in the simulations, only the first three models are acceptable. 
In the case of a constant Ly$_{\rm scale}$ parameter the four and two best-fit models are acceptable when diffuse gas stripping is
included or not. In the tail region only the eight best-fit models without diffuse gas stripping are acceptable.

We conclude that the best-fit models based on the total goodness are pre-peak ram pressure models if diffuse gas stripping is included,
otherwise the preferred timesteps are post-peak ram pressure (see Table~\ref{tab:molent}).

\subsection{The NUV and H{\sc i} morphologies \label{sec:hinuv}}

The observed H{\sc i} emission stems from warm ($\sim 8000$~K) gas with densities $\ga 1$~cm$^{-3}$.
The NUV is emitted by the photospheres of massive hot stars. It traces the star formation on timescales $< 300$~Myr.

The $48$ simulation snapshots of Table~\ref{tab:goodness} yield the highest total goodnesses for each kind of simulation
(with or without diffuse gas stripping, with a variable or constant Ly$_{\rm scale}$ parameter). Models with total
goodnesses which exceed $g=32000$ for a variable Ly$_{\rm scale}$ parameter (4new\_47, 1cnew\_47, 4b\_67, and 2b\_80)
and $g=31650$ for a constant Ly$_{\rm scale}$ parameter (2b\_80) are acceptable 
fits to the spectro-photometric observations. However, the observed H{\sc i}, NUV, and H$\alpha$ morphologies have also to be reproduced 
by the model snapshots. To assess the quality of the fits, we compared the model and observed morphologies of the downturn and tail regions by eye. 
The results are presented in Table~\ref{tab:goodness}. The key features used for the quality
assessment are: H{\sc i}: truncated disk, extraplanar gas south of the galactic disk, southwestern tail;
NUV: disk emission extending further than the H{\sc i} emission, southwestern tail offset from the H{\sc i} tail;
H$\alpha$: truncated disk, southwestern tail offset from the NUV tail, low surface brightness north-south filaments.
The adopted signs are: +: reproduced; $\sim$: approximately reproduced; -: not reproduced.

\begin{figure*}[!ht]
  \centering
  \resizebox{16cm}{!}{\includegraphics{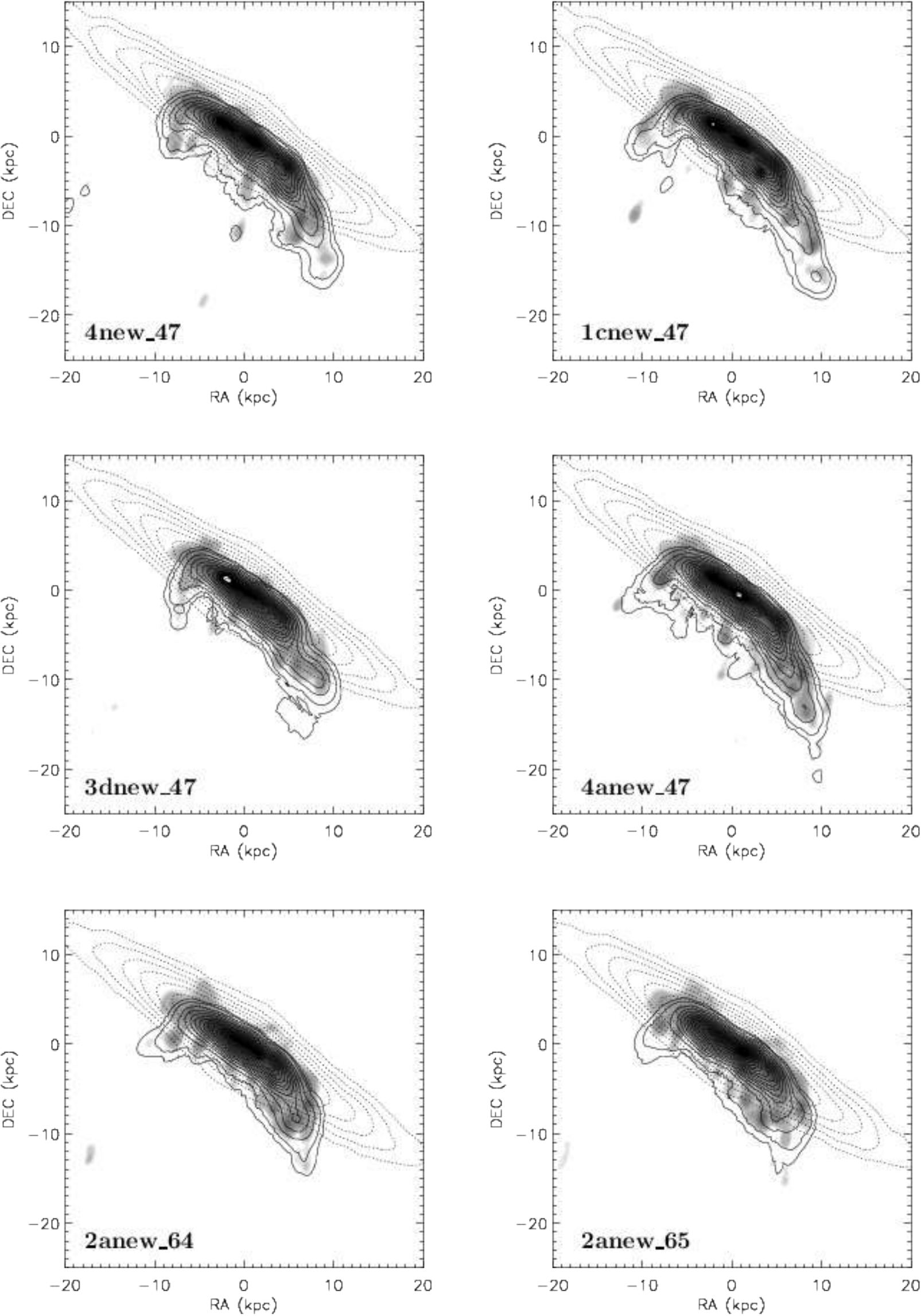}}
  \put(-300,630){\Large \bf best-fit}
  \caption{NUV (grayscale), H{\sc i} (solid contours), and stellar (dotted contours) images. Models with a diffuse ISM component. For the 
spectral fitting variable nuisance parameter Ly$_{\rm scale}$ was used.
  \label{fig:zusammen_nuv_newprofs-1}}
\end{figure*}

The model NUV and H{\sc i} distributions of the six model snapshots with highest goodnesses from the simulations
with a diffuse ISM component and a variable Ly$_{\rm scale}$ scale parameter are presented in Fig.~\ref{fig:zusammen_nuv_newprofs-1}.
They can be directly compared to the upper left panel of Fig.~\ref{fig:Halpha_HI}.

The observed truncation radius of the H{\sc i} disk of $7.5$~kpc as well as its symmetry along the major axis are well reproduced by all model snapshots. 
The observed strong asymmetry of the H{\sc i} distribution along the minor axis with a significant amount of gas pushed by
ram pressure downwind of the major axis is also reproduced by all model snapshots. The observed southwestern tail is present in all
model snapshots. However, the gas tails of models 4new\_47, 1cnew\_47, 3dnew\_47, and 4anew\_47 are somewhat more extended than the observed gas tail.
Moreover, the gas distribution of the eastern downturn region of all model snapshots are more extended in the downwind direction than it is observed.
The observed gas distribution of the downturn region is best reproduced by model 4new\_47.

In the downturn region the observed NUV distribution is more extended along the disk than the H{\sc i} distribution. 
This feature is reproduced by all model snapshots.
However, the extent of all six model NUV disks are smaller than the observed one. In addition, the NUV emission of all model snapshots
is extended in the downwind direction. Whereas the model NUV downturns filaments are perpendicular to the galactic disk in models 1cnew\_47,
4anew\_47, 2anew\_64, and 2anew\_65, they are further bent toward the galactic disk in models 4new\_47 and 3dnew\_47.
In this respect, the latter two model snapshots are most similar to observations. 

We observe an interesting feature in the NUV distribution of
models 2anew\_64, 2anew\_65: an extension in the upwind direction on the northwestern side of the galactic disk. 
This extent is caused by star created in the UV tail which have
the time to rotate into the downturn region. Since their angular momentum is that of the tail gas, these stars rotate in a plane which is tilted
with respect to the galactic plane leading to a position angle opposite to that of the tail in the downturn region. Because these stars 
need to have the time to fulfill about half of a galactic rotation, this effect occurs in simulation snapshots more than several $10$~Myr after peak
ram pressure. These model snapshots can thus be excluded. The observed bending of the southwestern NUV tail is
reproduced by all model snapshots. Its observed offset with respect to the H{\sc i} tail is only present in model 1cnew\_47.
We conclude that the observed H{\sc i} and NUV morphologies are best reproduced by model 4new\_47 followed by model 1cnew\_47.

\begin{figure*}[!ht]
  \centering
  \resizebox{16cm}{!}{\includegraphics{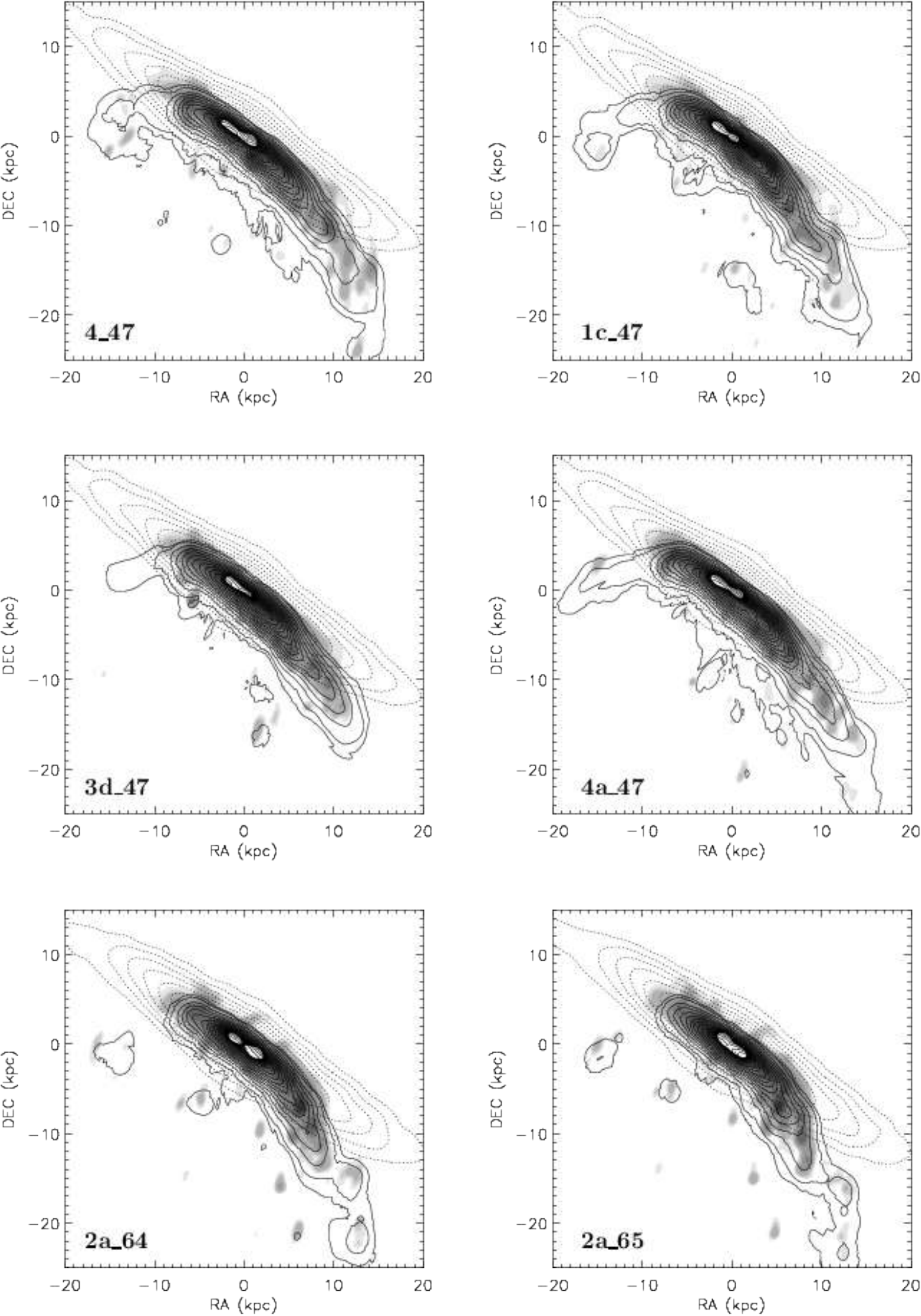}}
  \caption{Same as Fig~\ref{fig:zusammen_nuv_newprofs-1} but without a diffuse ISM component.
  \label{fig:zusammen_nuv_profs_newprofs-2}}
\end{figure*}

To demonstrate the effect of diffuse gas stripping on the model H{\sc i} and NUV distribution, we show the same model snapshots as in 
Fig.~\ref{fig:zusammen_nuv_newprofs-1} for simulation without diffuse gas stripping in Fig.~\ref{fig:zusammen_nuv_profs_newprofs-2}.
The resulting H{\sc i} and NUV distributions of all models snapshots are more extended than those of the snapshots of the corresponding
models with diffuse gas stripping and observations. Indeed, the snapshots with the highest goodnesses for the models without diffuse gas
stripping are not the same as those presented in Fig.~\ref{fig:zusammen_nuv_newprofs-1} and \ref{fig:zusammen_nuv_profs_newprofs-2}
(see Table~\ref{tab:goodness}). 

\begin{figure*}[!ht]
  \centering
  \resizebox{16cm}{!}{\includegraphics{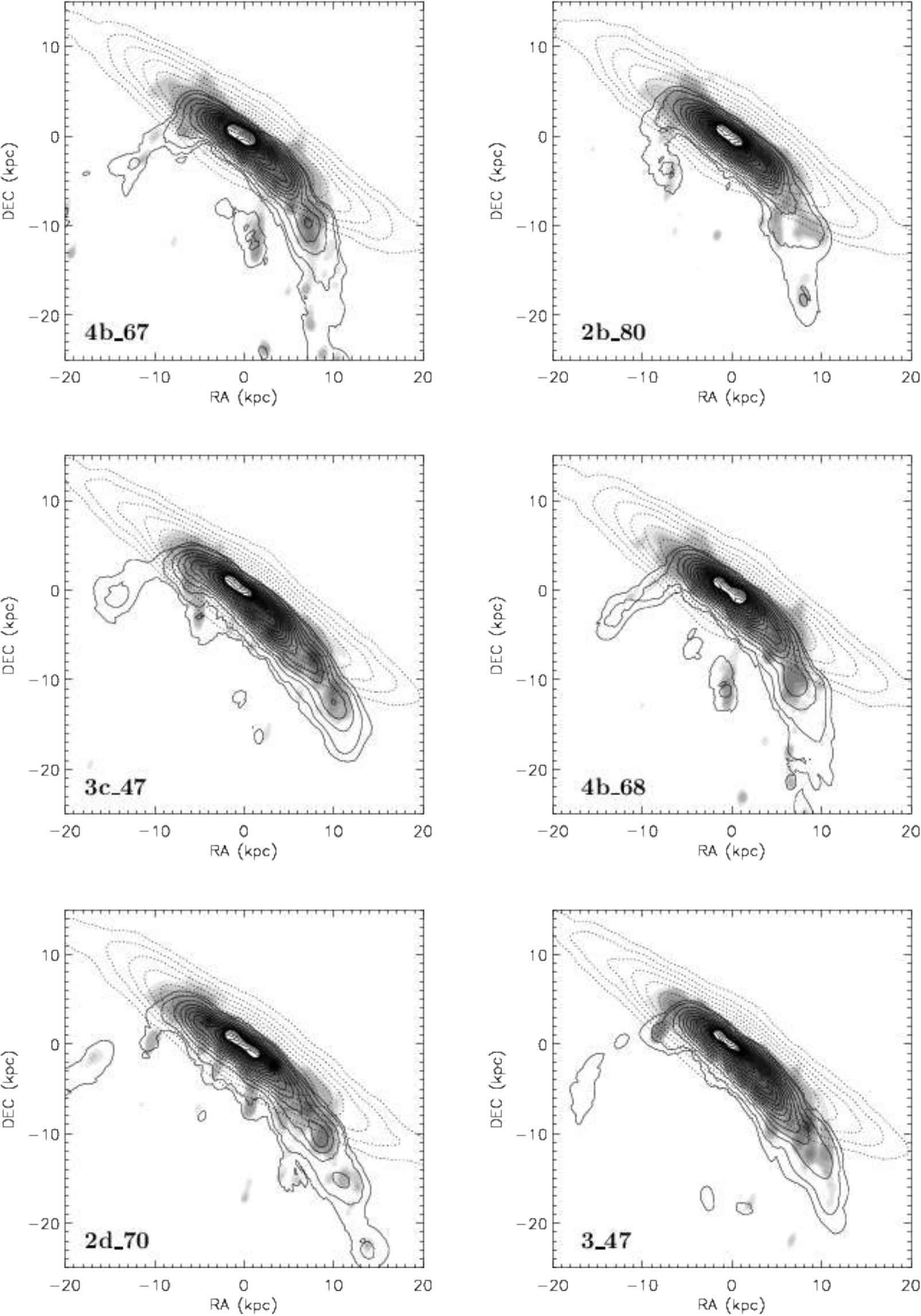}}
  \caption{NUV (grayscale), H{\sc i} (solid contours), and stellar (dotted contours) images. Models without a diffuse ISM component.
    For the spectral fitting a variable nuisance parameter Ly$_{\rm scale}$ was used.
  \label{fig:zusammen_nuv_profs-3}}
\end{figure*}

The snapshots with the highest goodnesses for the models without diffuse gas are presented in Fig.~\ref{fig:zusammen_nuv_profs-3}.
All model with high snapshot numbers (4b\_67, 2b\_80, 4b\_68, and 2d\_70) show extensions in the upwind direction 
on the northwestern side of the galactic disk and therefore
are excluded. The H{\sc i} and NUV tails of the two other models, 3c\_47 and 3\_47, are less bent than the ``best-fit'' models
with diffuse gas stripping (4new\_47 and 1cnew\_47). In addition, extended H{\sc i} tails in the downturn region of 3c\_47 and 3\_47
are not observed. We thus conclude that the models with diffuse gas stripping reproduce the observed H{\sc i} and NUV morphologies significantly 
better than the models without diffuse gas stripping.

\subsection{The H$\alpha$ morphology}

The model H$\alpha$ images consist of two components: (i) the H{\sc ii} regions ionized by young massive stars and (ii) diffuse ionized gas that is
ionized by the stellar UV radiation or possibly by strong shocks induced by ram pressure stripping.
The first component is modeled by the distribution of stellar particles with ages less than $20$~Myr. 
This is the only component used for the images of the models without diffuse gas stripping.
For the second component the emission measure for
all gas particles with temperatures smaller than $10^6$~K was calculated: $EM = \rho^2 \, d$, where $\rho$ is the gas density calculated
with SPH algorithm and the cloud diameter is given by $d=2\,(3\,M_{\rm cl}/(4\,\pi \rho))^{1/3}$ with the cloud mass $M_{\rm cl}$. 
In the absence of a model for the ISM ionization fraction
we assume a constant ionization for the diffuse and dense gas. We then added the H{\sc ii} region component with a heuristic normalization which
worked best to insure that extraplanar H{\sc ii} region are well visible within the emission of the diffuse ionized component.
In these images the surface brightness of the disk emission with respect to the extraplanar emission is realistic in a qualitative but not
in a quantitative sense.

Fossati et al. (2018) observed an H$\alpha$ tail whose high surface brightness part is clumpy and follows the NUV tail
(upper right panel of Fig.~\ref{fig:Halpha_HI}).
A diffuse component of low surface brightness extends downwind to the south. In addition, low surface brightness filaments 
extend further from the tail to the south. Another low surface brightness H$\alpha$ filament is detected close to the downturn in the northeast.
As noted by Fossati et al. (2018) all filaments indicate the direction of the motion of the galaxy. 
\begin{figure*}[!ht]
  \centering
  \resizebox{16cm}{!}{\includegraphics{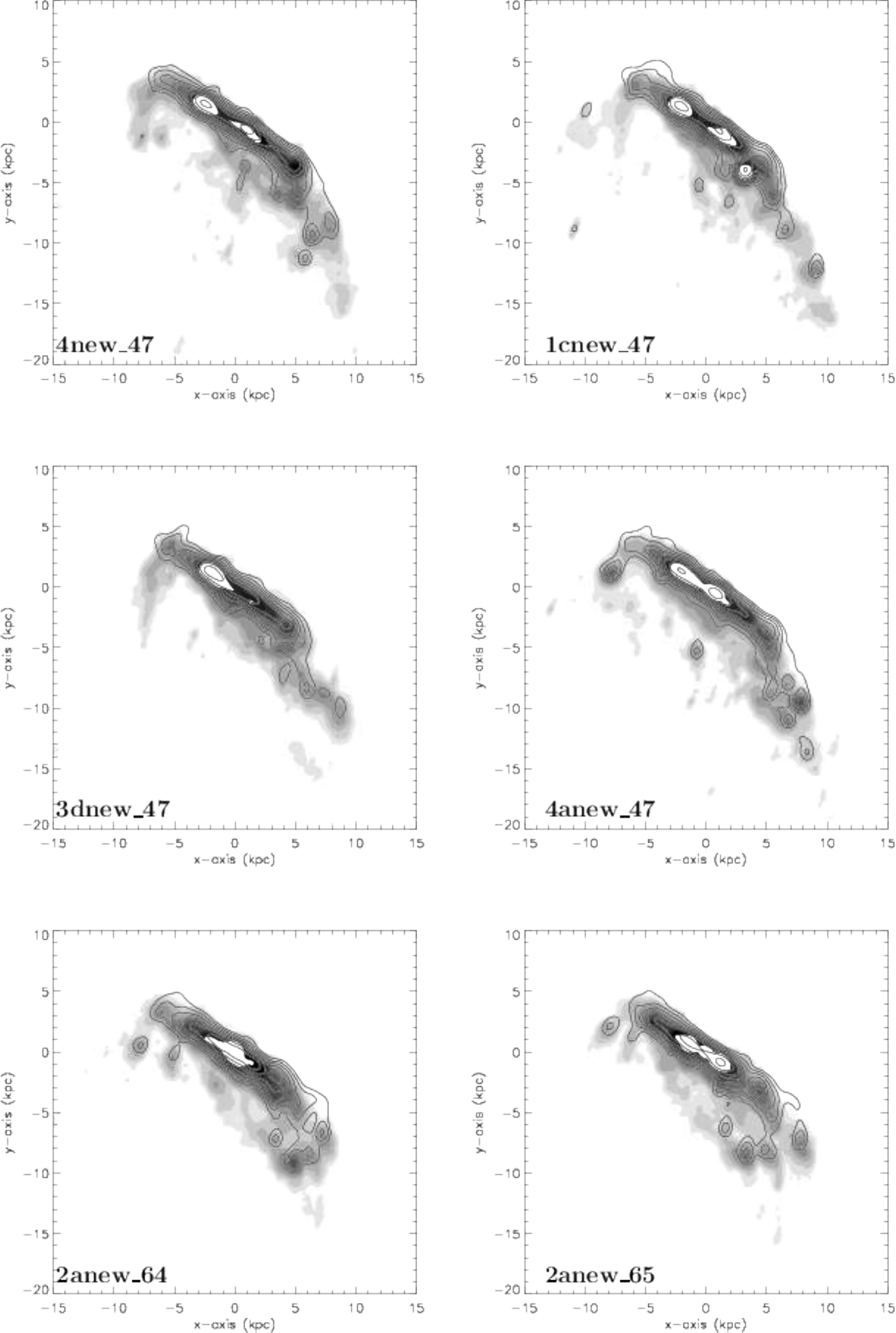}}
  \put(-300,650){\Large \bf best-fit}
  \put(-400,600){\Large 1}
  \put(-348,560){\Large 2}
  \put(-320,535){\Large 3}
  \put(-295,510){\Large 4}
  \caption{H$\alpha$ (grayscale) and NUV (contours) images. Models with a diffuse ISM component.
    For the spectral fitting a variable nuisance parameter Ly$_{\rm scale}$ was used.
  \label{fig:zusammen_ha_newprofs-1}}
\end{figure*}

The resulting model H$\alpha$ emission maps for the snapshots of highest goodness of the diffuse gas stripping model are presented in 
Fig.~\ref{fig:zusammen_ha_newprofs-1}. All model snapshots show an H$\alpha$ tail with the same extent as the NUV tail as it is observed.
Most of the models show extraplanar linear filaments in the downwind region of the galactic disk.
These filaments are thicker than the observed H$\alpha$ filaments (see Sect.~\ref{sec:iong}).
They are labeled for model 4new\_47 in the upper left panel of Fig.~\ref{fig:zusammen_ha_newprofs-1}.
Their lengths are comparable to those of the observed H$\alpha$ filaments.
All models show an extraplanar linear filament originating from the downturn region which extends out of the disk toward the south. 
This filament has a north-south direction in models 4new\_47, 1cnew\_47, and 3dnew\_47.
It has a position angle of $\sim 135^{\circ}$, i.e. it extends toward the southeast, in the other models. 
A second linear filament originating close to the galaxy center and extending into the downwind region
is observed in models 4new\_47, 1cnew\_47, and 4anew\_47. A third north-south filament  
is observed in the middle of the tail of model 4new\_47. A fourth north-south filament starting from the outer edge of
the H$\alpha$/NUV tail is present in models 4new\_47, 1cnew\_47, and 4anew\_47. 
The direct comparison between model 4new\_47 and the VESTIGE observations (Fig.~\ref{fig:Halpha_HI}) shows that the
latter filament might be interpreted as an extension of the gas tail. Because of the vertical orientation of its 
outermost part (compare to model 1cnew\_47), we rather interpret this feature as a filament.

Fossati et al. (2018) identified five extraplanar filaments in the downwind region: one originating close to the downturn 
region and four filaments in the tail region. Model 4new\_47 with its four H$\alpha$ filaments resembles most
the observed H$\alpha$ morphology. We therefore conclude that
model 4new\_47 reproduces best the deep VESTIGE H$\alpha$ observations (lower right panel of Fig.~\ref{fig:Halpha_HI}).

\begin{figure*}[!ht]
  \centering
  \resizebox{16cm}{!}{\includegraphics{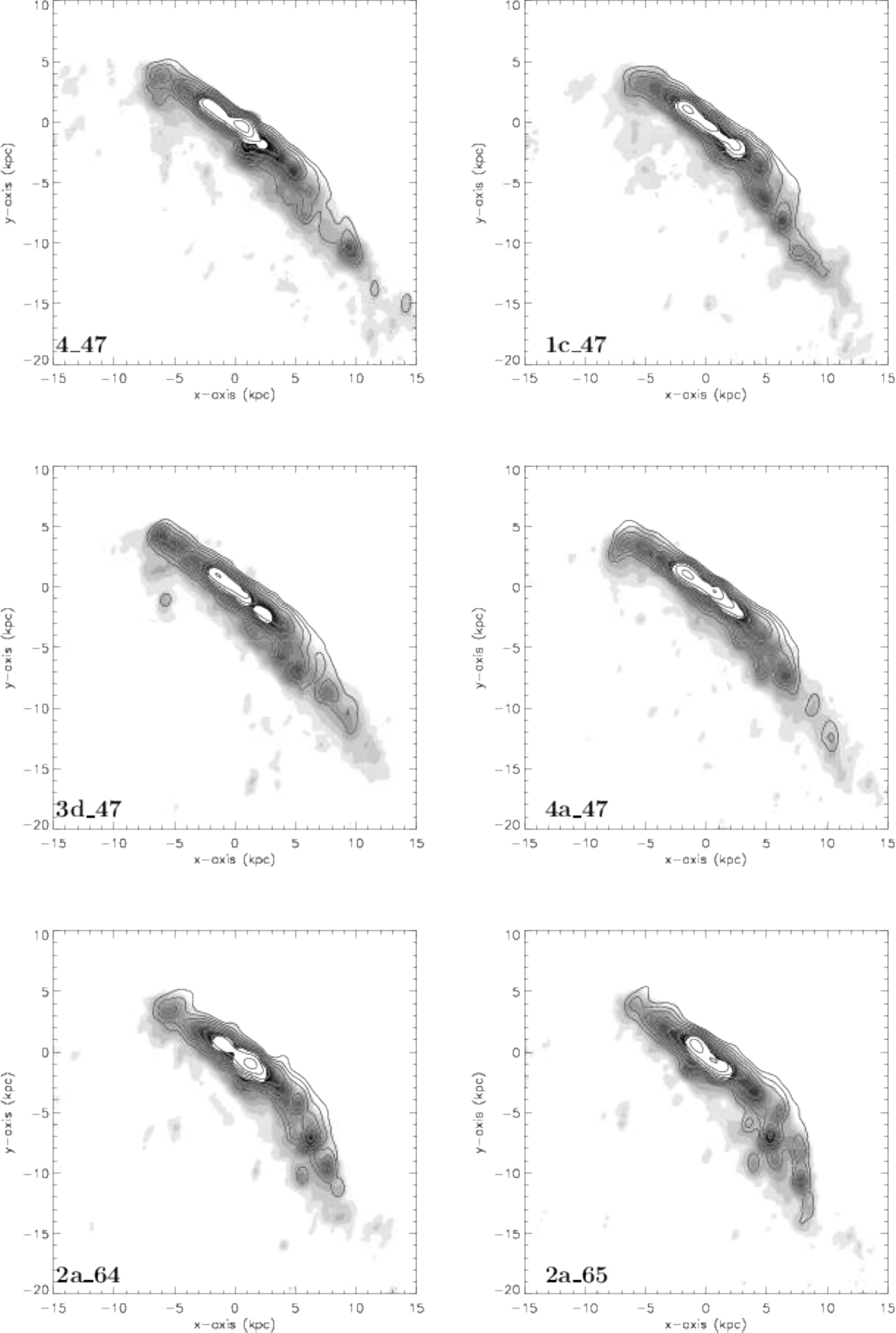}}
  \caption{Same as Fig.~\ref{fig:zusammen_ha_newprofs-1} but without a diffuse ISM component.
  \label{fig:zusammen_ha_profs_newprofs-1}}
\end{figure*}

To demonstrate the effect of diffuse gas stripping on the model H$\alpha$ emission distribution, we show the same model snapshots as in 
Fig.~\ref{fig:zusammen_ha_newprofs-1} for simulation without diffuse gas stripping in Fig.~\ref{fig:zusammen_ha_profs_newprofs-1}.
All model snapshots show an H$\alpha$ tail comparable to that of the simulations with diffuse gas stripping.
As expected, the number of extraplanar linear filaments is significantly decreased compared to the simulations with diffuse gas stripping.
Only models 4\_47, 1c\_47, and 3d\_47 show such a filament starting from the downturn region. However, its position angle  
differs significantly from zero (north-south direction). Vertical filaments along the projected direction of
ram pressure stripping are only present in the diffuse gas because of its increased stripping efficiency.
The ratio between tidal and ram-pressure acceleration is higher for the dense gas which is thus less
accelerated along the wind direction.
Based on the absence of north-south filaments we conclude that the presence
of diffuse gas stripping increases the number of extraplanar H$\alpha$ filaments in the downwind region and leads to a north-south alignment of these filaments. The simulations with diffuse gas stripping thus reproduce observations best.

\begin{figure*}[!ht]
  \centering
  \resizebox{16cm}{!}{\includegraphics{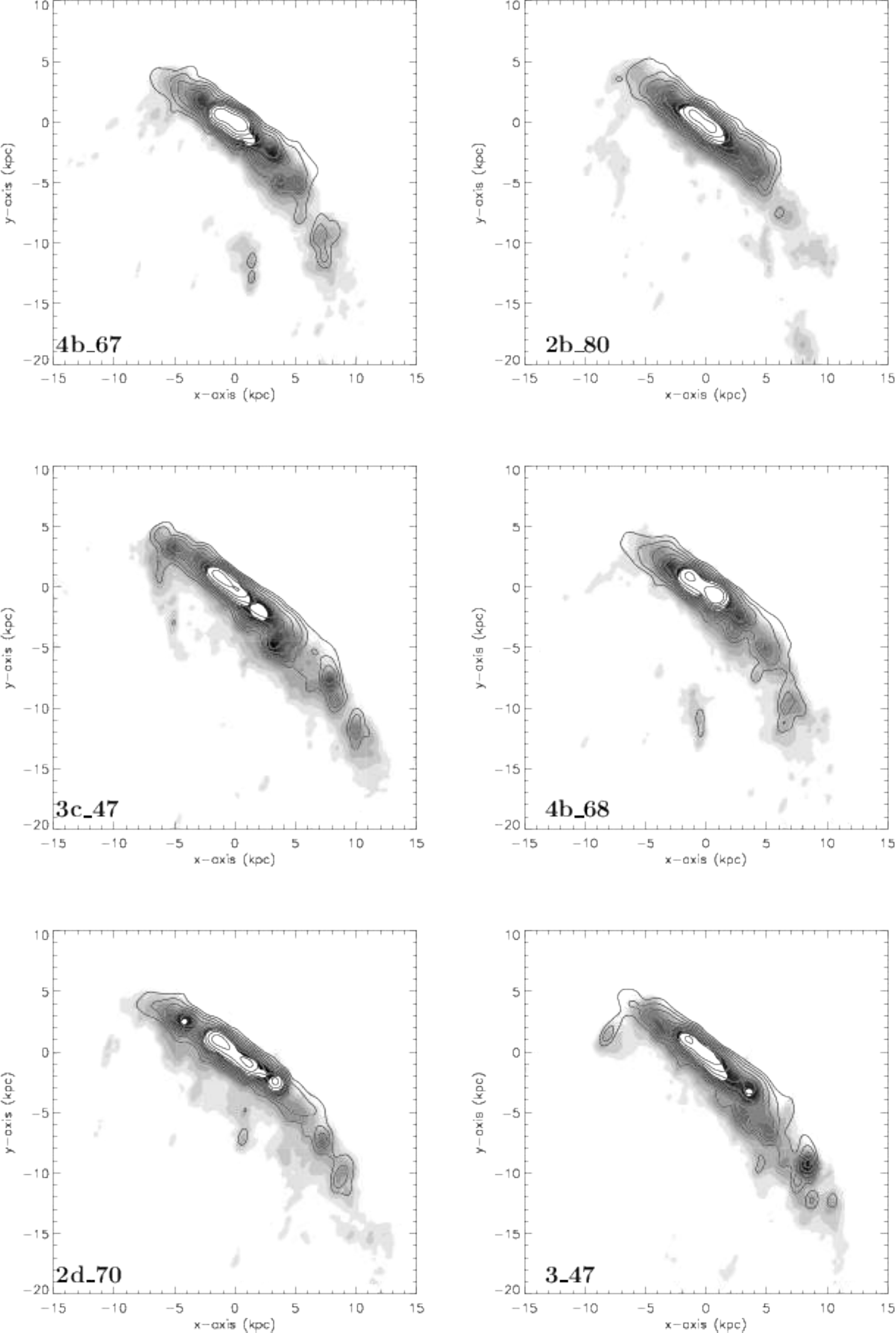}}
  \caption{H$\alpha$ (grayscale) and NUV (contours) images. Models without a diffuse ISM component.
    For the spectral fitting a variable nuisance parameter Ly$_{\rm scale}$ was used.
  \label{fig:zusammen_ha_profs-1}}
\end{figure*}

The snapshots with the highest goodnesses for the models without diffuse gas are presented in Fig.~\ref{fig:zusammen_ha_profs-1}.
The H$\alpha$ tail is comparable to that of the previously discussed models.
The extraplanar filaments starting from the downturn region are still present in all model snapshots.
However, their position angle is not zero as it is observed. Only model 2d\_70 shows north-south filaments starting from the galactic
disk and the tail region. Only the filament originating at the end of the tail deviates from the north-south alignment of the other 
filaments, in contrast to the corresponding filament $4$ in model 4new\_47 (upper left panel of Fig.~\ref{fig:zusammen_ha_newprofs-1}). 
We think that this difference is significant.

We conclude that the existence of north-south H$\alpha$ filaments is a natural consequence of the presence of diffuse gas stripping.
However, in rare cases, such filaments can also be created in models without diffuse gas stripping. 
Models 4new\_47 and 2d\_70 reproduce best the deep VESTIGE H$\alpha$ observations.

\subsection{The ``best-fit'' model \label{sec:bestfit}}

The results of the comparison between the model and observed H{\sc i}, NUV, and deep H$\alpha$ emission distributions are
presented in Table~\ref{tab:goodness}. Model 4new\_47 reproduces the H{\sc i} and H$\alpha$ emission
distribution in the downturn and tail region in a satisfactory way. Moreover, the overall morphology of the observed NUV emission
is qualitatively reproduced by the model. However, the NUV disk in the tail region is less extended than observed.
This is mainly an effect of NUV surface brightness, which is lower in the model NUV tail compared to observations
(see also the UV photometry in Figs.~\ref{fig:Region1_vollmer_summary_4_47} to \ref{fig:Region3_vollmer_summary_4_47}).
As a result, the offset between the H$\alpha$ and NUV tails is much less pronounced than
it is observed. Moreover, the model NUV emission of the downturn region shows too much
extraplanar emission on the downwind side compared to the GALEX observations. The observed offset between the 
NUV and the H{\sc i} tail is absent in this model. This offset is only present in model 1cnew\_47 and the observed NUV
emission distribution is best reproduced by this model snapshot. However, model 1cnew\_47 shows too much extraplanar H{\sc i} emission in the
downturn region. Based on the comparison of all observed characteristics with our model snapshots (Table~\ref{tab:goodness}) we 
conclude that model 4new\_47 fits the available observations including the optical spectra best.

\subsection{The optical spectra \label{sec:spectrum}}

\begin{figure*}[!ht]
  \centering
  \resizebox{\hsize}{!}{\includegraphics{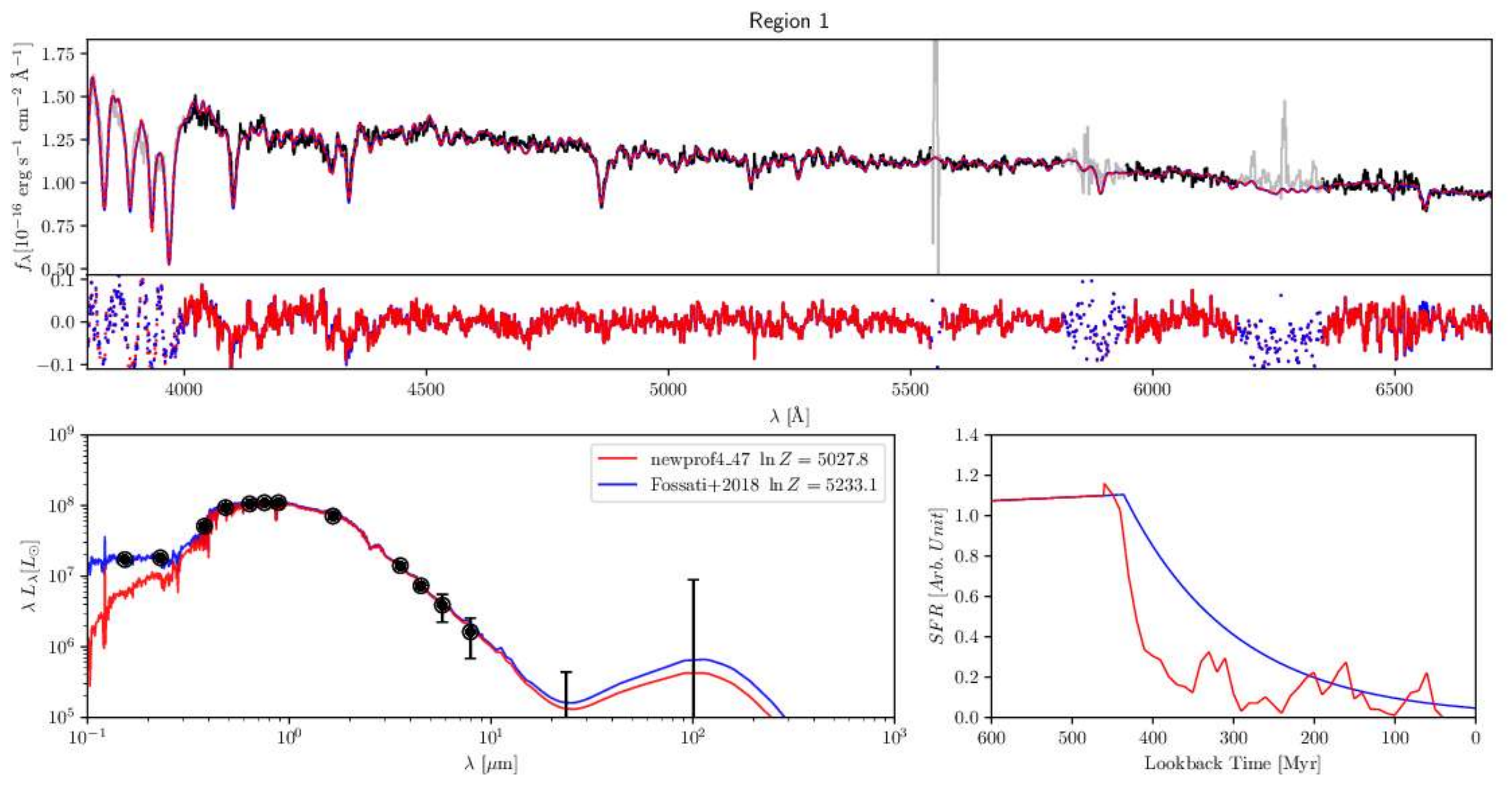}}
  \caption{Region 1: upper panel: FORS2 spectrum (black), Fossati et al. model (blue), model 4new\_47 with a diffuse ISM component and a variable Ly$_{\rm scale}$ scale parameter (red). Middle panel: fit residuals. Lower left panel: observed SED (black), Fossati et al. model (blue), model 4new\_47 (red). Lower right panel: star formation history; Fossati et al. model (blue), model 4new\_47 (red).
  \label{fig:Region1_vollmer_summary_4_47}}
\end{figure*}

\begin{figure*}[!ht]
  \centering
  \resizebox{\hsize}{!}{\includegraphics{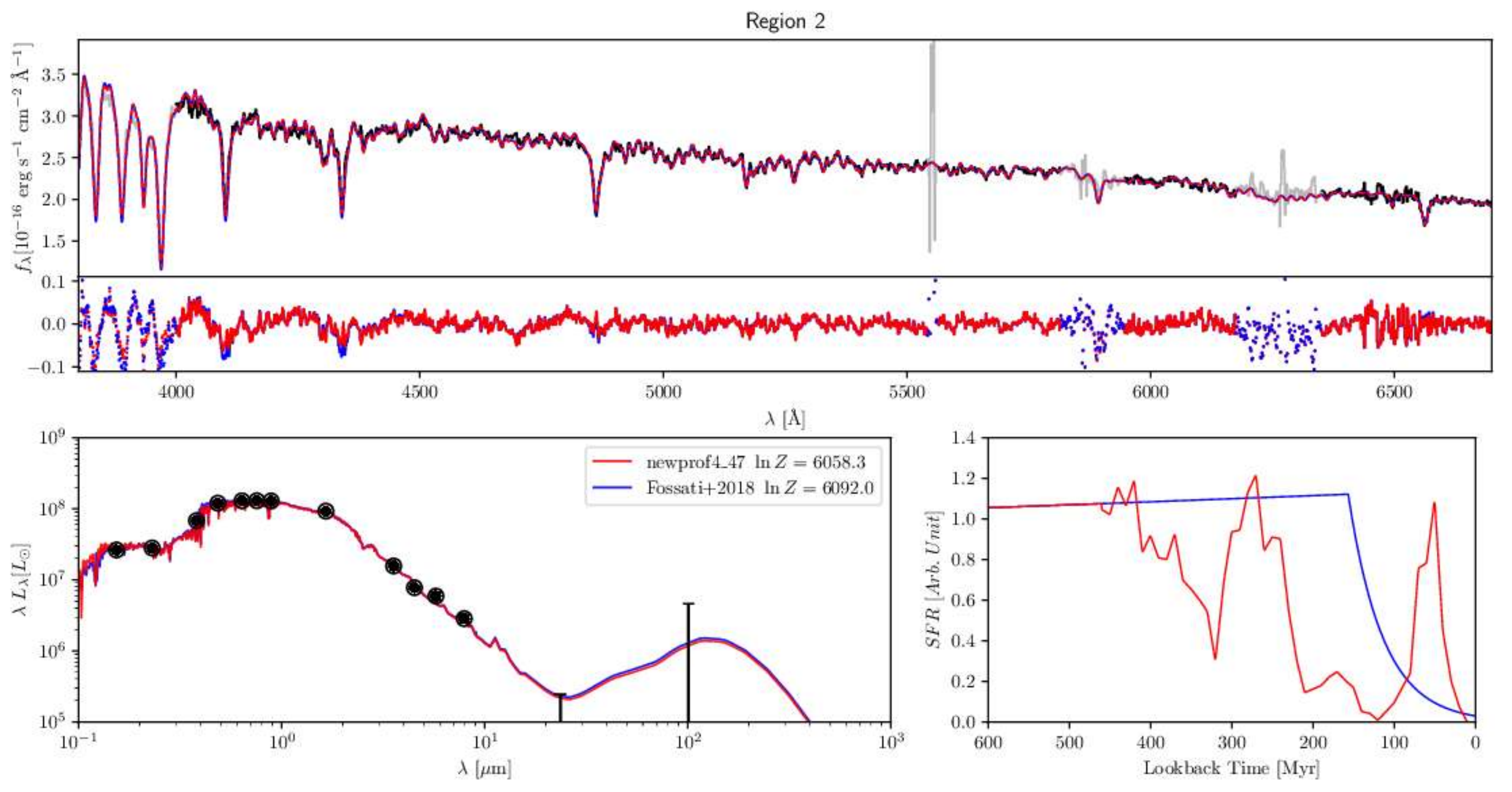}}
  \caption{Same as Fig.~\ref{fig:Region1_vollmer_summary_4_47} for region 2.
  \label{fig:Region2_vollmer_summary_4_47}}
\end{figure*}

\begin{figure*}[!ht]
  \centering
  \resizebox{\hsize}{!}{\includegraphics{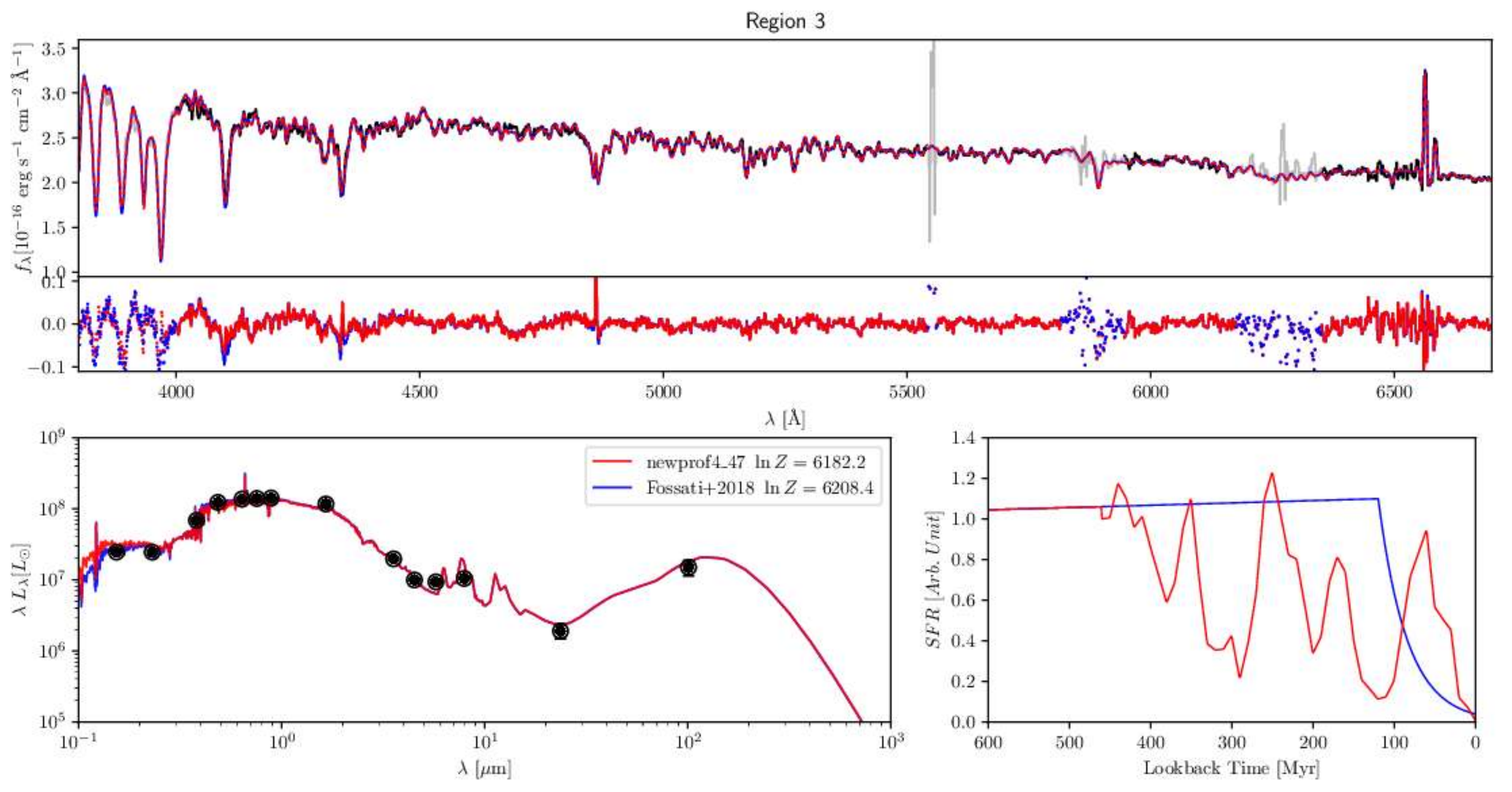}}
  \caption{Same as Fig.~\ref{fig:Region1_vollmer_summary_4_47} for region 3.
  \label{fig:Region3_vollmer_summary_4_47}}
\end{figure*}

\begin{figure*}[!ht]
  \centering
  \resizebox{\hsize}{!}{\includegraphics{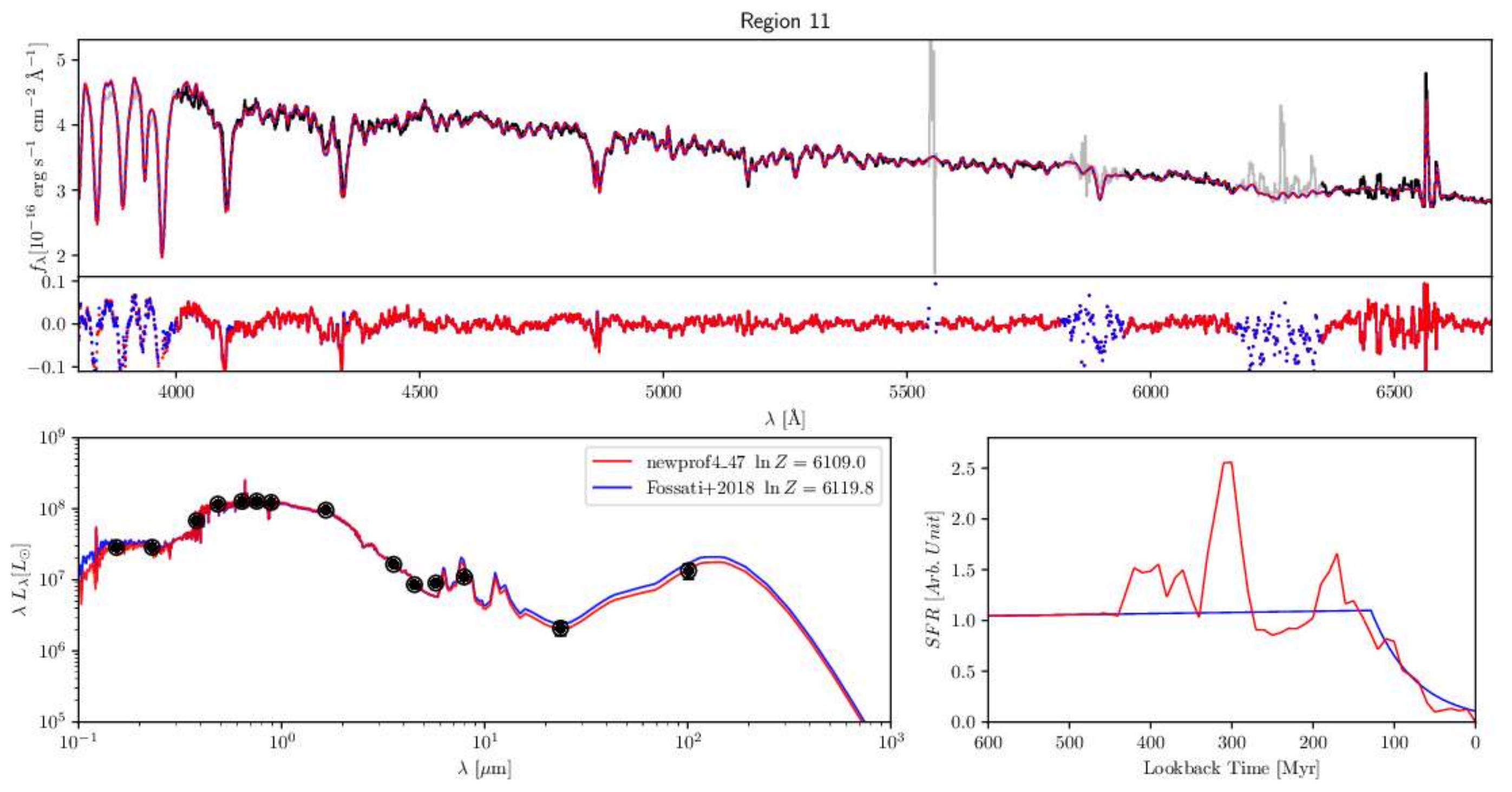}}
  \caption{Same as Fig.~\ref{fig:Region1_vollmer_summary_4_47} for region 11.
  \label{fig:Region11_vollmer_summary_4_47}}
\end{figure*}

\begin{figure*}[!ht]
  \centering
  \resizebox{\hsize}{!}{\includegraphics{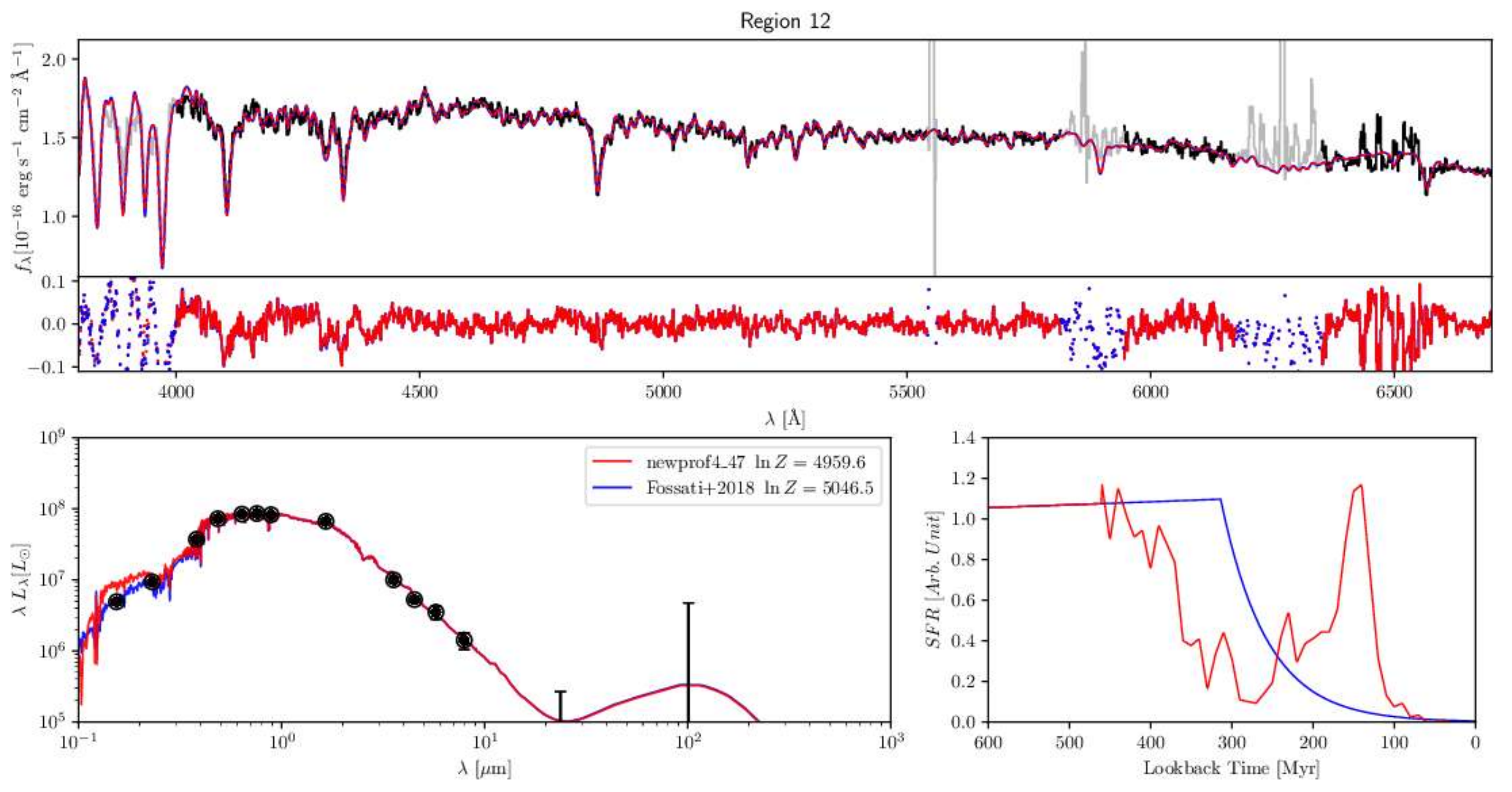}}
  \caption{Same as Fig.~\ref{fig:Region1_vollmer_summary_4_47} for region 12.
  \label{fig:Region12_vollmer_summary_4_47}}
\end{figure*}

\begin{figure*}[!ht]
  \centering
  \resizebox{\hsize}{!}{\includegraphics{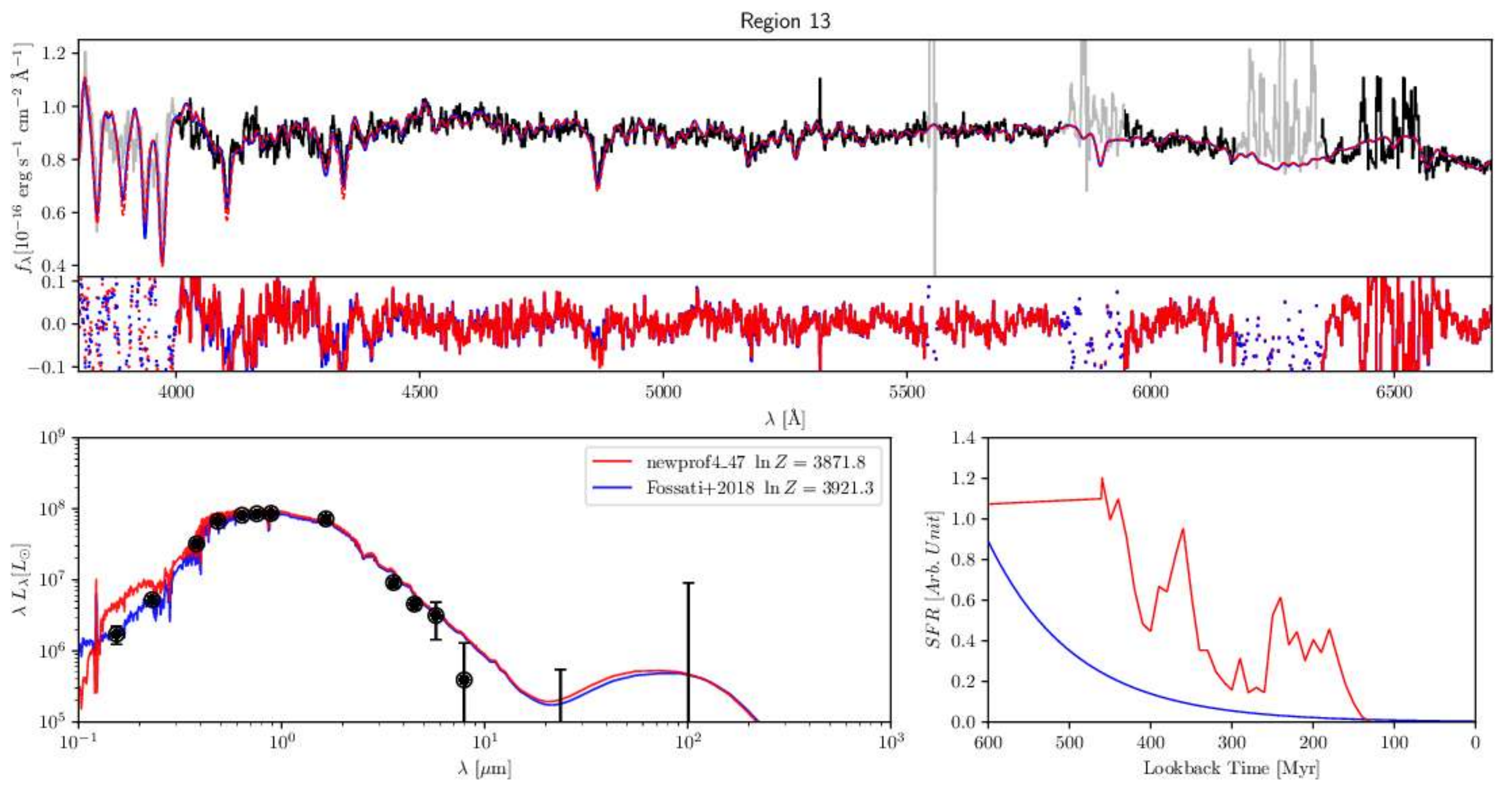}}
  \caption{Same as Fig.~\ref{fig:Region1_vollmer_summary_4_47} for region 13.
  \label{fig:Region13_vollmer_summary_4_47}}
\end{figure*}

The model optical spectra  and the model spectral energy distribution (SED) for the ``best-fit'' model 4new\_47 
(including diffuse gas stripping and a variable nuisance parameter Ly$_{\rm scale}$) are presented
in Figs.~\ref{fig:Region1_vollmer_summary_4_47} to \ref{fig:Region13_vollmer_summary_4_47} together with the star formation history for 
the regions labeled in Fig.~\ref{fig:N4330_fig3_NUV}. 
The optical to NIR SED only depends on the SFH prior to the onset of the simulations and is always
well reproduced by the models. The FUV and NUV flux densities are well reproduced by the model in regions~2, 3, and 11.
They are somewhat underestimated by the model in downturn regions~12 and 13 and overestimated in the outer tail region~1. 
The optical spectra of all positions are well reproduced by the model. For a seamless comparison we added the results of 
Fossati et al. (2018) for each position to Figs.~\ref{fig:Region1_vollmer_summary_4_47} to \ref{fig:Region13_vollmer_summary_4_47}
\footnote{The spectra of regions~1, 3, 11, 12, and 13 are not shown in Fossati et al. (2018).}. 
It becomes clear that the good resemblance between the model and observations for 
all regions is caused by the good resemblance between the observed and modeled star formation histories.
Whereas the SFHs of Fossati et al. (2018) are analytical with exponentially declining SFR after gas stripping,
the model SFHs show more variations. If the model SFHs show a late peak of the SFR after the general decline of the SFR
(region~2 and 12), the onset of the decline occurs earlier than that derived with a simple analytical function
as in Fossati et al. (2018) because this additional young stellar
population has to be compensated by a removal of stellar populations created just before the onset of SFR quenching.
We conclude that our ``best-fit'' model 4new\_47 reproduces the observed optical spectra and SED in a satisfactory way.   

\subsection{A constant nuisance parameter Ly$_{\rm scale}$ \label{sec:nuisance}}

For the calculation of the model goodness the nuisance parameter Ly$_{\rm scale}$ was varied systematically between $1/3$ and $3$ 
to search for the value which lead to the highest goodness. 
The parameter Ly$_{\rm scale}$ takes into account that either some UV flux from the massive stars located in a given
aperture escapes the region without ionizing the gas or that massive star outside the aperture ionize gas located inside the aperture. 
In addition, ram pressure induced shock can in principle lead to the ionization
of the ambient gas. In the first case Ly$_{\rm scale}$ is smaller than one, in the two latter cases it exceeds unity. 
The resulting Ly$_{\rm scale}$ parameters for Fossati et al. (2018) model SFH and the SFHs derived from the dynamical models with the highest goodnesses 
are presented in Table~\ref{tab:lalpha}. 
\begin{table}[!ht]
      \caption{Nuisance parameter Ly$_{\rm scale}$.}
         \label{tab:lalpha}
      \[
         \begin{array}{lcccccc}
           \hline
           {\rm Region\ (Fig.~\ref{fig:N4330_fig3_NUV})}     &              {\rm \ R1\ }  &  {\rm \ R2\ }  &  {\rm \ R3\ }  &  {\rm \ R11\ } &  {\rm \ R12\ }  & {\rm \ R13\ } \\
           \hline
           {\rm Fossati\ et\ al.} &  0.34 & 0.35 & 0.75 & 0.60 & 0.50 & 0.49 \\
           {\rm 4new\_47}  &  1.17 & 1.21 & 1.83 & 1.60 & 0.90 & 1.42 \\
           {\rm 4anew\_60} &  0.73 & 1.20 & 1.39 & 0.57 & 0.84 & 0.91 \\
           {\rm 4b\_67}    & 1.39 & 1.13 & 1.39 & 1.28 & 1.09 & 0.89 \\
           {\rm 2b\_80}    & 0.64 & 0.34 & 1.21 & 0.62 & 0.84 & 0.97 \\
           \noalign{\smallskip}
           \hline
         \end{array}
         \]
\end{table}
The analytical SFHs of Fossati et al. (2018) lead to Ly$_{\rm scale}$ smaller than one for all regions. Its values range from $0.33$ to $0.75$.
The Ly$_{\rm scale}$ parameter found for the dynamical models can exceed unity ranging from $0.33$ to $1.83$. Especially our ``best-fit'' model
4new\_47 leads to Ly$_{\rm scale} > 1$ in five out of six regions. Since these values are not exceedingly high, we think that this model
is well acceptable. Values exceeding unity could imply that ram pressure induced shock probably play a role in ionizing the
ambient ISM. However, the high flexibility of the fitting model and the PSF mismatch between the spectrum and the
photometry prevent such a conclusion.

All spectra and SEDs were recalculated using the values of the Ly$_{\rm scale}$ parameter found by Fossati et al. (2018).
The resulting model snapshots with the highest goodnesses are presented in Table~\ref{tab:goodness} for the simulations
with and without a diffuse gas stripping. As stated previously, all goodnesses of the models without diffuse gas
stripping are smaller than those of the model with diffuse gas stripping. In addition, a fixed Ly$_{\rm scale}$ parameter for
each region lead to highest goodnesses for snapshots at later times than those for the models with a variable Ly$_{\rm scale}$ parameter.

The resulting model H{\sc i}, NUV, and H$\alpha$ images are presented in Fig.~\ref{fig:zusammen_nuv_newprofs-1a} to \ref{fig:zusammen_nuv_profs-4a}.
Whereas the model H{\sc i} distributions are similar to the models with a variable Ly$_{\rm scale}$ parameter, the snapshots from the
models including a fixed Ly$_{\rm scale}$ parameter all show extraplanar NUV emission in the upturn region on the upwind side
(Figs.~\ref{fig:zusammen_nuv_newprofs-1a} and \ref{fig:zusammen_nuv_newprofs-3a}). 
This NUV emission, which is not observed, is due to massive stars created in the tail region which had the time to rotate in to the upturn region 
(see Sect.~\ref{sec:hinuv}).
In addition, none of the models with a fixed Ly$_{\rm scale}$ parameter shows the vertical low surface brightness H$\alpha$ filaments on the downwind side of the
NUV/H$\alpha$ tail.
We therefore conclude that the models with a fixed Ly$_{\rm scale}$ parameter should be discarded
{and this parameter has primarily a nuisance role in the fit rather than a physical meaning.}

\section{Discussion\label{sec:discussion}}

The inclusion of diffuse gas stripping significantly changed the results of our simulations. Since a variable Ly$_{\rm scale}$ parameter leads to the
highest goodnesses, we do not consider the models with a constant  Ly$_{\rm scale}$ parameter in the following analysis.
When the tail region is considered separately, the timesteps of interest are $\sim 50$~Myr earlier before peak ram pressure when diffuse gas stripping is included
in the simulations (Table~\ref{tab:goodness1}).

This is expected, because the gas which is pushed  into the tail becomes diffuse and is stripped with a higher efficiency than the denser gas.
Unexpectedly, the opposite trend is found for the downturn region: the timesteps of interest are $\sim 70$~Myr earlier before peak ram pressure when diffuse gas stripping 
is not included. The reason for this behaviour is the following: as expected, the diffuse gas is stripped rapidly out of the galactic disk on the windward 
side when diffuse gas stripping is included. At the same time, the gas which remains within the disk is compressed to higher densities compared to the
simulations without diffuse gas stripping. This results into a more compact gas distribution
at $t \ga 200$~Myr after the onset of the simulations, i.e. the surface density contrast of the gas located within the galactic disk is enhanced.
Since more time is needed to push the outer dense arms to smaller galactic radii, the timesteps of interest for the downturn (windward) side are delayed 
when diffuse gas stripping is included. The bulk of the star formation occurs in the dense gas arms.

The situation changes when both, the downturn and tail regions, are fitted together (Table~\ref{tab:goodness}). 
Generally, the H$\alpha$ and NUV emission distributions of the downturn region are more often reproduced when diffuse gas stripping is not included in the simulation.
Overall, the H{\sc i} distribution of the downturn region is hard to reproduce (4new\_47 and 4\_d70).
The best reproductions of the observed H{\sc i} distribution of the tail are only found in models with diffuse gas stripping.
The H$\alpha$ emission distribution of the tail region is as often reproduced when diffuse gas stripping is not included, but only
for peak and post-peak models. Most frequently, post-peak models are not allowed, because they show a NUV horn feature in the leeward side of the downturn 
region (see Sect.~\ref{sec:hinuv}).

Whereas the two acceptable models with diffuse gas stripping 
according to the total goodness (4new\_47 and 1cnew\_47) have $\Delta t_{\rm peak}=t-t_{\rm peak} \sim -170$~Myr, i.e. they are pre-peak models, the two 
acceptable models without diffuse gas stripping  (4b\_67 and 2b\_80) are post-peak models ($\Delta t_{\rm peak} > 0$~Myr). 
Based on the comparison
between the model and observed H{\sc i}, NUV, and H$\alpha$ distributions the pre-peak models 4new\_47 and 1cnew\_47 are clearly preferred (Sect.~\ref{sec:bestfit}).

\subsection{The ``best-fit'' model}

\begin{figure}[!ht]
  \centering
  \resizebox{\hsize}{!}{\includegraphics{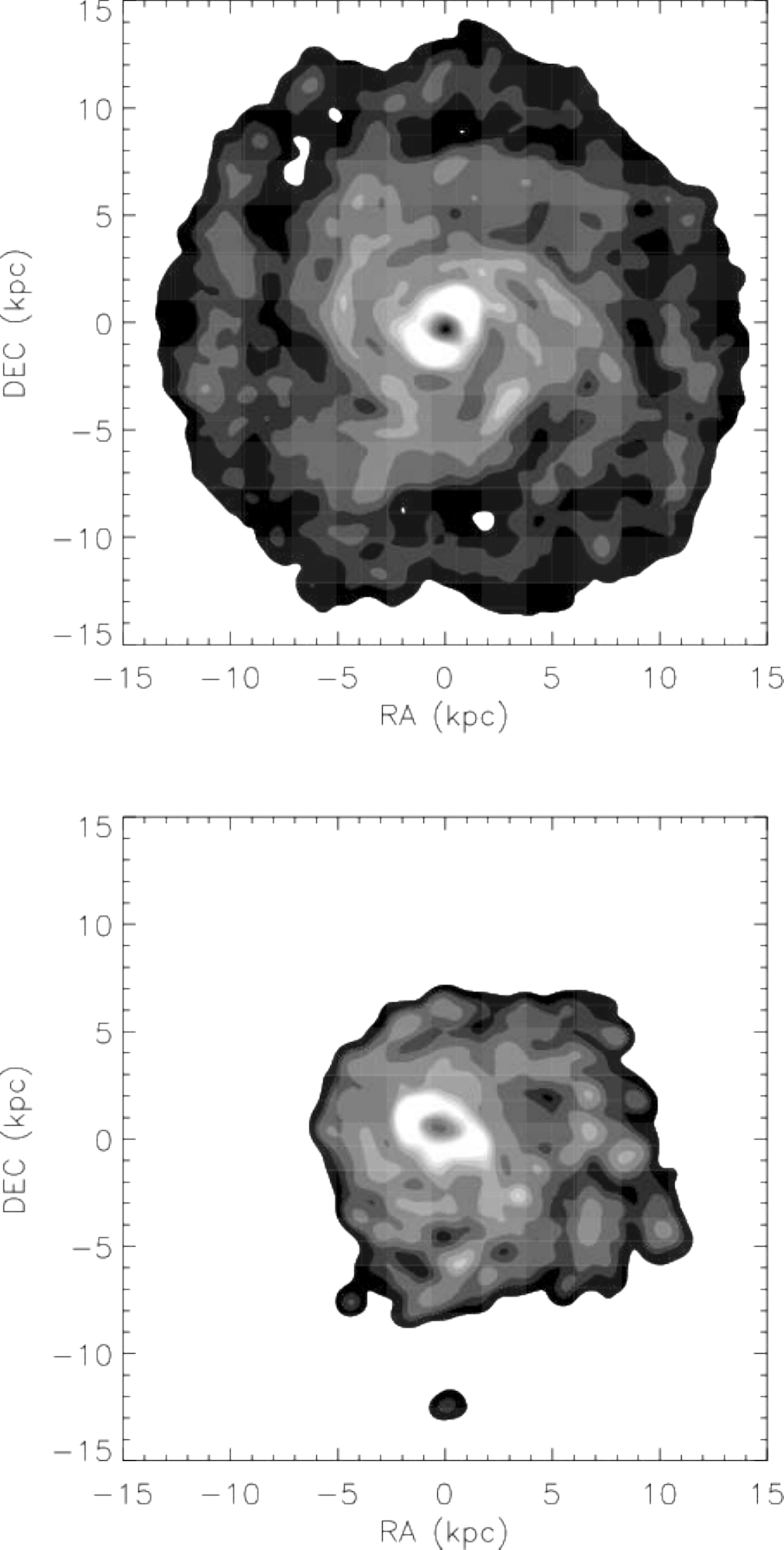}}
  \caption{Face-on projection of the distribution of the total gas surface density at the beginning of the simulation (upper panel) and 
    at $\Delta t = -140$~Myr (lower panel). The contours correspond to 1,3,5,10,20,30,40,50~M$_{\odot}$pc$^{-2}$. The resolution is 820~pc.
  \label{fig:inidist}}
\end{figure}

The ``best-fit'' model is 4new\_47 with a maximum ram pressure of $p_{\rm max}=6000$~cm$^{-3}$(km\,s$^{-1}$)$^2$, a width of the Lorentzian temporal
profile of $t_{\rm HW}=200$~Myr, and a time to peak ram pressure of $\Delta t = t - t_{\rm peak}=-140$~Myr. 
The face-on projection of the distribution of the total gas surface density at the beginning of the simulation and at $\Delta t = -140$~Myr
is presented in Fig.~\ref{fig:inidist}. The initial gas distribution of the galaxy has multiple spiral arms with surface densities between 
10 and 20~M$_{\odot}$pc$^{-2}$. During the ram pressure stripping event we observe a highly non-linear interaction between these arms and the 
induced gas compression. The gas distribution at $\Delta t = -140$~Myr has a compressed windward side and a gas tail with a multi-arm
structure on the opposite side.

The model includes the stripping of diffuse gas and
a variable Ly$_{\rm scale}$ parameter for the calculation of the model optical spectra and thus the model goodness. 
The interaction parameters are close to those found by Vollmer et al. (2012): $p_{\rm max}=5000$~cm$^{-3}$(km\,s$^{-1}$)$^2$, $t_{\rm HW}=100$~Myr, and
$\Delta t = -100$~Myr. The Vollmer et al. (2012) was not able to reproduce the UV color in the stripped gas-free regions, because the
width of the Lorentzian profile was too small. A larger width permits an earlier gas stripping which is consistent with the available
UV observations and the optical spectra. In addition, the inclusion of diffuse gas stripping is essential. Only if this
ingredient is added to the simulations the gas leaves the galactic disk early enough to lead to FUV/NUV flux densities and optical spectra
consistent with observations. The diffuse gas stripping leads to the formation of linear filaments of diffuse ionized gas which
is present in deep VESTIGE H$\alpha$ observations. The model filaments are thicker than the observed filaments
because of the coarse spatial resolution of the diffuse gas particles.
It should be noted that pronounced filamentary  structures  are  naturally produced 
in  simulations whenever magnetic fields are taken into account (Ruszkowski et al. 2014; Tonnesen \& Stone 2014).
It has to be shown for other galaxies undergoing ram pressure stripping if this effect is ubiquitous.

\subsection{Ram pressure stripping of cosmic ray electrons}

\begin{figure}[!ht]
  \centering
  \resizebox{\hsize}{!}{\includegraphics{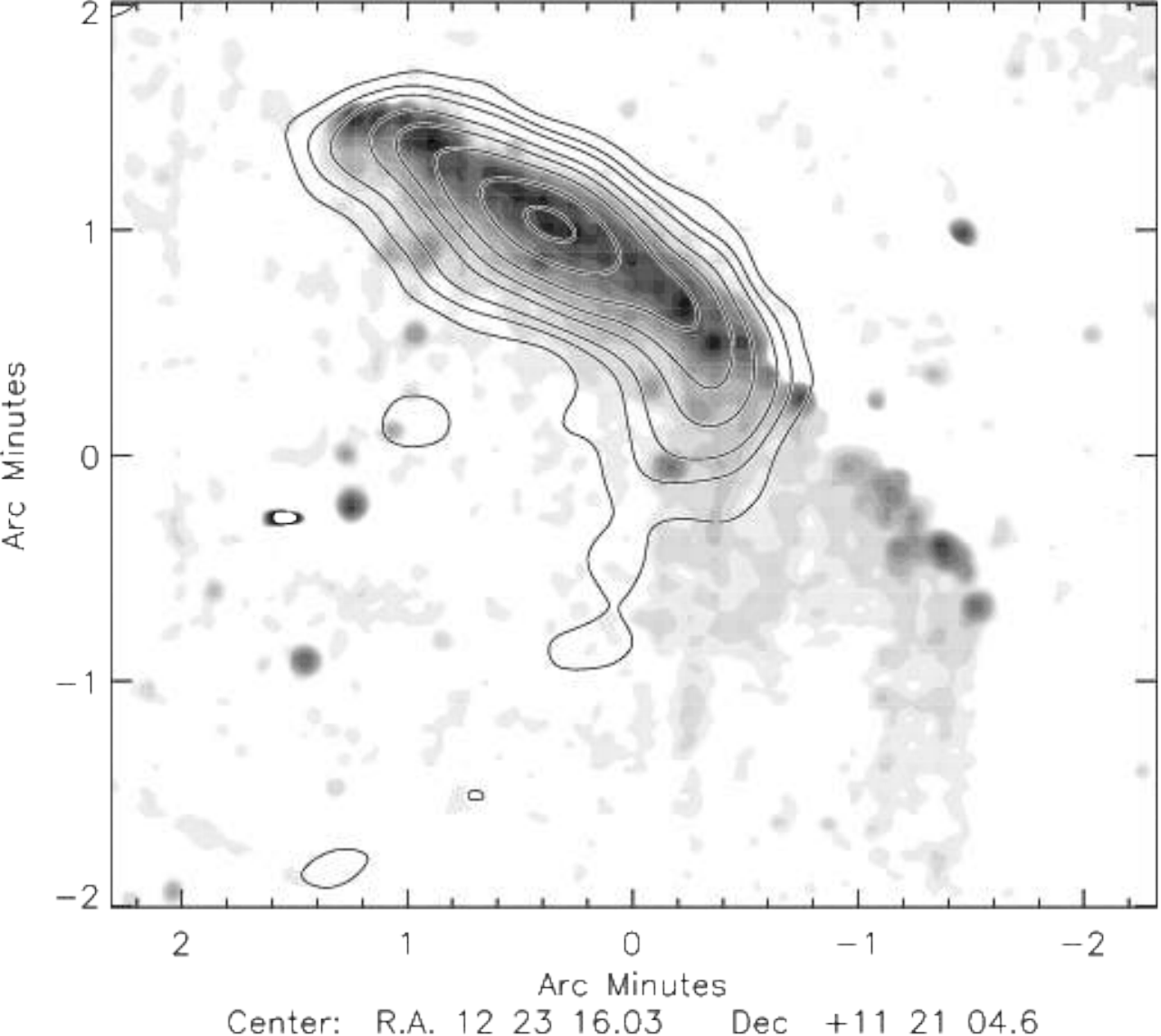}}
  \resizebox{\hsize}{!}{\includegraphics{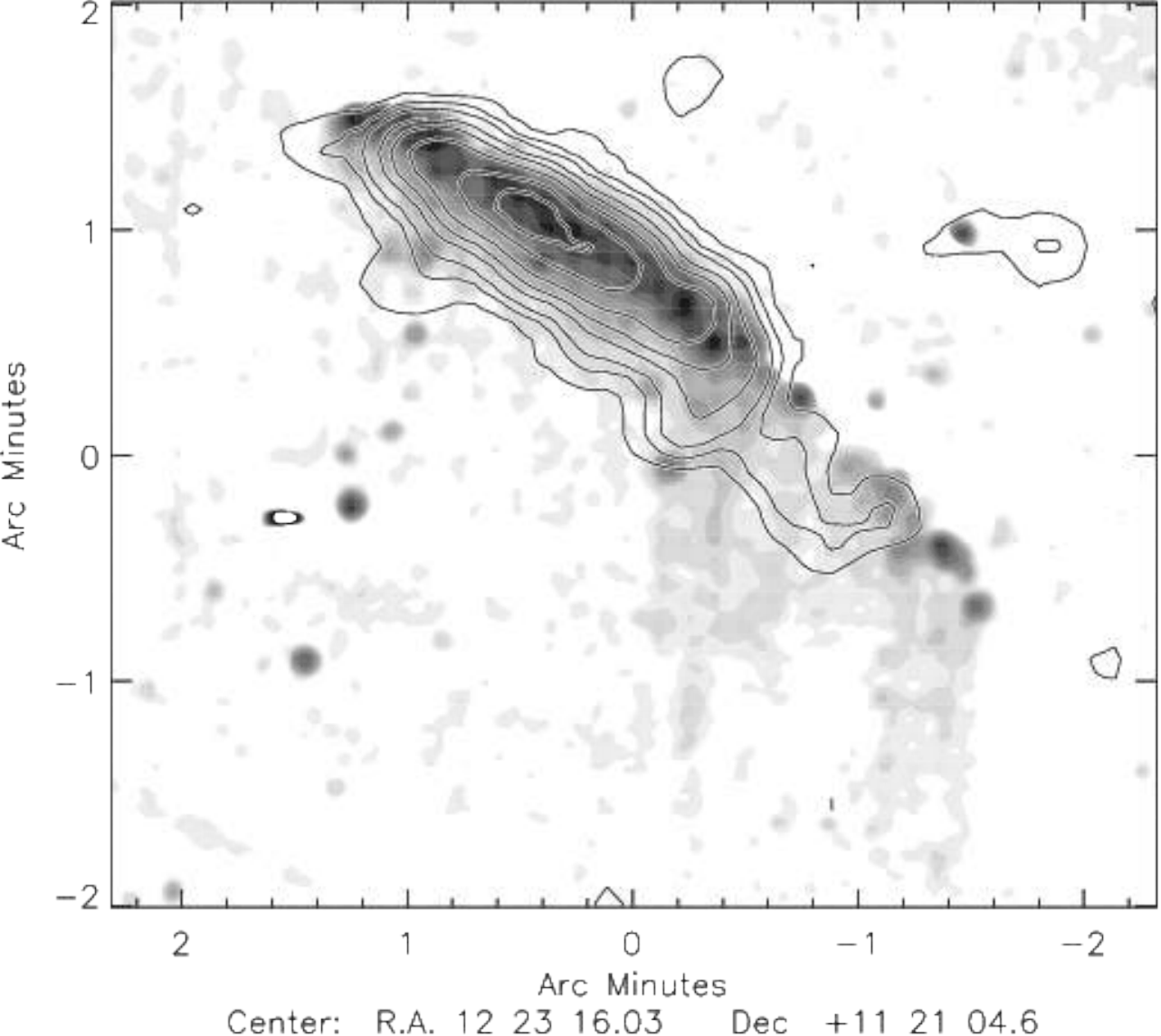}}
  \caption{Radio continuum emission (contours) on the VESTIGE H$\alpha$ map. Upper panel: emission at 6cm. 
The contour levels are $(1,2,3,5,7,10,15,20,30,50,100) \times 50$~$\mu$Jy/beam. The resolution is $(22'' \times 22'')$. 
Lower panel: emission at 20cm from VIVA (Chung et al. 2009). The contour levels are $(1,2,3,5,7,10,15,20,30,50,100) \times 75$~$\mu$Jy/beam.
The resolution is $(19.4'' \times 15.4'')$.
  \label{fig:Halpha_RC}}
\end{figure}

NGC~4330 does not show an asymmetric ridge of polarized emission as other ram pressure stripped galaxies (Vollmer et al. 2012).
This is due to the NGC~4330's projection which places the region of compressed gas along the line of sight.
On the other hand, Vollmer et al. (2012) discovered a tail structure bending southward in their 6~cm radio continuum data 
(upper panel of Fig.~\ref{fig:Halpha_RC}). This tail structure is also visible in the VIVA $20$~cm data (lower panel of Fig.~\ref{fig:Halpha_RC}).
In addition, an outer tail of low surface brightness extending toward the southwest is detected at $20$~cm.
The tail, which is detected at $6$ and $20$~cm, bends to the south at the southwestern edge of the starforming disk. 
It coincides with the northern part of the low surface
brightness diffuse H$\alpha$ emission. The cosmic ray electrons produced in the luminous H{\sc ii} regions of the galactic disk
are most probably pushed to the south by ram pressure. This hypothesis is corroborated by the vertical magnetic field lines
detected within the radio continuum tail (Vollmer et al. 2013). The spatial coincidence of the radio continuum and diffuse
H$\alpha$ tails suggests that both gas phases are stripped together. 

We do not observe a steepening of the spectral index in the $6$ and $20$~cm tail as expected for an aging cosmic
electron population. The spectral index is given by $\alpha=\ln(f_6/f_{20})/\ln(\nu_6/\nu_{20})$, where $f$ is the surface brightness or
flux density. 
For a purely aging population of cosmic ray electrons the evolution of the flux density is given by $f=f_0 \times \exp(-t/t_{\rm syn})$.
The time which is needed for a steepening of the spectral index from an initial value $\alpha_{\rm in}$ to a final value
$\alpha_{\rm end}$ can be calculated in the following way:
\begin{equation}
t_{\rm SI}=\frac{(\alpha_{\rm in}-\alpha_{\rm end})\,\ln(\nu_6/\nu_{20})}{1/t_{\rm syn,6cm}-1/t_{\rm syn,20cm}}\ .
\end{equation}
Based on equipartition between the energy densities of the cosmic ray electrons and the magnetic field we estimate the magnetic field strength 
to be $B \sim 6$-$7~\mu$G in the tail region assuming an extent along the line-of-sight of $l \sim 4$~kpc. 
The associated synchrotron timescale at $6$~cm is then
\begin{equation}
t_{\rm syn}= 4.5 \times 10^7 \big( \frac{B}{10~\mu{\rm G}} \big)^{-3/2} \nu_{\rm GHz}^{-1/2}~{\rm yr} \sim 40~{\rm Myr}\ .
\end{equation}
If we assume that $\alpha_{\rm in}-\alpha_{\rm end} \la 0.2$ in the $6$ and $20$~cm tail the timescale for the steepening
of the spectral index is $t_{\rm SI} \la 24$~Myr.
With a north-south extent of $3.5$~kpc the southward bulk velocity is $\ga 140$~km\,s$^{-1}$ within the sky plane.

Assuming that the outer radio continuum tail is barely detected at a $2\sigma$ level at $6$~cm, we estimate a spectral
index between $6$ and $20$~cm of $\sim -1.5$. 
With an initial spectral index of $\alpha_{\rm in}=-0.7$ to $-0.5$, a final spectral index $\alpha_{\rm in}=-1.5$, 
and a magnetic field strength of $B=6~\mu$G we obtain $t_{\rm SI} \sim 90$ to $120$~Myr.
If the cosmic ray electrons were produced in the normally starforming disk before gas stripping $t_{\rm SI}$ should be comparable to
the star formation quenching time derived for that region which corresponds to region~2 in Fossati et al. (2018).
Whereas Fossati et al. (2018) determined a quenching time of $175$~Myr based on their simple analytical star formation histories,
our preferred model shows that a more recent star formation episode (about $50$~Myr ago; Fig.~\ref{fig:Region2_vollmer_summary_4_47}) 
is also consistent with the data.
It is not clear if such an episode can create enough cosmic ray electrons to make the synchrotron emission detectable. 
On the other hand, a somewhat lower magnetic field strength of $B=4$-$5~\mu$G would lead to $t_{\rm SI} \sim 175$~Myr. 
We thus conclude that a scenario in which the cosmic ray electrons were produced in the healthy disk before gas stripping
and were then pushed southward together with the diffuse ionized gas is consistent with the available data.

It is surprising that there is no $20$~cm emission associated with the upper part of the
outer H$\alpha$ tail where we see compact emission, most probably H{\sc ii} regions.
Either these H{\sc ii} regions are too young and did not produce cosmic ray electrons yet or the production of
cosmic ray electrons by these regions is low. We suggest that the present star formation in the outer tail is sporadic and low level
and this explains the absence of a significant amount of cosmic ray electrons there.

\subsection{How to ionize the stripped diffuse gas}

It is surprising that the surface brightness of the H$\alpha$ filaments does not systematically decrease southward. 
These filaments therefore need to be confined by the ambient ICM during $\sim 60$~Myr without recombining, i.e.
there must be a continuous ionization mechanism. In the following we argue that the stripped diffuse gas is ionized by thermal conduction.
With a recombination timescale $t_{\rm rec}=0.1\,(n_{\rm e}/1~{\rm cm}^{-3})^{-1}$~Myr=$60$~Myr this yields an electron density of 
$n_{\rm e} \sim 2 \times 10^{-3}$~cm$^{-3}$. 
The mean electron density of the H$\alpha$ filaments can be estimated by the relation
\begin{equation}
L(H\alpha)=n_{\rm e}n_{\rm p}\alpha_{{\rm H}\alpha}^{\rm eff}V\,\phi_{\rm V}\,h\,\nu_{{\rm H}\alpha}\ ,
\end{equation}
(Osterbrock \& Ferland 2006) where $n_{\rm p}=n_{\rm e}$ is the proton density, $\alpha_{{\rm H}\alpha}^{\rm eff}=1.17 \times 10^{-13}$~cm$^3$s$^{-1}$
is the H$\alpha$ effective recombination coefficient, $V$ is the volume of the emitting region, $\phi_{\rm V}$ its volume filling factor,
$h$ the Planck constant, and $\nu_{{\rm H}\alpha}$ the frequency of the H$\alpha$ transition.
The volume $V$ is the product of the emitting surface and the extent along the line-of-sight $l \sim 4$~kpc.
Following Boselli et al. (2016) we assume $\phi_{\rm V}=0.1$ and [N{\sc ii}]/H$\alpha$=0.5.
With a mean H$\alpha$+[N{\sc ii}] surface brightness of $4 \times 10^{-18}$~erg\,cm$^{-2}$s$^{-1}$arcsec$^{-2}$ (Fossati et al. 2018)
we obtain a mean electron density of $n_{\rm e}=0.05$~cm$^{-3}$. With an emitting surface of $25$~kpc$^2$ the mass of ionized gas
in the filaments is about $10^7$~M$_{\odot}$.

Since the derived mean electron density of the filaments exceeds that derived from the recombination timescale,
the hydrogen atoms in the filaments have to be continuously ionized.
Boselli et al. (2016) reached the same conclusion for the long, low surface brightness H$\alpha$ tails of NGC~4569, another
ram-pressure stripped galaxy in the Virgo cluster.
The source of ionization might be (i) collisional ionization through shocks, (ii) collisional ionization through the thermal electrons of 
the ambient ICM which confines the filaments, 
or (iii) radiative ionization by embedded massive stars. We discard the latter hypothesis because the gas density in the filaments
is not high enough to permit continuous star formation. Hypothesis (i) is improbable because the direction of filaments is aligned 
with the ram pressure wind and it is difficult to perceive how ram pressure induced shocks can pervade the entire filaments.  

We are thus left with hypothesis (ii): if we assume that the timescale for evaporation by the ICM equals the
collisional ionization timescale by ICM-ISM collisions $t_{\rm ICM-ISM}=t_{\rm evap}=10\,(N_{\rm H}/10^{20}~{\rm cm}^{-2})$~Myr (Vollmer et al. 2001) and 
if the recombination timescale is $t_{\rm rec}=0.1\,(0.3~{\rm cm}^{-3}/1~{\rm cm}^{-3})^{-1}$~Myr=$0.3$~Myr, we obtain a hydrogen column density of
$N_{\rm H} \sim 3 \times 10^{18}$~cm$^{-2}$. This is well below the VIVA detection limit ($\sim 2 \times 10^{19}$~cm$^{-2}$; Abramson et al. 2011).
We thus suggest that the most probable mechanism for the ionization of the H$\alpha$ filaments are collisions between the ICM
thermal electrons and the hydrogen atoms of the filaments 
(as in ionized gas filaments in cooling flow clusters; see e.g. Mc Donald et al. (2010).

\section{Summary and conclusions \label{sec:conclusions}}

The edge-on Virgo spiral galaxy NGC~4330 shows signs of ongoing ram pressure stripping: a truncated H{\sc i} disk together with a
one-sided tail (Chung et al. 2007, 2009), a UV tail associated with the H{\sc i} tail (Abramson et al. 2011), a truncated H$\alpha$
disk (Abramson et al. 2011), and an H$\alpha$ tail with vertical low surface density filaments (Fossati et al. 2018; upper panels of Fig.~\ref{fig:Halpha_HI}).
There is no asymmetric ridge of polarized radio continuum emission as observed in other ram-pressure stripped Virgo galaxies
because the compression is occurring in the sky plane (Vollmer et al. 2012). A previous dynamical model could reproduce all observed
characteristics except the UV color of the stripped gas-free regions of the outer galactic disk (Vollmer et al. 2012).
Since the UV color depends on the star formation history of the observed region, the stripping timescale of the dynamical model was not
consistent with observations.

Fossati et al. (2018) obtained VLT FORS2 deep optical spectra along the major axis of NGC~4330. In addition, they collected literature photometry 
in $15$ bands from the far-UV to the far-IR. Star formation histories in apertures along the major axis of the galaxy were constructed
by means of stellar population fitting to the joint observational data, SED and optical spectrum. 
Based on an analytical model of quenched star formation histories, they found that 
the outermost radii of the galactic disk were stripped $\sim 500$~Myr ago, while the stripping reached the inner $5$~kpc in the last $100$~Myr. 

We present new dynamical simulations to reproduce all observed characteristics of NGC~4330: the H{\sc i}, UV, and H$\alpha$ morphologies
together with the SEDs and the optical spectra. The dense gas is modeled by a sticky particle scheme (Sect.~\ref{sec:model}).
The influence of galactic structure, i.e. spiral arms, on the gas stripping process was studied by
delaying the ram pressure peak with respect to the beginning of the simulations. Most importantly, we introduce a new recipe for
the stripping of diffuse gas with a low density and a high volume filling factor (Sect.~\ref{sec:iong}):

\noindent
Warm diffuse gas: once a gas particle has left the galactic disk and its density falls below a critical density $n_{\rm crit}^{\rm warm}$,
ram pressure is increased by ($0.044$cm$^{-3}/n_{\rm ISM})^{2/3}$. 

\noindent
Hot ($T > 10^6$~K) diffuse gas: temperatures are assigned to the gas particles. Once a gas particle has left the galactic disk and its density falls 
below a critical density $n_{\rm crit}^{\rm hot}$, ICM-ISM mixing sets in and increases the temperature immediately to $9 \times 10^6$~K.
If the density of the mixed gas increase through compression it can cool through X-ray emission according to its density
and temperature. The acceleration due to ram pressure is increased by a factor of ten
once the temperature of a particle exceeds $T=10^5$~K. Moreover, the equation of motion for the hot gas particles was modified 
by adding the acceleration caused by the gas pressure gradient. We do not solve an explicit energy equation and the calculation of
hydrodynamic effects is approximate. 

\noindent
We made $50$ simulations with five different Lorentzian temporal ram-pressure profiles, five different delays between the simulation onset and
peak ram pressure, with (newprof) and without (prof) diffuse gas stripping (Table~\ref{tab:molent}).
Model photometry and spectra were calculated for all timesteps ($\Delta t=10$~Myr) of the $50$ simulations. In addition, we decided to handle the 
Ly$_{\rm scale}$ nuisance parameter (Sect.~\ref{sec:observations}) in two different ways: (i) we set it to the fixed values found by Fossati et al. (2018) and
(ii) we treated it as a free parameter and kept the value that lead to the best fit to observations.
We then searched for the $12$ timesteps with the highest goodnesses for the downturn, tail, and both (downturn+tail) regions
(Table~\ref{tab:goodness}). The individual regions (downturn and tail) were generally not best fitted by the same model nor by the same timestep.
This is due to the limitations of our model where the gas stripping is treated with rather simple recipes (see Sect.~\ref{sec:iong})
and the limited number of temporal ram pressure profiles.
Our analysis is based on the combined goodness of both regions.

Based on the direct comparison between our model snapshots and the available observations (Table~\ref{tab:goodness}) we conclude that
the inclusion of diffuse gas stripping:
\begin{enumerate}
\item
changes significantly the gas, NUV (Figs.~\ref{fig:zusammen_nuv_newprofs-1} and \ref{fig:zusammen_nuv_profs_newprofs-2}), 
and H$\alpha$ morphologies (Figs.~\ref{fig:zusammen_ha_newprofs-1} and \ref{fig:zusammen_ha_profs_newprofs-1}) of the simulations.
\item
naturally leads to vertical low surface density filaments in the downwind region of the galactic disk 
(Fig.~\ref{fig:zusammen_ha_newprofs-1}).
These filaments occur less frequently in the simulations without diffuse gas stripping (Figs.~\ref{fig:zusammen_ha_profs_newprofs-1} and
\ref{fig:zusammen_ha_profs-1}).
\item
leads to better joint fits to the SEDs and optical spectra (Table~\ref{tab:goodness}).
In general, the SEDs and optical spectra of the downturn and tails regions fitted separately are best reproduced by different model snapshots 
(Table~\ref{tab:goodness}). We based our selection of the best-fit model on the goodness of the joint fit of the downturn and tail regions.
\item
significantly improves the resemblance between the model and observations.
\end{enumerate}
The H{\sc i}, NUV, and H$\alpha$ morphologies of the model snapshots which best reproduce the SEDs and optical spectra are sufficiently different
to permit a selection of best-fit models.
Model 4new\_47 is our preferred model (Table~\ref{tab:goodness}, lower panels of Fig.~\ref{fig:Halpha_HI}) 
with a maximum ram pressure of $p_{\rm max}=6000$~cm$^{-3}$(km\,s$^{-1}$)$^2$, a width of the Lorentzian temporal
profile of $t_{\rm HW}=200$~Myr, and a time to peak ram pressure of $\Delta t = -140$~Myr. The model includes the stripping of diffuse gas and
a variable Ly$_{\rm scale}$ parameter for the calculation of the model optical spectra. 
A variable Ly$_{\rm scale}$ nuisance parameter leads not only to higher goodnesses, i.e. better joint fits to the SEDs and optical spectra, as
expected but also to a much better reproduction of the observed H$\alpha$ and NUV images.
The interaction parameters are close to those found by Vollmer et al. (2012) who found a time to peak ram pressure of $\sim 100$~Myr.

The radio continuum morphology of the H$\alpha$ tail is twofold: (i) a tail of relatively high surface brightness detected at $6$~cm and $20$~cm 
originating from the southwestern part of the galactic disk which still actively forms stars and (ii) a low surface brightness tail associated 
with the outer H$\alpha$ tail with a steep spectral index only detected at $20$~cm (Fig.~\ref{fig:Halpha_RC}).  
The magnetic field strength of both radio continuum tails is estimated to be $B \sim 6$-$7$~$\mu$G.
The spatial coincidence of the radio continuum and diffuse H$\alpha$ tail suggests that both gas phases are stripped together. 
The southward bulk motion of the $6$~cm and $20$~cm is about $150$~km\,s$^{-1}$ with a timescale of $\sim 25$~Myr.
For the outer part of the H$\alpha$ tail a scenario in which the cosmic ray electrons were produced in the healthy disk before gas stripping
and were then pushed southward together with the diffuse ionized gas is consistent with the available data.
We suggest that the star formation in the outer tail is sporadic and low level and this explains the absence of a significant amount of cosmic ray electrons there.

Since the surface brightness of the H$\alpha$ filaments does not systematically decrease southward (upper right panel of Fig.~\ref{fig:Halpha_HI}), 
they need to be confined by the ambient ICM during 
$\sim 60$~Myr without recombining. We suggest that the mixed ISM is ionized by collisions with the thermal electrons of the ambient ICM which confines the filaments.

The modeling of diffuse gas stripping represents an important step to understand the reaction of the multiphase ISM to ram pressure.
The inclusion of diffuse gas stripping improves the resemblance between the model and observations (SED and optical spectra) and 
gives naturally rise to extended low surface density filaments oriented along the wind direction (diffuse H$\alpha$ emission). More detailed simulations of cluster galaxies affected by
ram pressure are needed to confirm our findings.

\begin{acknowledgements}
BV would like to thank R.~Beck for useful discussions.
We are grateful to the CFHT team who assisted us in the observations: T. Burdullis, D. Devost, B. Mahoney, N. Manset, A. Petric, S. Prunet, K. Withington.
The VESTIGE project is financially supported by the ``Programme National de Cosmologie and Galaxies'' (PNCG). 
MF has received funding from the European Research Council (ERC) under the European Union’s Horizon 2020 research and innovation programme (grant agreement No 757535).
We would like to thank the anonymous referee for her/his comments which helped to improve the article.
\end{acknowledgements}

\clearpage

\appendix

\section{ICM-ISM penetration length\label{sec:plength}}

\begin{figure}[!ht]
  \centering
  \resizebox{\hsize}{!}{\includegraphics{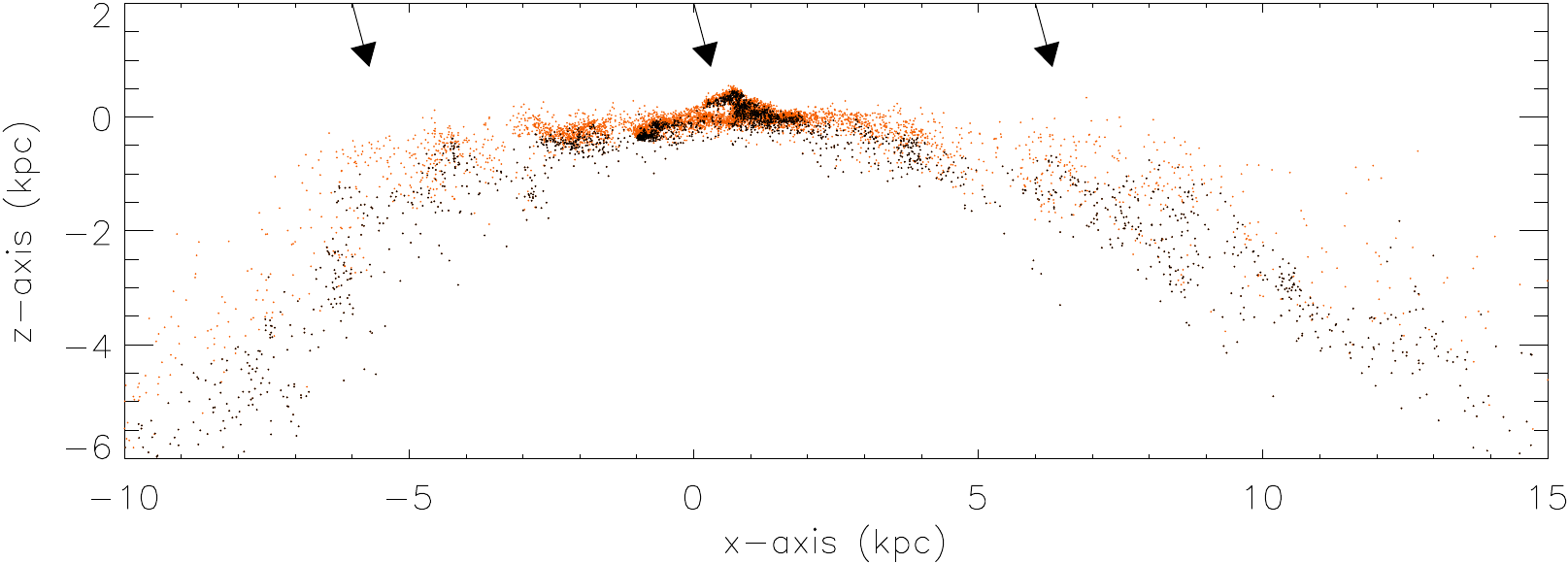}}
  \caption{Edge-on projection of a slice ($-2~{\rm pc} < y < 2~{\rm pc}$) of model~1 at a time to peak ram pressure $\Delta t_{\rm peak}=-100$~Myr.
    Black dots: gas clouds that are protected against the ram pressure wind by other clouds.
    Grey dots: gas clouds that are not protected against the ram pressure wind by other clouds.
    The arrows indicate the wind direction which is opposite to the direction of the galaxy's motion within the ICM.
  \label{fig:cloudprotect}}
\end{figure}

To estimate the model ICM penetration length into the ISM, we consider a slice ($-2~{\rm pc} < y < 2~{\rm pc}$) of model~1
at a time to peak ram pressure $\Delta t_{\rm peak}=-100$~Myr, when the gas tail is in place. 
In the chosen projection the wind direction has no component perpendicular to the image plane. It is therefore possible
to estimate the penetration length which by definition depends on the density of the gas clouds.
From Fig.~\ref{fig:cloudprotect} we estimate the penetration length to be about $100$~pc, $200$-$300$~pc, and about $1$~kpc 
for regions of highest, medium, and low density, respectively. The penetration length, which is the mean free path between clouds,
depends on the number of particles and the mass-size relation of the gas clouds. We used the relation found by Vollmer et al. (2001)
which lead to realistic gas stripping radii.

\section{Total goodness}

\begin{figure}[!ht]
  \centering
  \resizebox{8cm}{!}{\includegraphics{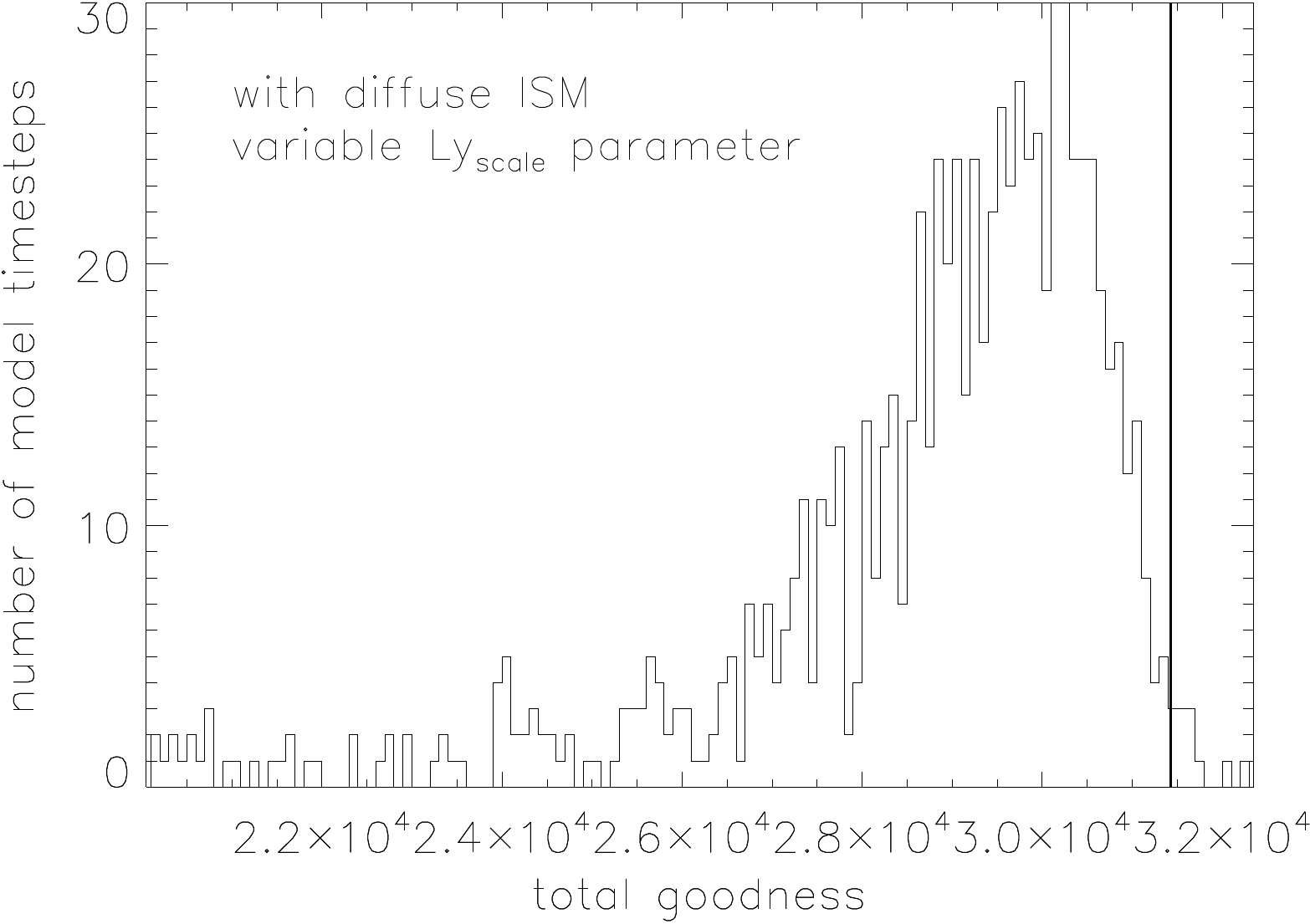}}
  \resizebox{8cm}{!}{\includegraphics{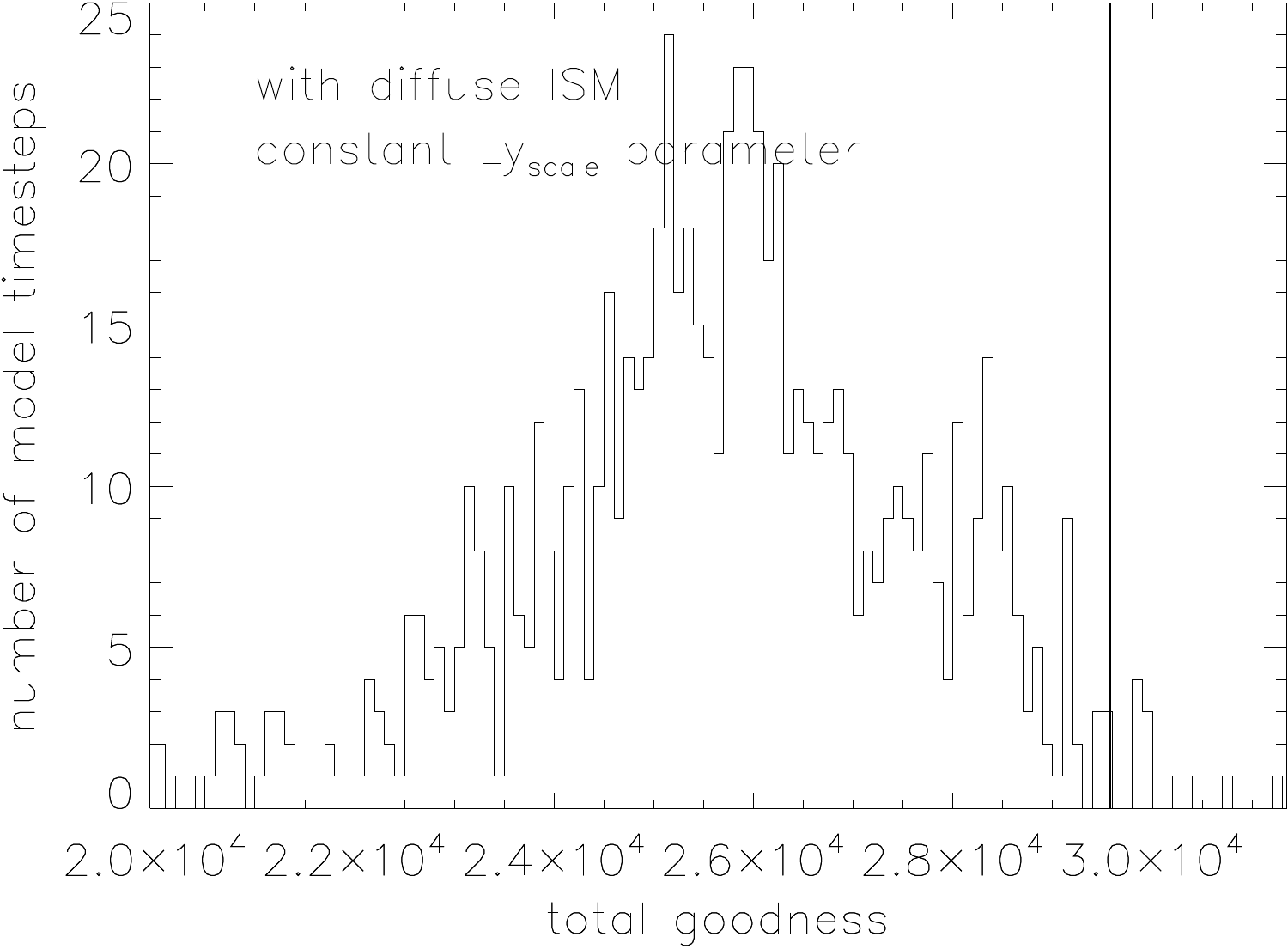}}
  \resizebox{8cm}{!}{\includegraphics{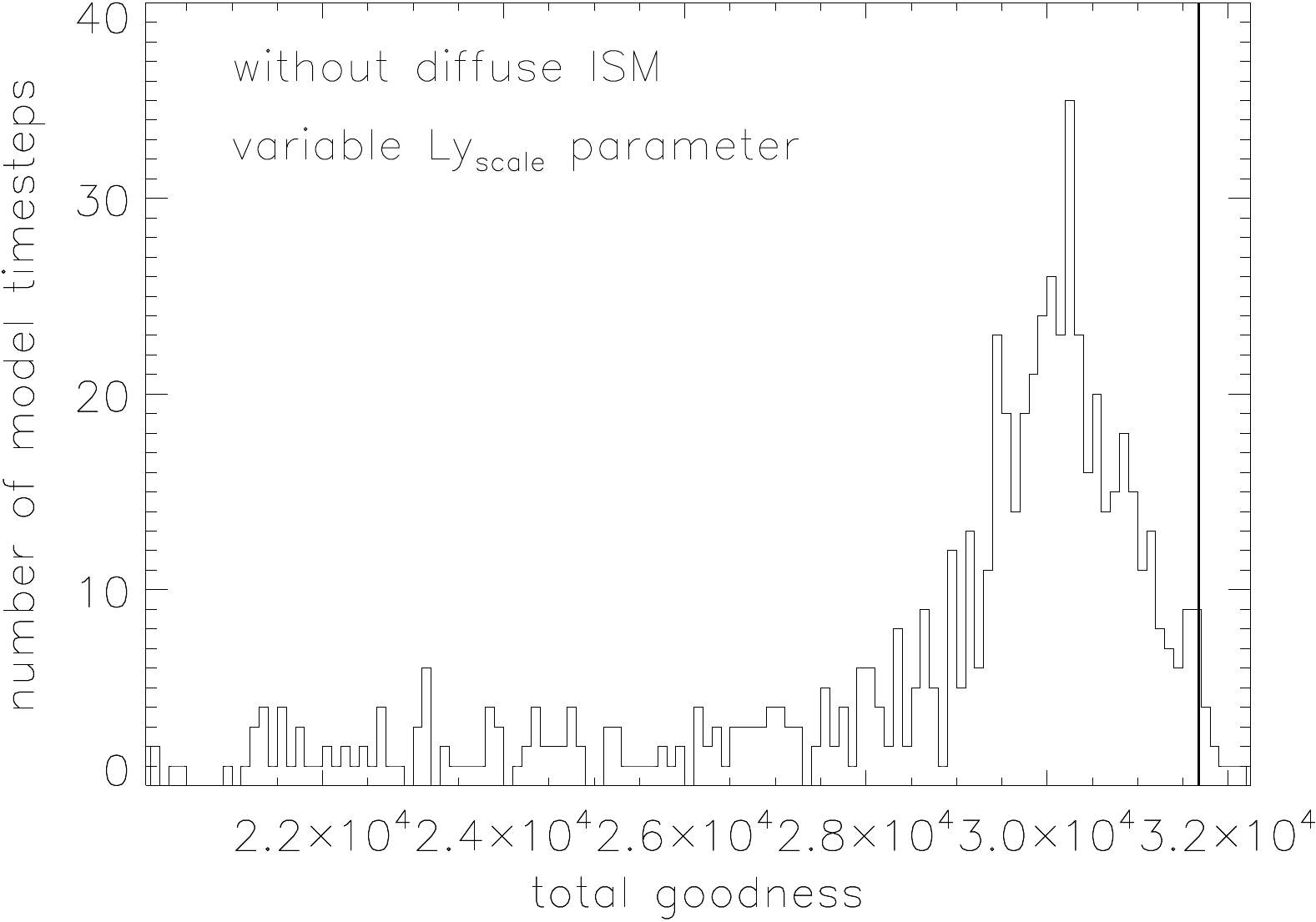}}
  \resizebox{8cm}{!}{\includegraphics{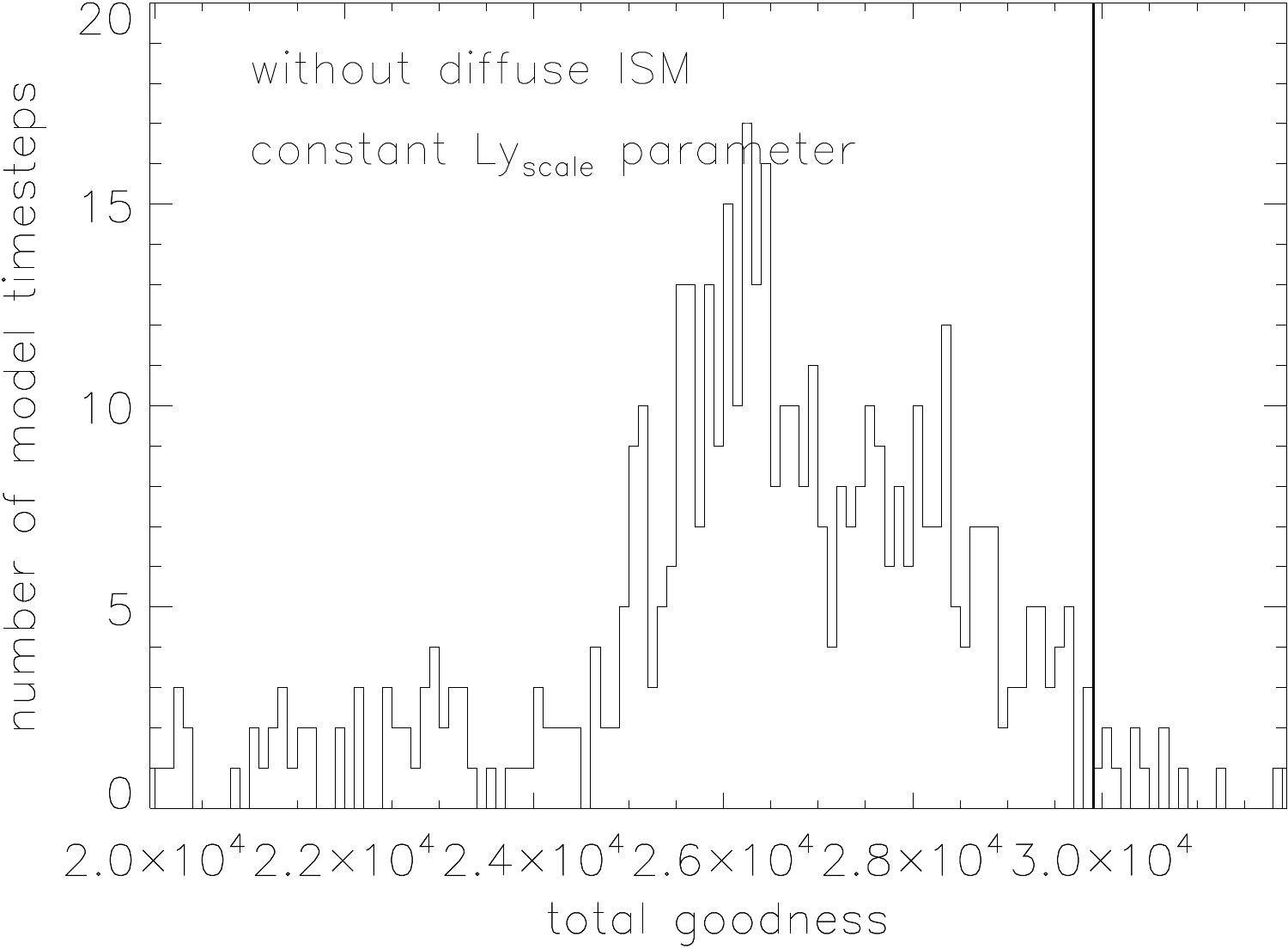}}
  \caption{Goodness distributions of the different simulations. The vertical lines delimit the $12$ models with the highest goodnesses
    (see Table~\ref{tab:goodness}).
  \label{fig:goodnessdist}}
\end{figure}

The distributions of the total goodnesses of the simulations with a variable and a constant Ly$_{\rm scale}$ parameter and
with and without diffuse gas stripping are presented in Fig.~\ref{fig:goodnessdist}. All distributions show a tail to high
goodnesses which are comprised of $\sim 20$ models.

\begin{figure*}[!ht]
  \centering
  \resizebox{\hsize}{!}{\includegraphics{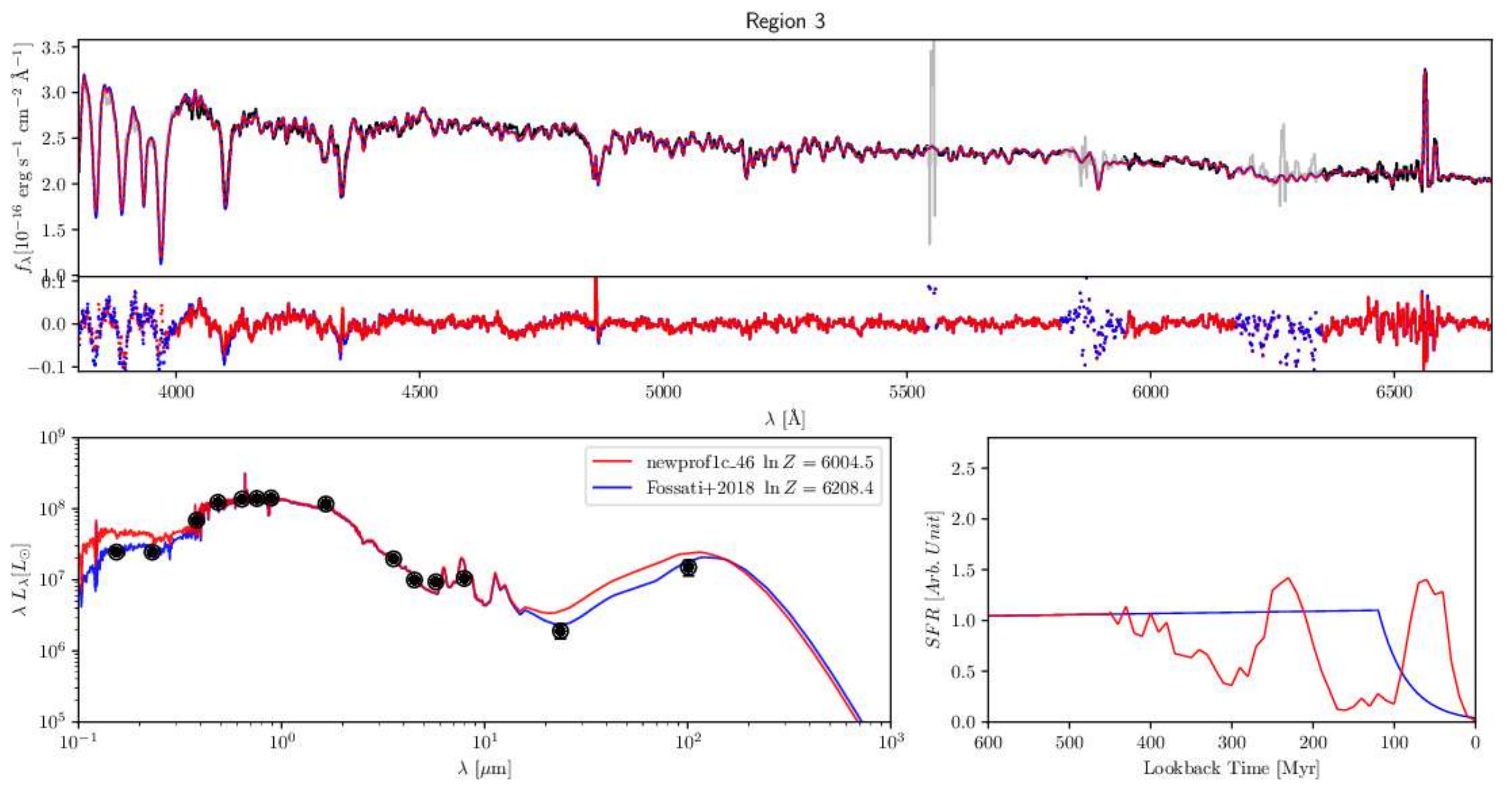}}
  \caption{Model 1cnew\_46, region~3. The total goodness is $g=6004$, $\Delta g=178$ less than the total goodness of 
    model 4new\_47 (Fig.~\ref{fig:Region3_vollmer_summary_4_47}). Upper panel: FORS2 spectrum (black), Fossati et al. model (blue), 
    model 1cnew\_46 (red). Middle panel: fit residuals. Lower left panel: observed SED (black), Fossati et al. model (blue), model 1cnew\_46 (red). 
    Lower right panel: star formation history; Fossati et al. model (blue), model 1cnew\_46 (red).
  \label{fig:goodcomp}}
\end{figure*}

To give an impression on the quality of the fits with respect to the total goodness $g$, Fig.~\ref{fig:goodcomp} shows two models
with $\Delta g = 178$. The model with the higher goodnesses reproduces the observed UV and, as a consequence, FIR fluxes significantly better.

\section{Model fitting with a constant Ly$_{\rm scale}$ scale parameter}

The model results obtained with a constant nuisance parameter Ly$_{\rm scale}$ are presented for the 
H{\sc i} and NUV emission distributions and H$\alpha$ emission distribution.
For the models with the highest total goodnesses (Table~\ref{tab:goodness}) the corresponding observed emission
distribution are less reproduced by these models compared to the models with a variable nuisance parameter.

\begin{figure*}[!ht]
  \centering
  \resizebox{16cm}{!}{\includegraphics{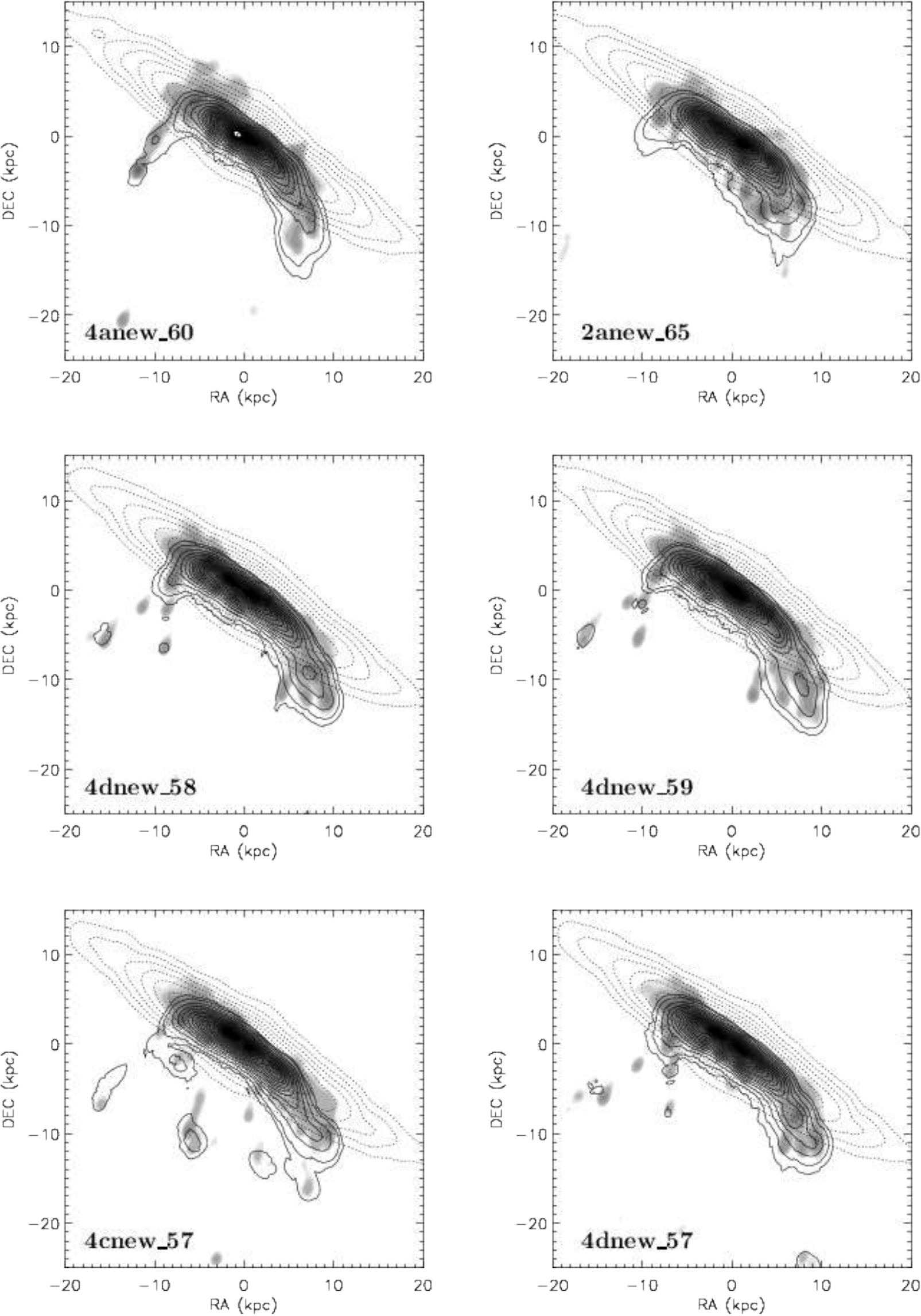}}
  \caption{NUV (grayscale), H{\sc i} (solid contours), and stellar (dotted contours) images. Models with a diffuse ISM component.
    For the spectral fitting a constant nuisance parameter Ly$_{\rm scale}$ was used.
  \label{fig:zusammen_nuv_newprofs-1a}}
\end{figure*}

\begin{figure*}[!ht]
  \centering
  \resizebox{16cm}{!}{\includegraphics{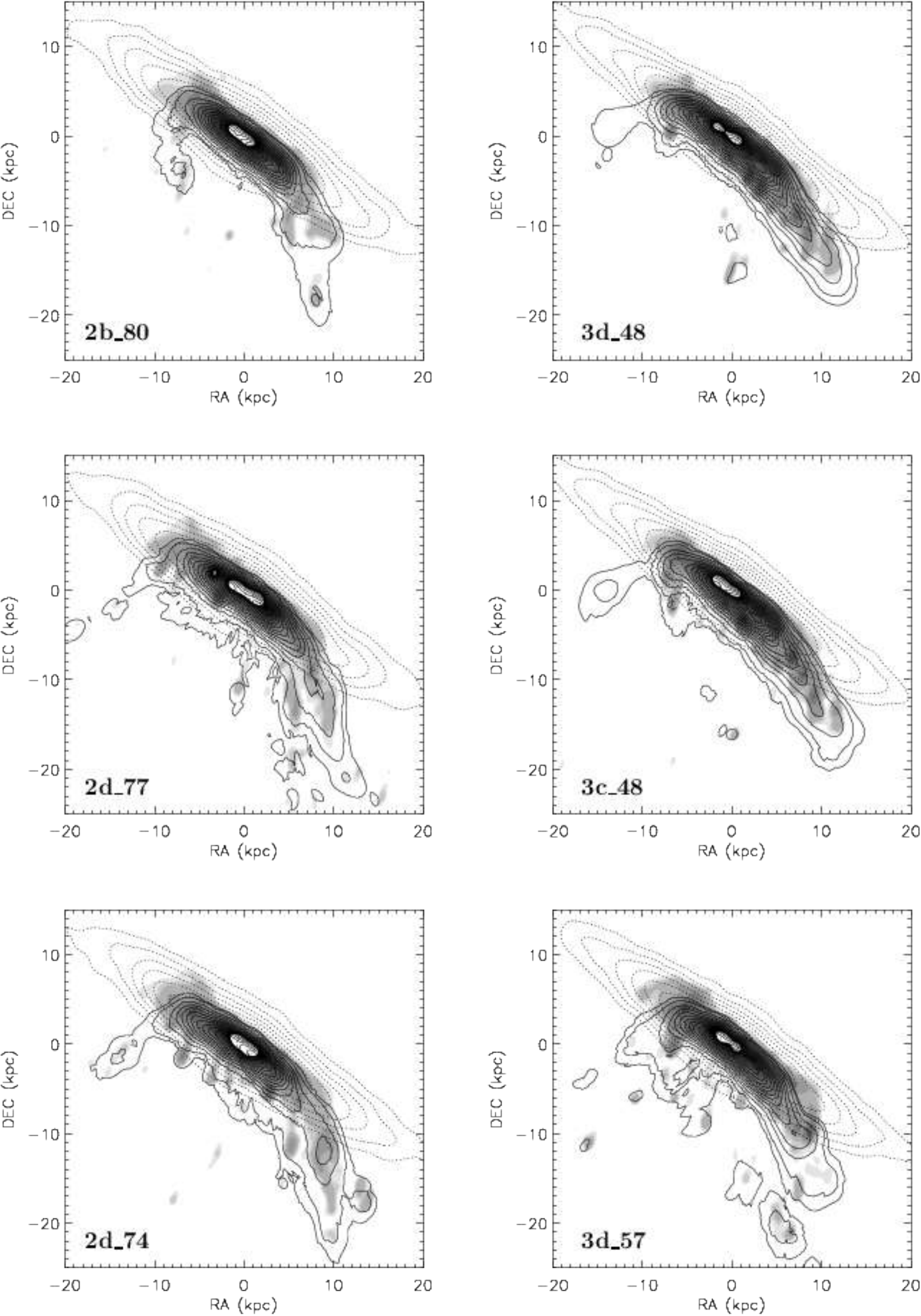}}
  \caption{NUV (grayscale), H{\sc i} (solid contours), and stellar (dotted contours) images. Models without a diffuse ISM component.
    For the spectral fitting a constant nuisance parameter Ly$_{\rm scale}$ was used.
  \label{fig:zusammen_nuv_profs-2a}}
\end{figure*}

\begin{figure*}[!ht]
  \centering
  \resizebox{16cm}{!}{\includegraphics{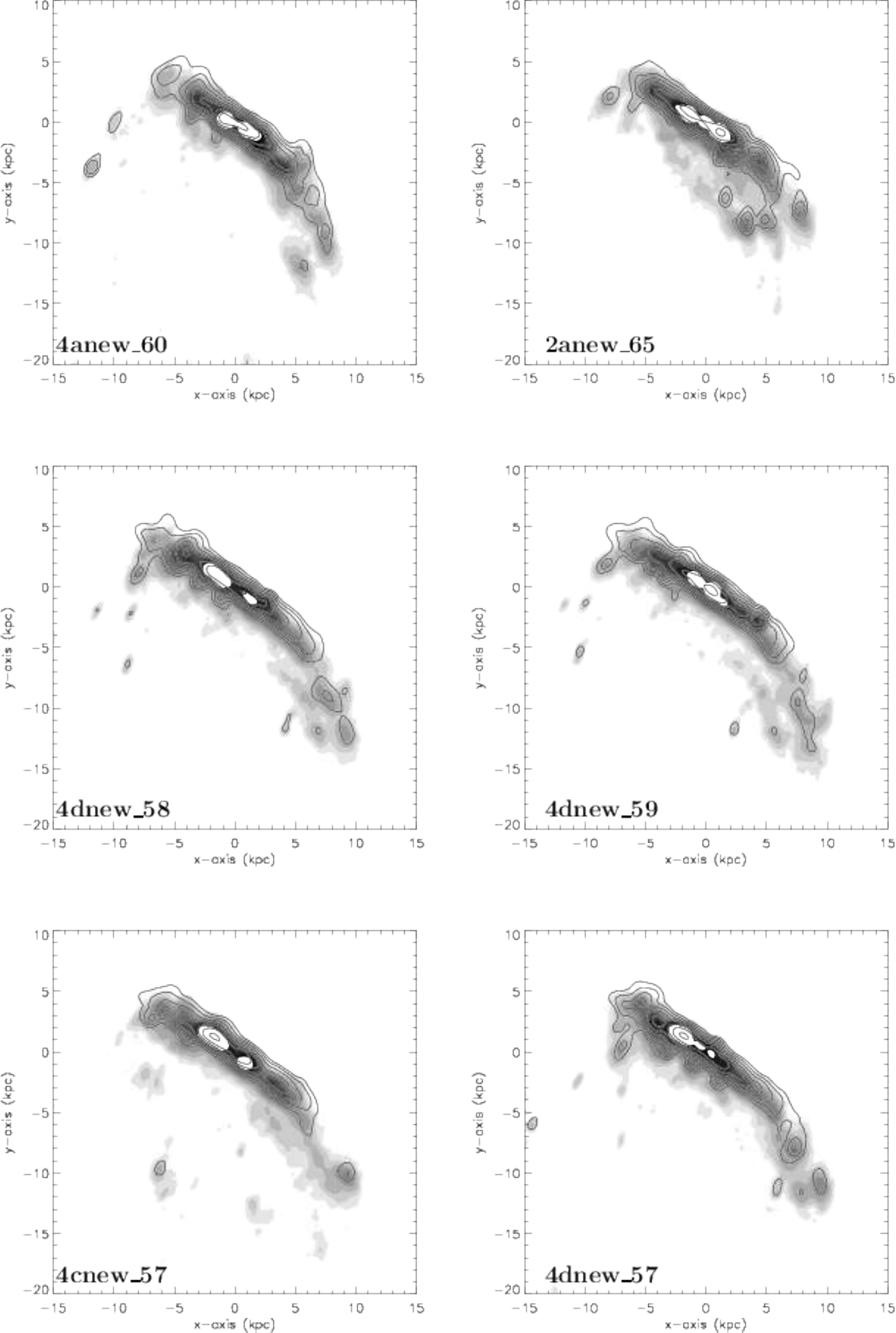}}
  \caption{H$\alpha$ (grayscale) and NUV (contours) images. Models with a diffuse ISM component.
    For the spectral fitting a constant nuisance parameter Ly$_{\rm scale}$ was used.
  \label{fig:zusammen_nuv_newprofs-3a}}
\end{figure*}

\begin{figure*}[!ht]
  \centering
  \resizebox{16cm}{!}{\includegraphics{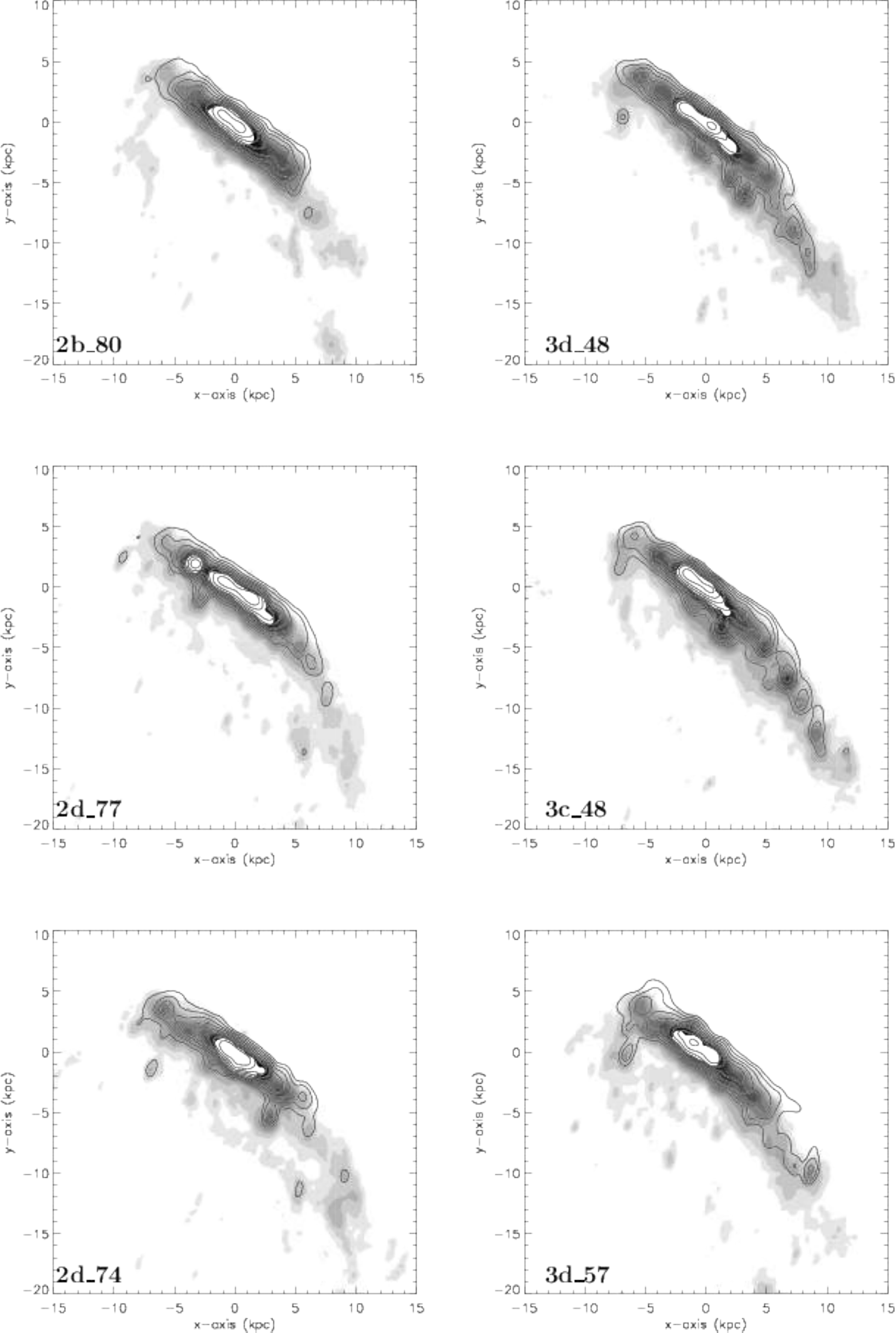}}
  \caption{H$\alpha$ (grayscale) and NUV (contours) images. Models without a diffuse ISM component.
    For the spectral fitting a constant nuisance parameter Ly$_{\rm scale}$ was used.
  \label{fig:zusammen_nuv_profs-4a}}
\end{figure*}

\end{document}